\documentclass[useAMS,usenatbib]{mn2e}
\usepackage{aas_macros}
\usepackage{graphicx}
\usepackage{deluxetable}

\title[{A New Grid of Synthetic Spectra for [WC] CSPNe}]
{A New Grid of Synthetic Spectra for the Analysis of [WC]-type Central Stars of Planetary Nebulae}

\author[Keller et al.]
{Graziela R. Keller$^{1,2}$\thanks{E-mail:
graziela@astro.iag.usp.br}, James E. Herald$^2$,
Luciana Bianchi$^2$, 
\newauthor 
Walter J. Maciel$^1$ and Ralph C. Bohlin$^3$\\
$^1$Instituto de Astronomia,
Geof\'{\i}sica e Ci\^{e}ncias Atmosf\'{e}ricas, Universidade de S\~{a}o Paulo, Cidade Universit\'{a}ria, S\~{a}o Paulo/SP, Brazil.\\
$^2$Department of Physics and
Astronomy, The Johns Hopkins University, 3400 N. Charles Street, Baltimore, MD 21218, USA.\\
$^3$Space Telescope Science Institute, 3700 San Martin Drive, Baltimore, MD 21218, USA.}

\date{Released 2011 Xxxxx XX}

\pagerange{\pageref{firstpage}--\pageref{lastpage}} \pubyear{2011}

\begin{document}

\maketitle

\begin{abstract}
We present a comprehensive grid of synthetic stellar-atmosphere spectra, suitable for
the analysis of high resolution spectra of hydrogen-deficient post-Asymptotic Giant
Branch (post-AGB) objects hotter than 50000 K, migrating along the constant luminosity
branch of the Hertzsprung-Russell diagram (HRD). The grid was calculated
with CMFGEN, a state-of-the-art stellar atmosphere code that
properly treats the stellar winds, accounting for expanding
atmospheres in non-LTE, line blanketing, soft X-rays, and wind clumping. We include many ionic
 species that have been previously
neglected. Our uniform set of models fills a niche in an
important parameter regime, i.e., high effective temperatures, high
surface gravities, and a range of mass-loss values. The grid
constitutes a general tool to facilitate determination of the
stellar parameters and line identifications and to interpret morphological changes of the stellar spectrum as
stars evolve through the central star of planetary nebula (CSPN)
phase. We show the effect of major physical
parameters on spectral lines in the far-UV, UV, and optical regimes. We analyse
UV and far-UV spectra of the central star of NGC 6905 using the grid to constrain
its physical parameters, and proceed to further explore other parameters not taken in consideration in the grid. 
This application shows that the grid can be used 
to constrain the main photospheric and wind parameters, as a first
step towards a detailed analysis. The full grid of synthetic spectra, comprising far-UV, UV, optical, and IR spectral regions, 
is available on-line.
\end{abstract}


\section{Introduction}

Evolutionary models predict that low and intermediate mass stars
(of initial masses between 1 and 8 M$_{\odot}$) leave the AGB and evolve through the constant luminosity track of the 
HRD up to very high effective temperatures
($\sim200000$ K, for CSPNe of about $0.6$ M$_{\odot}$),
then turn on to the White Dwarf (WD) cooling track, where their luminosities fade while their effective temperatures decrease 
and, finally, they end their lives as WDs \citep[see, for example,][and references therein]{Vassiliadis1994,2005ARA&A..43..435H}.
As the pos-AGB stars evolve toward higher temperatures, they
 ionize the surrounding hydrogen-rich material previously ejected by the AGB precursors,
giving rise to bright planetary nebulae (PNe). Because these objects expel
the majority of their initial mass prior to settling on the WD
cooling sequence, they are a prime source of chemically processed
material for the interstellar medium (ISM) and thus, a fundamental
ingredient of galactic chemical evolution \citep[e.g.][and references therein]{Marigo2001,Bianchi2011,Karakas2010}. In addition, their
intense UV radiation fields and fast winds during the post-AGB phase can influence the dynamics of
the ISM.

Approximately 20 per cent of the CSPNe are hydrogen deficient, a condition thought to be due to thermal pulses
that occur after the star leaves the AGB. These pulses can happen at 
different epochs of the evolution, resulting in stars with
different properties. If the thermal pulse occurs after the star has
already entered the WD cooling track, it is termed a very late thermal
pulse or VLTP \citep[from
the `born-again' scenario of ][]{1983ApJ...264..605I}. The late
thermal pulse or LTP occurs during the CSPN phase while hydrogen
burning is still ongoing. In both cases, the star returns towards the
AGB. The star can also undergo an AGB final thermal pulse or AFTP, which happens at the end of the AGB phase
\citep{2001Ap&SS.275....1B, 2001Ap&SS.275...15H}. The amount of remaining
hydrogen varies among these scenarios. A VLTP produces hydrogen-free stars \citep{2006PASP..118..183W}, while in the LTP case, the
hydrogen content at the star surface is decreased to a few per cent
by mass. If the star undergoes an AFTP instead, it is left with a
relatively high hydrogen content. A measurement of the nitrogen abundance can help 
distinguish whether a VLTP or a LTP have occurred, because the evolution of the stellar nitrogen content is different 
in each of these scenarios, resulting in an abundance of about $0.1$ per cent by mass 
in the case of a LTP event and up to a few per cent in the case of a VLTP \citep{Werner2006}. The three scenarios 
also predict somewhat different helium, carbon and oxygen abundances, therefore, measurements of these quantities 
help to determine if the star underwent a LTP, VLTP, or a AFTP 
\citep[see, for example,][and references therein]{2001Ap&SS.275....1B}. 

The H-deficient CSPNe are commonly divided into three main classes,
which are thought to constitute an evolutionary sequence. These are:
[WC], showing spectra very similar to those of Population I
Wolf-Rayet (WR) stars, with strong carbon and helium emission lines;
the PG1159-type, occupying the region at the top of the WD cooling track in
the HRD, and characterized by absorption lines of highly ionized He,
C and O, besides UV wind lines much weaker than the ones
seen in [WC] stars; [WC]-PG1159, that are
believed to be transition objects between the two other classes. The
[WC] stars are further divided into early ([WCE]) and late-type
([WCL]) objects, showing lines from ions of higher and lower ionization stages,
respectively.
The evolution of the H-deficient CSPNe might
proceed in the following way: AGB $\rightarrow$ [WCL] $\rightarrow$
[WCE] $\rightarrow$ [WC]-PG1159 $\rightarrow$ PG1159 $\rightarrow$
WD
\citep{1991A&A...247..476W, 2000A&A...362.1008G, 2001A&A...367..983P}.
According to this scenario, the [WC] CSPNe would evolve from the AGB
at an almost constant luminosity, towards higher temperatures.
As the stars evolve, their radii decrease until the nuclear burning
ceases and the stars progress quickly as
PG1159 onto the WD cooling track, while luminosity and mass-loss
decrease and the wind reaches very high terminal velocities.

A solid determination of the photospheric and wind parameters, as
well as of the chemical composition of CSPNe, is crucial to address
questions concerning the stellar evolution, the connection among the
classes of H-deficient central stars, the wind driving mechanism and
its effect on the ISM, and to understand the surrounding nebulae.
Modelling the observed spectrum with stellar atmosphere codes is the
most direct method for the derivation of stellar parameters, and
constructing model grids covering the relevant range of parameters
is the best way to accomplish a systematic analysis, and constrain the major parameters.

The majority of CSPNe spectral analyses in the past have been carried out in
an ad-hoc fashion, that is, sets of models being calculated for each observed CSPN. These models vary in
sophistication, with more or less ions being included in the calculations, allowing or not for wind clumping, 
etc \citep[see, for example,][]{1996A&A...312..167L,1997A&A...320...91K,1998A&A...330..265L,1998A&A...330.1041K,
2001MNRAS.328..527D,2004ApJ...609..378H,2005ApJ...627..424H,2007ApJ...654.1068M}. To date, only a few Galactic CSPNe are well studied.
With the GALEX surveys \citep{2005ApJ...619L...1M, 2009Ap&SS.320...11B},
thousands of Milky Way post-AGB objects, including CSPNe and WDs, are being
measured photometrically in the UV \citep{2005ApJ...619L..27B, 2007ApJS..173..659B, Bianchi2011, Bianchi2011b}.
Therefore, a general tool to facilitate determination of the stellar
parameters will allow researchers to exploit the many-fold
increase of samples of these objects and to finally clarify the final stages of
stellar evolution.  We constructed a large grid of models covering the CSPN parameter space, 
which can be used to quickly constrain the main parameters and limit the need of `ad-hoc' computed models in each case. 

Several grids of stellar atmosphere models have been calculated to
help analyse data for massive early type stars. Examples include:
for main-sequence, giants, and supergiant stars, the ``Kurucz''  (in LTE and hydrostatic equilibrium)  models
\citep{1991sabc.conf..441K}; for WN WR stars, the non-LTE grids of
\citet{2004A&A...427..697H}, which account for winds; for O-stars, the non-LTE, hydrostatic equilibrium 
models TLUSTY OSTAR 2002 \citep{2003ApJS..146..417L}, the 
non-LTE, accounting for expanding atmospheres, WM-basic grids of \citet{BianchiandGarcia} and, 
in a more limited fashion, \citet{2001A&A...375..161P}; for B-stars, the non-LTE, hydrostatic 
equilibrium TLUSTY BSTAR 2006 models \citep{2007ApJS..169...83L}. For CSPNe, \citet{1987MNRAS.228..759C}
computed a grid of non-LTE, plane-parallel model atmospheres, in
hydrostatic equilibrium, with H and He opacity sources. Also, \citet{Rauch2003} 
presented a grid of non-LTE, plane-parallel, line blanketed models calculated in hydrostatic equilibrium, useful for the analysis of H-rich CSPNe and their PNe. 
Only relatively recent advances in both computing power and stellar
atmosphere codes made it possible to generate models that adequately reproduce most of the 
wind features in the spectra of CSPNe with winds. As a
consequence, grids of non-LTE models, accounting for line blanketed expanding 
atmospheres and clumping, spanning this important phase of 
post-AGB evolution do not yet exist. The grid presented here is the first step towards filling this gap.

In this paper, we present a grid of synthetic spectra for the analysis of
spectroscopic data from [WC] CSPNe hotter than 50000 K. The paper
is organized as follows. Section \ref{sec:thecode}
describes the stellar atmosphere code. We describe the grid in
section \ref{sec:thegrid} and present illustrative results in section
\ref{sec:spectraldiagnostics}. In section \ref{sec:ObservedSpectra}, we apply the
grid to derive physical parameters of NGC 6905 in order to
illustrate the use of the model grid. The paper finishes
with conclusions in section \ref{sec:conclusions}.

\section{The code}
\label{sec:thecode}

[WC]-type CSPNe are very hot, with temperatures ranging from $\sim$ $20000$ to $200000$ K
and extended atmospheres, which expand reaching supersonic velocities. Line blanketing affects their atmospheric structure and
emergent spectra. Not all CSPNe have dense winds, since the presence of intense mass loss, if driven by radiation pressure, is related to the 
closeness of the star to the Eddington limit \citep{Pauldrach1988}. For example, as the PG1159 central stars approach the WD cooling track, their winds fade and their 
spectral lines are less and less conspicuous \citep[see examples in][]{Werner2010}. The winds of 
CSPNe are believed to be radiatively driven
and as such, subjected to instabilities, which are theoretically predicted to lead to the formation of
clumping and shocks emitting soft X-rays \citep{Owocki1988}. Such effects were
quantified observationally in the winds of massive O-type stars
\citep{Bianchi2002, Garcia2004a, Bianchi2009}. The grid of
models was calculated using the non-LTE radiative
transfer code CMFGEN
\citep{1998ApJ...496..407H, 2005ApJ...627..424H}, which solves
iteratively for the level populations and radiation field, assuming
radiative equilibrium, in a spherically symmetric expanding outflow.
It accounts for wind clumping, soft X-rays, and line blanketing through a ``super-level"
approach. CMFGEN was originally developed to
model the winds of massive WR stars and has been adapted
for objects with weaker winds such as O-stars and CSPNe \citep[as
described by][]{2003ApJ...588.1039H, 2005ApJ...627..424H}. Details about the code
are given in the references above. Here, we briefly describe the more important aspects.

The radiative luminosity is taken to be constant throughout the wind
and is given by
\begin{equation}
L=4 \pi R_{\ast}^{2} \sigma T_{\ast}^{4},
\end{equation}
where $R_{\ast}$ and $T_{\ast}$ are defined at a Rosseland mean
optical depth of 20.

CMFGEN does not solve the dynamical equations of the wind, requiring
the mass-loss rate and the velocity law to be supplied. Here,
we have adopted a standard velocity law, where
\begin{equation}
v(r)=v_{\infty}\left(1-\frac{r_{0}}{r}\right)^{\beta},
\end{equation}
with $\beta=1$ which is the value usually adopted in modelling these
objects. $r_{0}$ is roughly equal to $R_{\ast}$.

Currently, CMFGEN requires the density structure to be given.
The procedure adopted here was to attach the wind velocity law to the underlying 
hydrostatic structure of the star \citep[as described in][]{Hillier2001}, which we obtain using the
TLUSTY code \citep{1995ApJ...439..875H}. TLUSTY computes 
plane-parallel, non-LTE atmospheres in hydrostatic equilibrium and
requires surface gravity, temperature, and abundances as inputs.

Instabilities in the stellar winds of hot stars are thought to generate density inhomogeneities, stochastically distributed
throughout the wind, which are usually referred to as clumping
\citep[see, for example,][]{2008cihw.conf...17M}. In CMFGEN, the radiative
transfer in an inhomogeneous medium is
implemented assuming that the clumps are small compared to the mean free
path of the photons and that the interclumping medium is void (this treatment of clumping
is sometimes referred to as microclumping approximation).
The degree of clumping in the wind is parametrized by the clumping filling
factor $\textit{f}$, with the density inside the clumps being a factor $\textit{f}^{-1}$
higher than the density of the homogeneous wind model with same
mass-loss rate.
Since, according to this approximation, there is no material in between the clumps, the statistical
equations are solved for the intraclump medium. From the ion-level populations obtained, the non-LTE opacities and emissivities for the clump
material are calculated. The transfer equation is then solved
substituting the homogeneous-wind opacities and emissivities with the ones
calculated for the intraclump material multiplied by the filling
factor.

The main effect of treating clumps in the above
approximation is that it reduces the empirical mass-loss rate by a factor
$\sqrt{\textit{f}^{-1}}$ when using diagnostics that depend on the
square of the density, such as emission lines and radio thermal
emission. There is evidence of clumping in the winds of
[WC] stars, but the quantification of the clumping factor and its radial variation 
is made difficult by the weaker electron
scattering wings of intense emission lines of these stars in
comparison with their massive WR stars counterparts. Despite
that, \citet{2008cihw.conf..251T}, based on the electron
scattering wings of emission lines of three early-type [WC] stars, NGC 6751, NGC 5189,
and NGC 1501, found evidence for clumping with $\textit{f} > 0.25$. Further evidence
for the presence of inhomogeneities in these winds comes from the work of
\citet{2000A&A...364..597G, 2001A&A...370..513G}, who interpreted
moving features seen on the top of emission lines of the [WC] NGC 40
and BD +30 3639 as larger scale outflowing blobs. CMFGEN accounts for clumping
using an exponential law,
\begin{equation}
\textit{f}(v) = \textit{f}_{\infty} + (1 - \textit{f}_{\infty})\exp{(-v(r)/v_{\mathrm{clump}})},
\label{f}
\end{equation}
in which the radiative instabilities are supposed to be damped at
low wind velocities and, as a consequence, the wind is assumed smooth until a
certain wind velocity, $v_{\mathrm{clump}}$, is achieved. Beyond this velocity, the degree of
clumping increases outwards until the clumping filling factor finally reaches its
terminal value, $\textit{f}_{\infty}$, which, in this work, is assumed to be $0.1$. In equation
\ref{f}, $\textit{f}(v)$ and $v(r)$ are the clumping filling factor and the
velocity of the wind at a given radius, respectively. Here, $v_{\mathrm{clump}}$
was taken to be $200$ km s$^{-1}$.

CMFGEN accounts for line blanketing, which alters the ionization and
temperature structure of the model
atmosphere. In particular, the opacity by, e.g., iron lines redistributes the
UV radiation field. Neglecting it would lead to an overestimation
of the ionization of the atmosphere. The back warming and
surface cooling effects of the line blanketing allow lower ions in
the outer layers to coexist with the higher ones from the deeper
layers. Inclusion, in the models, of this phenomenon was noticed to better reproduce the spectra
of massive WR stars \citep{2004A&A...427..697H}. In CMFGEN,
the inclusion of the line blanketing is done through the
superlevel approximation, in which several energy levels are grouped
into a smaller number of superlevels and all the levels within a
superlevel share the same departure coefficient.

CMFGEN can also include, in the ionization calculations, the effect of soft X-rays, which would be
created in shocks distributed throughout the wind. These shocks are
believed to originate from the instability of the line driving
mechanism. X-rays may considerably alter the ionization structure in the
atmosphere, since they produce higher ionization stages of certain ions as, for example, O VI  and N V 
\citep[see, for example,][and references therein]{Bianchi2002, Garcia2004a, Martins2005, Bianchi2007a}.
In CMFGEN, the X-rays are taken into consideration
in the calculations, by specifying the temperature and velocity of the shocks.
Also, a volume filling factor for the
distribution of X-rays sources has to be given in order to set the
level of the emission. The X-ray emissivity is taken from the X-ray
code of \citet{1977ApJS...35..419R}, which at
present is available only for solar abundances. In section \ref{subsec:xrays},
we describe the effect on the model spectra, of including soft X-rays in the calculations.

In order to calculate the detailed emerging spectra we adopted a
microturbulence velocity which varies with depth such that
\begin{equation}
v_{\mathrm{turb}}(r)=v_{\mathrm{min}}+(v_{\mathrm{max}}-v_{\mathrm{min}})\frac{v(r)}{v_{\infty}},
\end{equation}
where $v_{\mathrm{min}}$ and $v_{\mathrm{max}}$ are the minimum and maximum
microturbulence velocities. We assumed $v_{\mathrm{min}}=10$ km s$^{-1}$ and
$v_{\mathrm{max}}=50$ km s$^{-1}$ for all the models. The models resolution varies across the wavelength range, with lines being better 
sampled than continuum regions. As an example, in the model B150.M70.v2500 (see Table \ref{M06}), 
the central regions of the emission components of the O VI $\lambda \lambda$ $1031.9$, $1037.6$ $\mathrm{\AA}$ and C IV $\lambda \lambda$ $1548.2$, $1550.8$
$\mathrm{\AA}$ lines are sampled by 81 and 46 flux points in an 1 $\mathrm{\AA}$ interval, respectively.

\section{The grid}
\label{sec:thegrid}

\begin{figure}
\includegraphics[angle=90,scale=0.40]{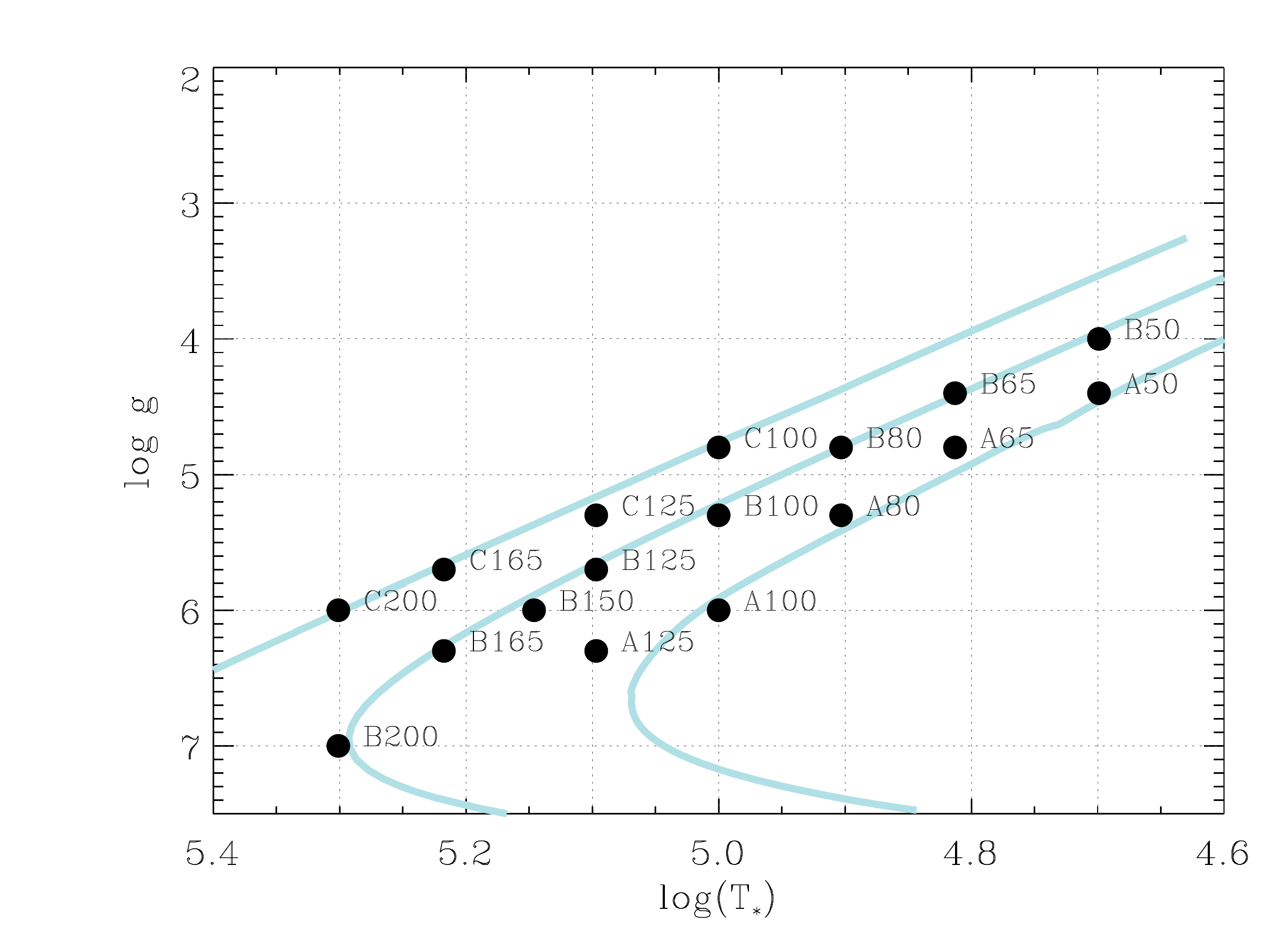}
\caption{Grid points in the $\log T_{\ast}$-$\log g$ plane. The evolutionary
tracks of \citet{2006A&A...454..845M} are also shown (continuous lines) for CSPNe with 
0.5, 0.6, and 0.9 M$_{\odot}$, from right to
left. Each point shown is identified by a label which corresponds to a group of
models with the same temperature, surface gravity, and radius, but with a range of
mass-loss rates and wind's terminal velocities, as can be seen in Tables
\ref{M05}, \ref{M06} and \ref{M08}.} \label{HRgrid}
\end{figure}

The grid presented here is intended to comprehensively cover the parameter space
of [WC] CSPNe. The combination of parameters chosen for each of the grid models approximately follow
the evolutionary calculations of
\citet{2006A&A...454..845M} for CSPNe with final masses of $0.5$, $0.6$
and $0.9$ M$_{\odot}$ (which correspond to initial masses of $1.0$, $3.1$, and
$5.5$ M$_{\odot}$, respectively),
as shown in Fig. \ref{HRgrid}. The stellar parameters $L$, $R_{\ast}$, $T_{\ast}$, $\log
g$, and $M_{\ast}$ of the grid models are inside the range of values
predicted for these stars. A chosen model temperature corresponds to a different model radius for each of the evolutionary tracks used, since 
the luminosity varies with mass on the constant luminosity track in the HRD. Also, for the models within one track, every temperature 
corresponds to a different radius in order to keep an almost constant luminosity. 
         
\begin{figure}
\includegraphics[angle=90,scale=0.4]{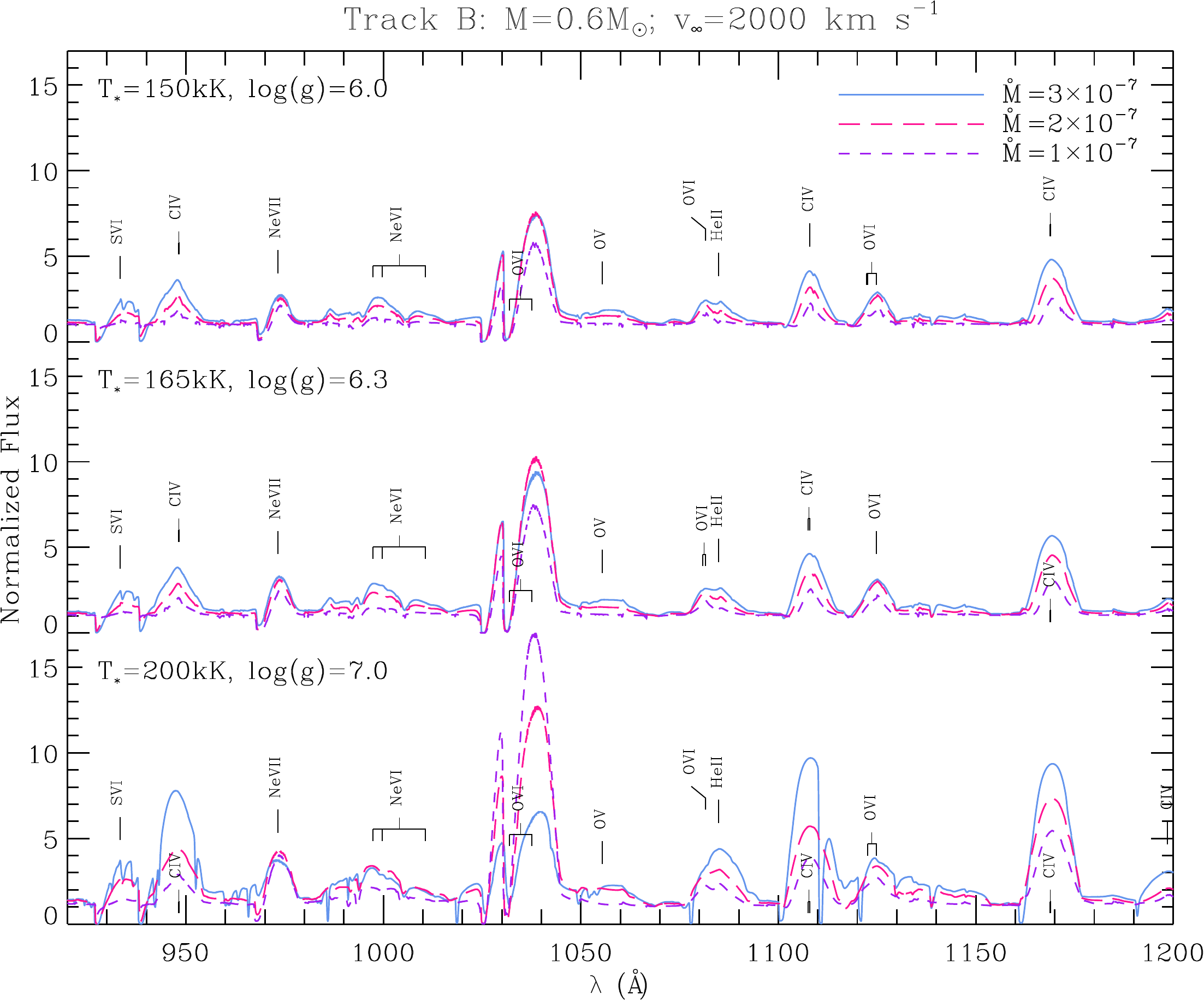}
\includegraphics[angle=90,scale=0.4]{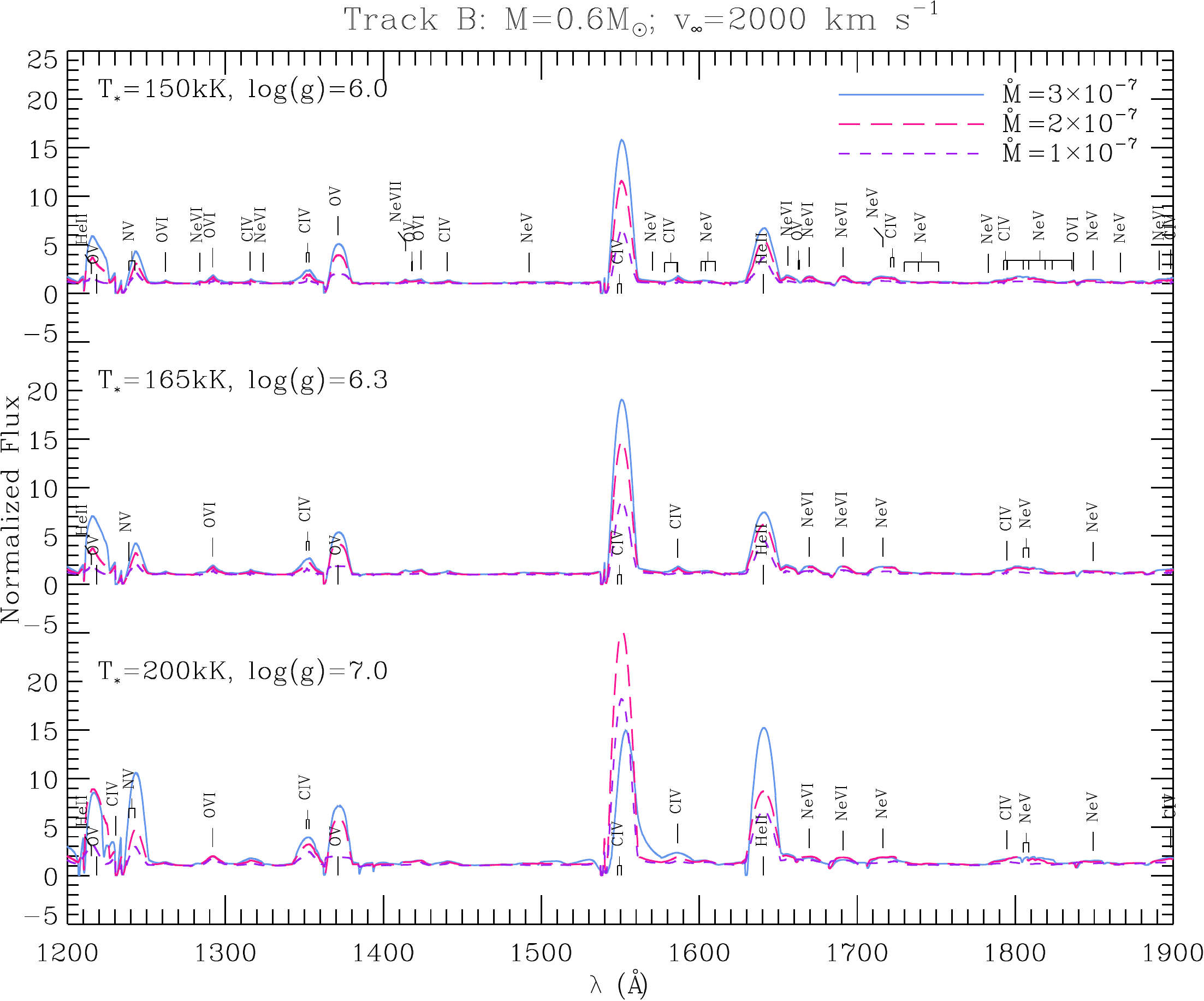}
\includegraphics[angle=90,scale=0.4]{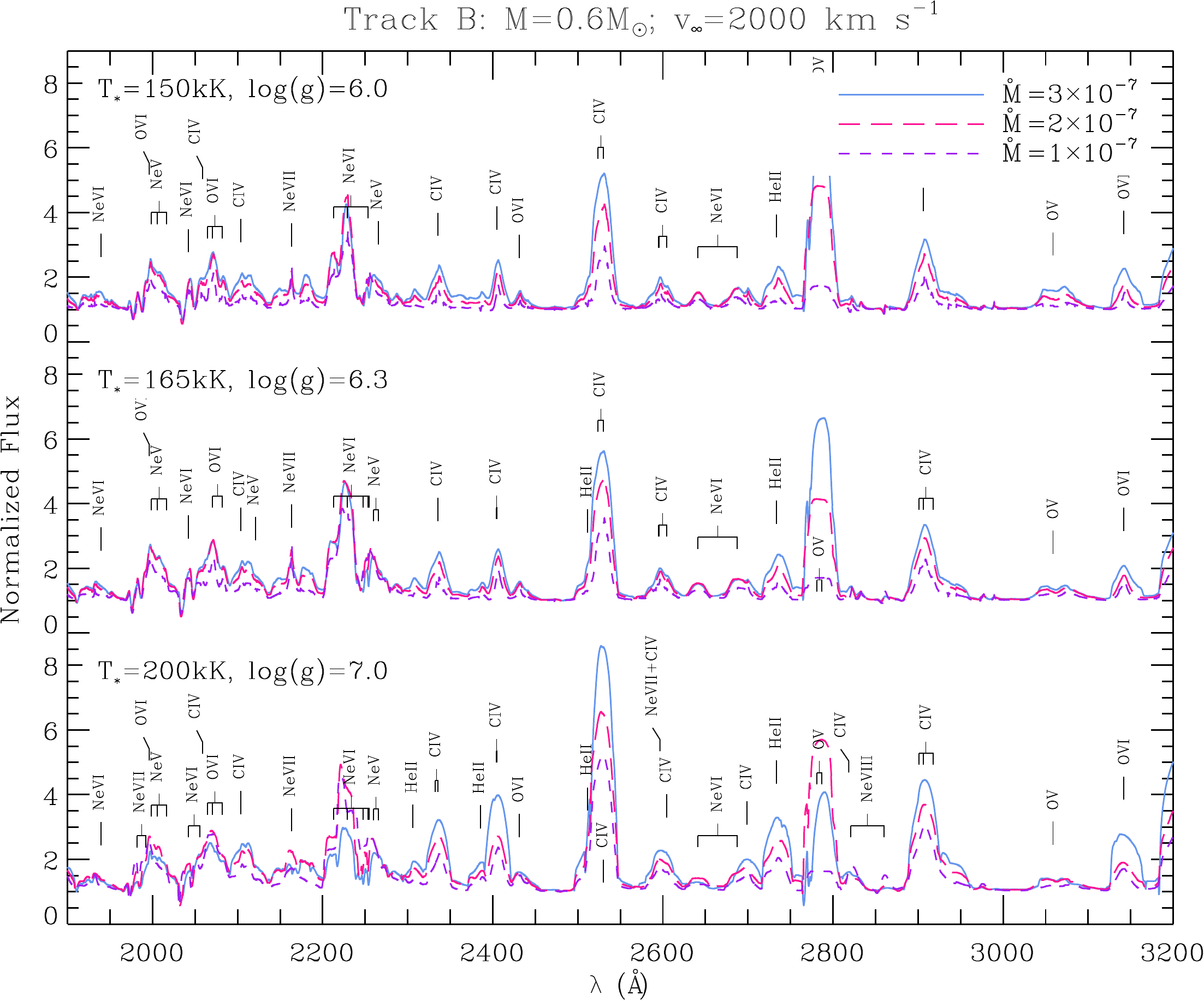}
\includegraphics[angle=90,scale=0.4]{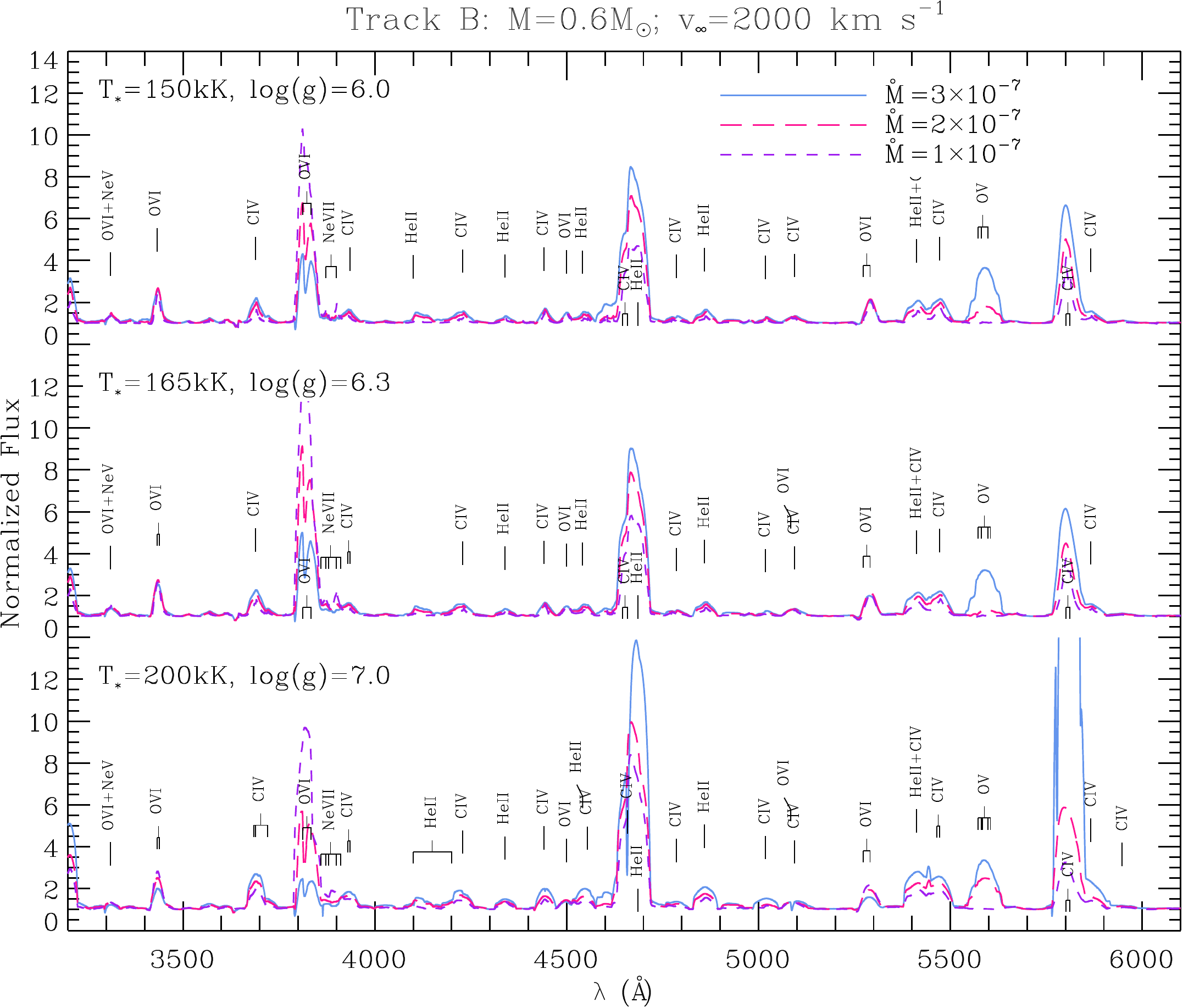}
\caption{Examples of plots comparing different grid models, from the far-UV to the optical range. Here and in 
subsequent figures, mass-loss rates are given in M$_{\odot}$ yr$^{-1}$. Similar plots for all grid models are available on-line.} \label{webplot}
\end{figure}

We have selected eight stellar temperature values ranging from $50000$ to $200000$ K and nine values of $\log
g$ between $4.0$ and $7.0$. For each combination of $\log g$ and $T_{\ast}$
seen in Fig. \ref{HRgrid}, models with different mass-loss rates and wind's terminal velocities were computed. The wind's terminal velocities and the mass-loss rates
were chosen to cover the typical values of these parameters found in the literature, taking into account that our models assume a 
clumped wind whereas some mass-loss rates were derived in the literature assuming a smooth wind. Additional models differing only in neon abundance were
also computed (discussed in section \ref{subsec:Neabundance}). The grid adds
up to a total of 199 models.
The list of models is given in Tables \ref{M05}, \ref{M06}, and
\ref{M08}. The grid is available on-line at
http://dolomiti.pha.jhu.edu/planetarynebulae.html. There, the user will have
access to the synthetic spectra and related documentation, along with plots
comparing the different models such as shown in Fig. \ref{webplot}.

By following three tracks for different stellar mass values, we 
covered different combinations of temperature and surface 
gravity/stellar radius. The [WC] CSPNe with temperatures 
above 50000 K, are characterized by the lack of strong absorption 
lines not affected by wind emission, which prevents the determination 
of their surface gravity through spectral analysis. When the distance 
is known, the derived stellar parameters can place the object on a 
specific evolutionary track, from which log g can be inferred. 
Otherwise, the analysis can only constrain the transformed radius 
value and mass and log g are not uniquely constrained. The transformed radius is defined as
\begin{equation}
R_{\mathrm{t}}=R_{\ast}\left[\frac{v_{\infty}/2500\ \mathrm{km}\ \mathrm{s}^{-1}}{\dot{M}/10^{-4}
\ \mathrm{M}_{\odot}\mathrm{yr}^{-1}}\right]^{2/3},
\end{equation}
and is a measure of how dense the wind is (smaller values of $R_{\mathrm{t}}$ translate into denser winds).
Models with the same temperature, differing only in mass-loss rate and in stellar radius,
but with the same transformed radius and the same wind's terminal velocity, are known
to produce very similar wind features in the UV range
\citep{1989A&A...210..236S, 1993A&A...274..397H}. 

\begin{figure}
\includegraphics[scale=0.435]{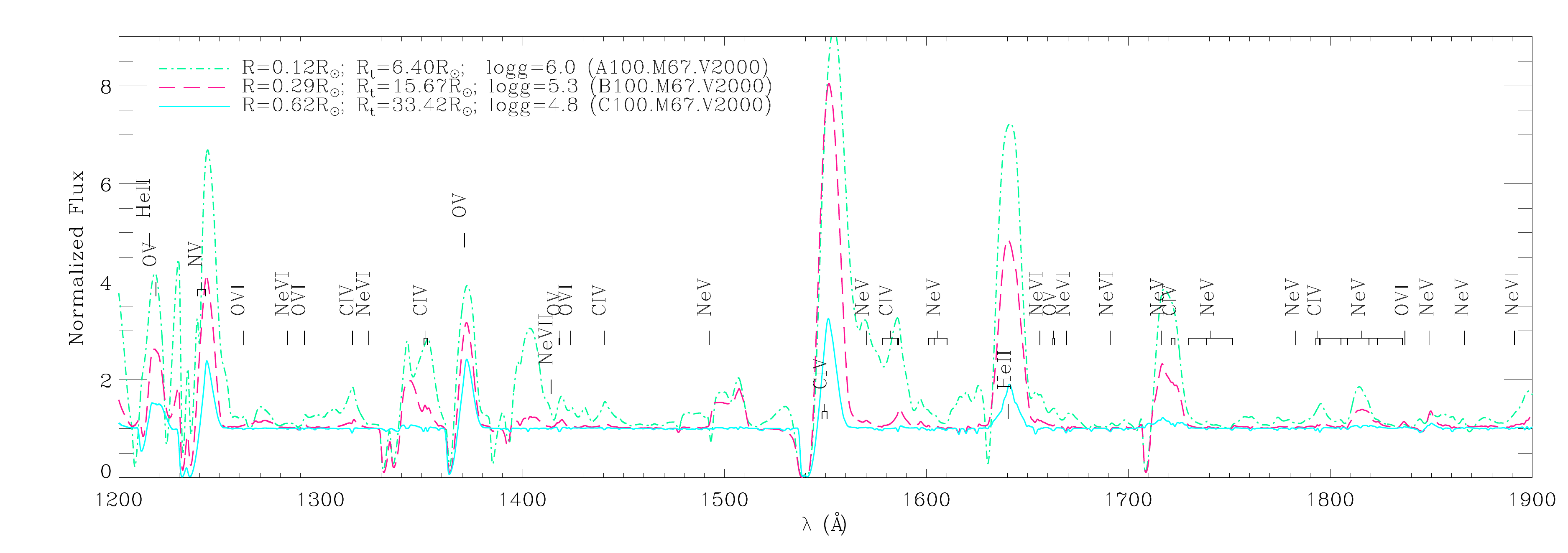}
\caption{Models A100.M67.V2000, B100.M67.V2000, and C100.M67.V2000, with $T=100$ kK, $\dot{M}=2\times10^{-7}$ M$_{\odot}$ yr$^{-1}$ 
and $v_{\infty}=2000$ km s$^{-1}$, differ in 
$R$, $R_{\mathrm{t}}$, and $\log g$. Wind features vary greatly due to the different wind densities.} \label{difftracks}
\end{figure}

Models having the same temperature, mass-loss rate and wind's terminal velocity, but from different mass tracks, 
differ in luminosity, radius, $\log g$, and transformed radius. Different $\log g$ result in differences on the photospheric lines. 
Differing $R_{\mathrm{t}}$ affects the wind lines. Fig. \ref{difftracks} shows models A100.M67.V2000, 
B100.M67.V2000, and C100.M67.V2000 that, despite having the same mass loss and wind's terminal velocity, have very different wind features 
as a result of their different transformed radii. 
These three models also differ in $\log g$, but the lack of spectral features free from wind effects makes it impossible to distinguish among 
different $\log g$ values. Surface gravity, however, influences the emission line profiles by affecting the underlying 
photospheric ones.

It is believed that [WCL] and [WCE] stars form an evolutionary
sequence due to their locations in the $\log T_{\mathrm{eff}}$-$\log g$ diagram. 
Carbon and helium are the main constituents of their atmospheres.
Spectral analyses have, however, suggested different C:He ratios
between the two subtypes
\citep{1997IAUS..180..114K, 1997A&A...320...91K}.
\citet{2001MNRAS.328..527D}, \citet{2003IAUS..209..243C}, and \citet{2007ApJ...654.1068M},
on the other hand, find no systematic
discrepancies between the C:He ratios between [WCL] and [WCE] stars.
Thus, since it is not yet clear if the two groups of [WC] CSPNe have
different carbon and helium abundance patterns, we adopted a
constant C:He ratio for all the models in the grid. Typical measured
values for elemental abundances are (by mass): X$_{\mathrm{He}}$=0.33-0.80,
X$_{\mathrm{C}}$=0.15-0.50 and X$_{\mathrm{O}}$=0.06-0.17 \citep{2001Ap&SS.275...27W}.
Here, we adopted a C:He ratio of $0.45:0.43$ and an oxygen abundance
of X$_{\mathrm{O}}$=0.08 by mass. Nitrogen abundances in these objects typically range from
undetectable to $\sim2$ per cent by mass, and we have adopted X$_{\mathrm{N}}$=0.01
\citep{1997A&A...320...91K, 1998A&A...330..265L}. We assumed a neon
abundance of $2$ per cent by mass, i.e., higher than
the solar value by a factor of 11.5, since strong overabundances of this
order have been reported for [WC] stars
\citep{1998A&A...330..265L,2005ApJ...627..424H}. Iron is expected to
be depleted through s-process \citep{2003IAUS..209...85H}, as was
indeed observed by \citet{2004A&A...413..329S}. Thus, the grid models
have an iron abundance a factor of $100$ lower than the solar value.
Solar abundances were adopted for all other elements present in the
models. 

The model calculations include many ionic species that have been previously
neglected. The elements and ionic
species considered in each model of the grid, along with the number
of levels and superlevels used, can be found in the on-line documentation.
The ionic species included in each model vary, since in many cases their number was limited to keep the models
within a workable size, or due to unavailability of
precise atomic data. The hotter models were calculated first and, as the
temperature was decreased for the calculation of the cooler models,
ionic species were introduced as needed based on the analysis of the ionization fractions.
Table \ref{GridIons} shows the species considered in the calculation of the grid models. 

\begin{figure}
\includegraphics[scale=0.435]{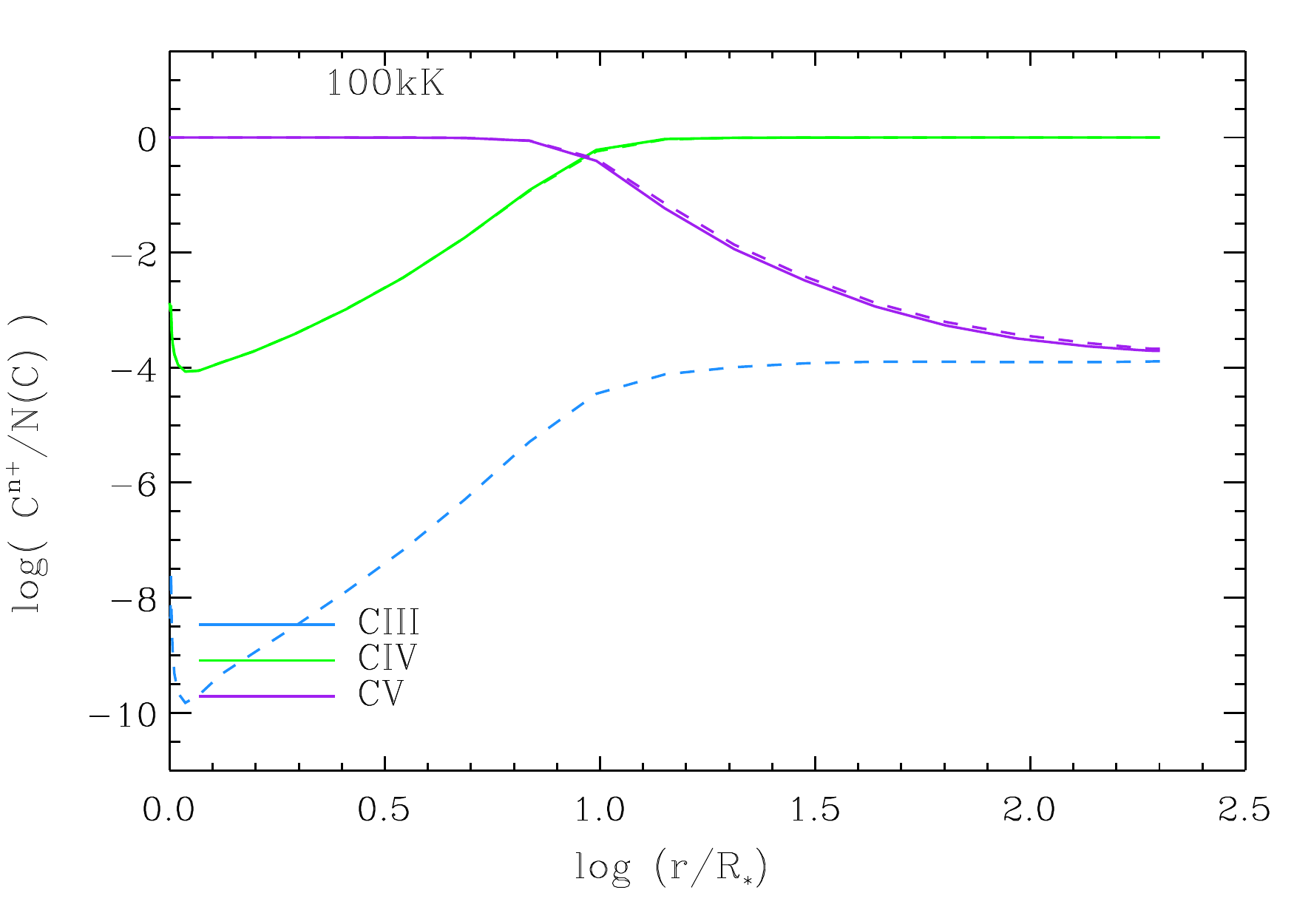}
\caption{The C ionization fractions of two T$_{\ast}$=100000 K models, with identical  stellar 
parameters, and the C III ion included (dashed lines) or not included (solid lines) in the calculations.}\label{CIIIC}
\end{figure}

As a general rule, we considered it not necessary to include a lower ionization stage when  
the ionization fraction of the lowest included ionization stage was $\leq10^{-2}$. 
Tests were performed on representative models in order to assess the necessity of including lower ionization stages.
As an example of the procedure, we give here more details about the C III, C IV, and C V ions. 
All models with temperatures up to 80000 K include C III. 
For temperatures of the order of 125000 K, C IV is much less abundant than C V; 
the bottom panel of Fig. \ref{Clines} illustrates the typical behaviour.
For some T$_{\ast}$=125000 K models showing relevant C IV ionization fractions, C III was 
included, but proved to be negligible and to have no important effect on the 
ionization fractions of C IV and C V. In models with T$_{\ast}$=100000 K, C IV may become comparable to, or more important than C V, 
at some radial distance from the star. In the T$_{\ast}$=100000 K 
models where this distance is not very large, we added C III in the calculations. 
As shown in Fig. \ref{CIIIC},  
the ionization fraction of C III is only $\sim$ 10$^{-4}$ in 
the outer parts of the wind and even lower in the inner wind layers.
Its inclusion caused no noticeable changes 
in the ionization fraction of the C IV ion, and was therefore  
considered not essential in this temperature regime.

\section{Spectral diagnostics of $\dot{M}$ and $T_{\ast}$}
\label{sec:spectraldiagnostics}

The preferred method of determining stellar temperatures is through
the ionization balance of He and CNO elements. The observation of
spectral lines from different ionization stages of the same element
avoids the need for assuming abundance ratios. Consistence is achieved
by the use of several diagnostic elements. However, in the
spectra of [WC] CSPNe, not many elements show lines from different
ionization stages. Therefore, the existence or absence of
features in the spectra due to ions of different ionization potentials,
along with the general appearance of the spectral lines, is also used as
temperature diagnostic.

In these stars, both photospheric and wind parameters affect the spectral
features. Since several parameters can affect a
spectral line, the identification of lines which are particularly
sensitive to one parameter in a given regime is important for constraining
its value. By examining the far-UV, UV and optical synthetic spectra produced
for the grid, we identified spectral features sensitive to mass-loss rate and stellar temperature 
and selected those most useful for placing constraints on their values. These are
discussed in sections
\ref{subsec:WCE} and \ref{subsec:WCL}.

\subsection{[WCE] types}
\label{subsec:WCE}

The far-UV, UV and optical spectra of [WCE] CSPNe are particularly poor in
lines of multiple ionization stages
of the same element. Only oxygen, neon and nitrogen show lines from multiple
stages, yet
only oxygen presents strong lines throughout the whole temperature
range considered.

Several other lines can be used as diagnostics of photospheric and
wind parameters. The He II lines $\lambda$ $1640.4$ and $\lambda$ $4685.7$ $\mathrm{\AA}$ (the last one appears blended with a strong C IV line
at $\lambda$ $4658.9$ $\mathrm{\AA}$, since in [WCE] stars, the wind's terminal velocities are high) are mostly sensitive to mass loss, and show little sensitivity to 
temperature in the regime between $T_{\ast}=100000$ and $165000$ K. This happens because, for these
high temperatures, helium is almost totally ionized. Examples of line profiles are shown in
Fig. \ref{HeIIlines}, along with the helium ionization fractions, in which
we see that the fraction of He II remains virtually unaltered in the hotter
models.

\begin{figure}
\includegraphics[scale=0.44]{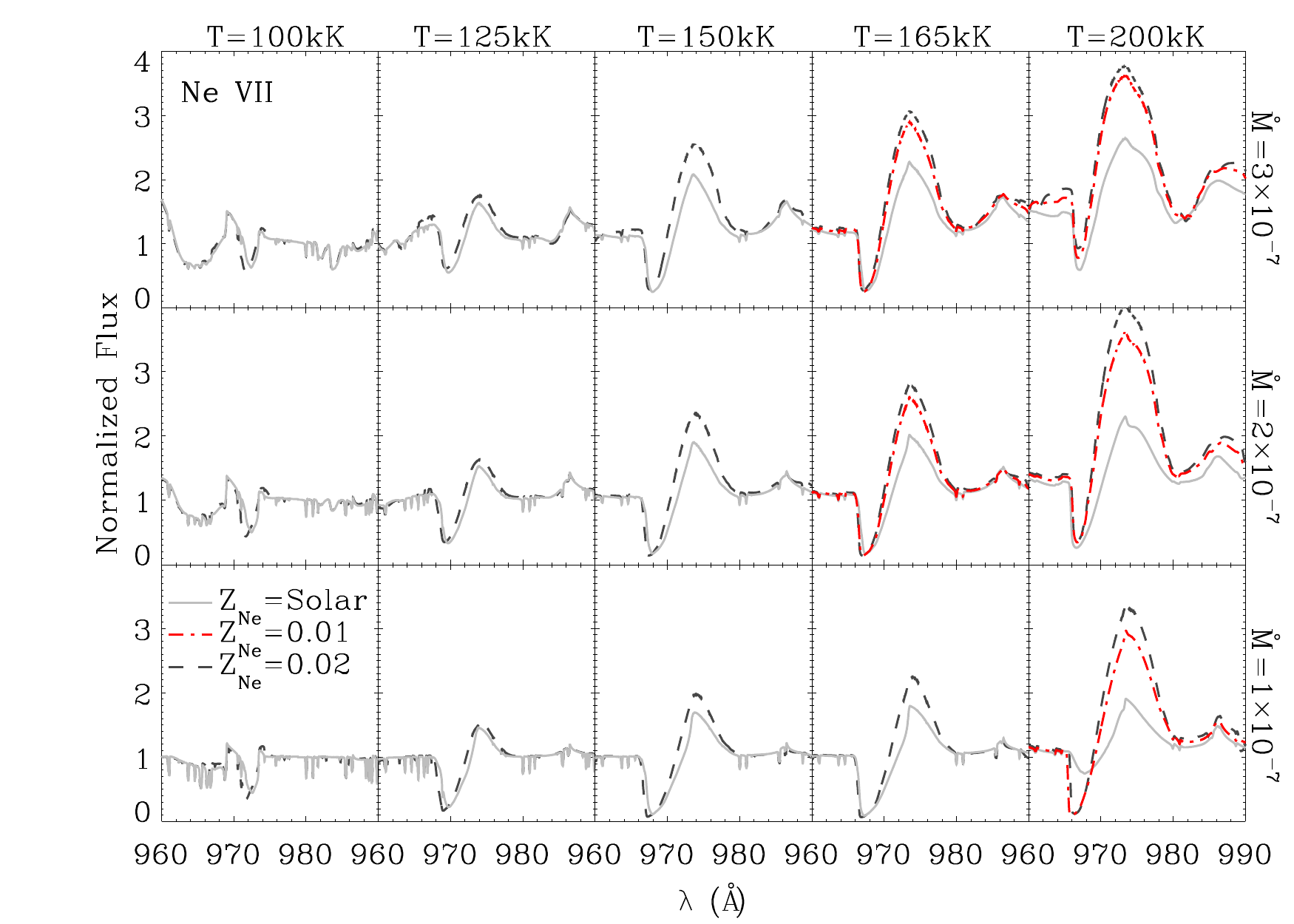}
\includegraphics[scale=0.44]{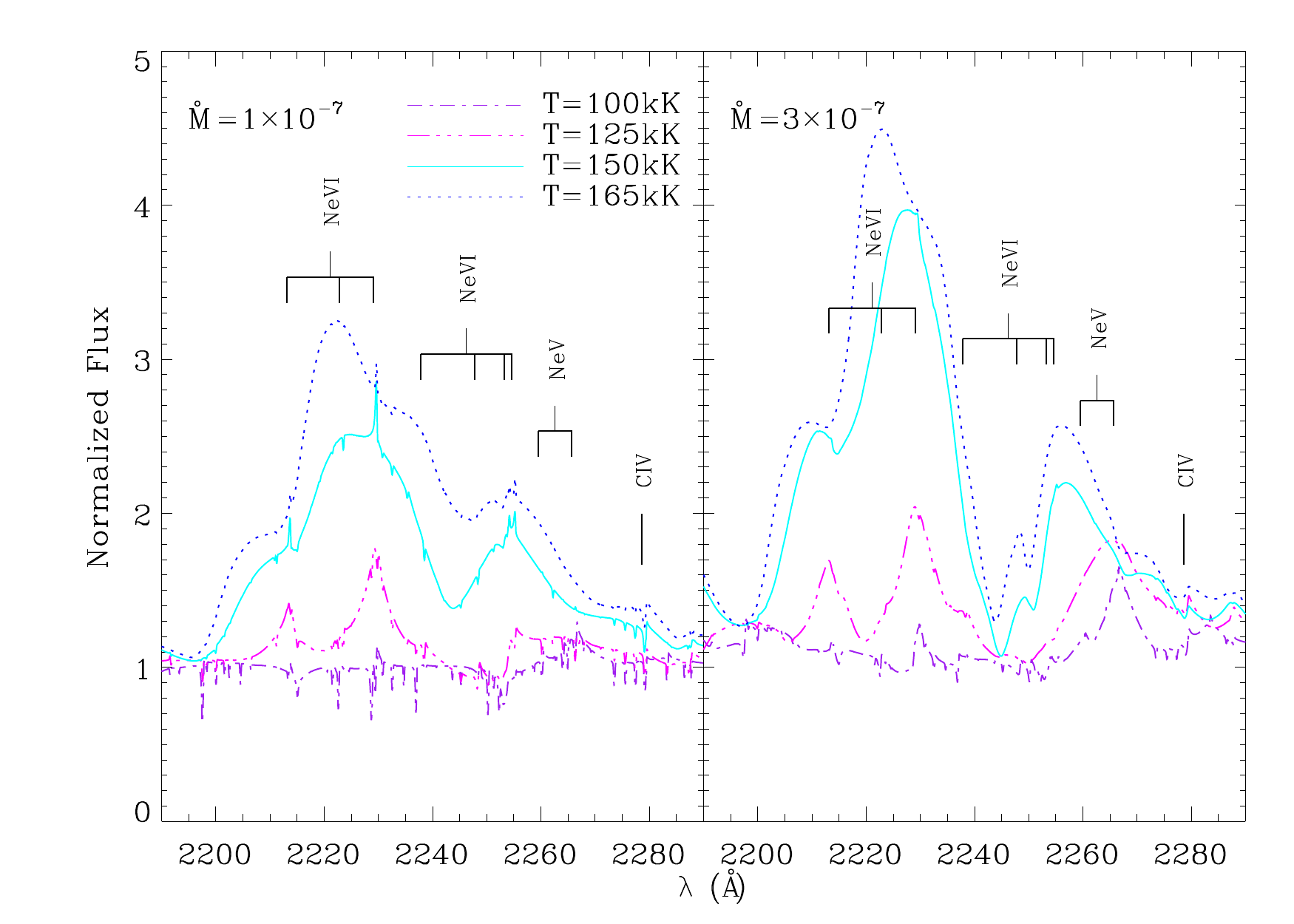}
\includegraphics[scale=0.44]{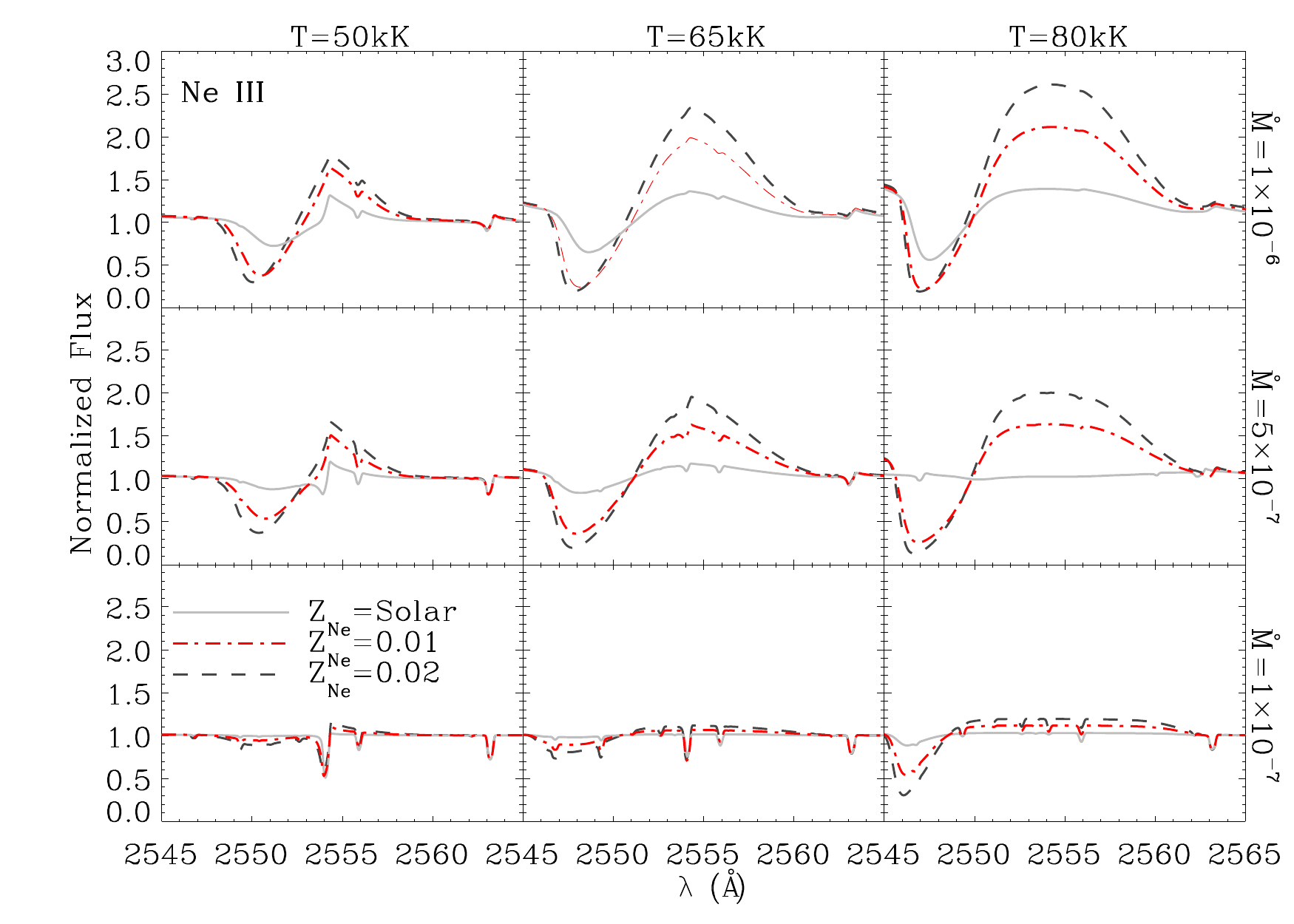}
\includegraphics[scale=0.44]{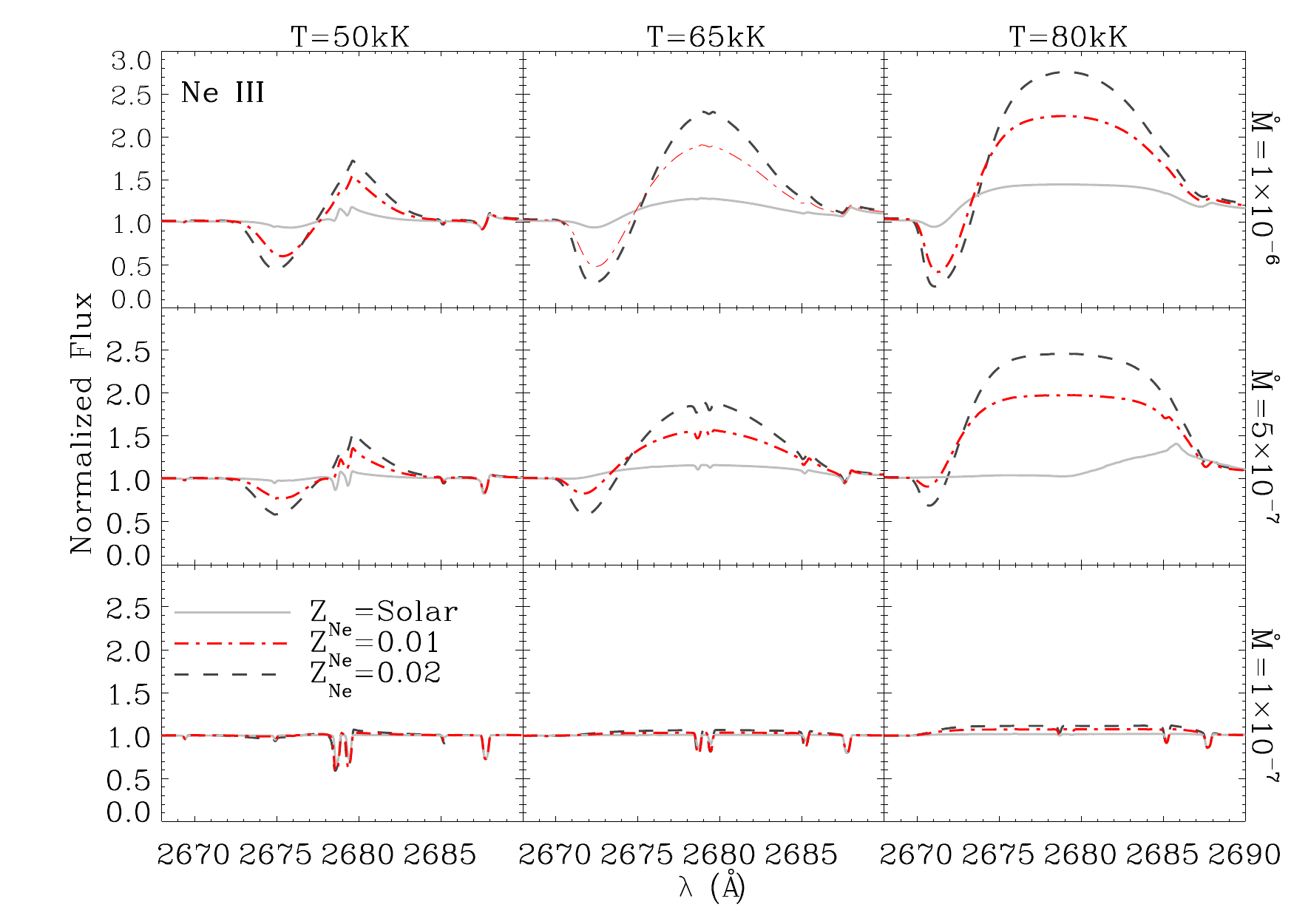}
\centering
\includegraphics[scale=0.43]{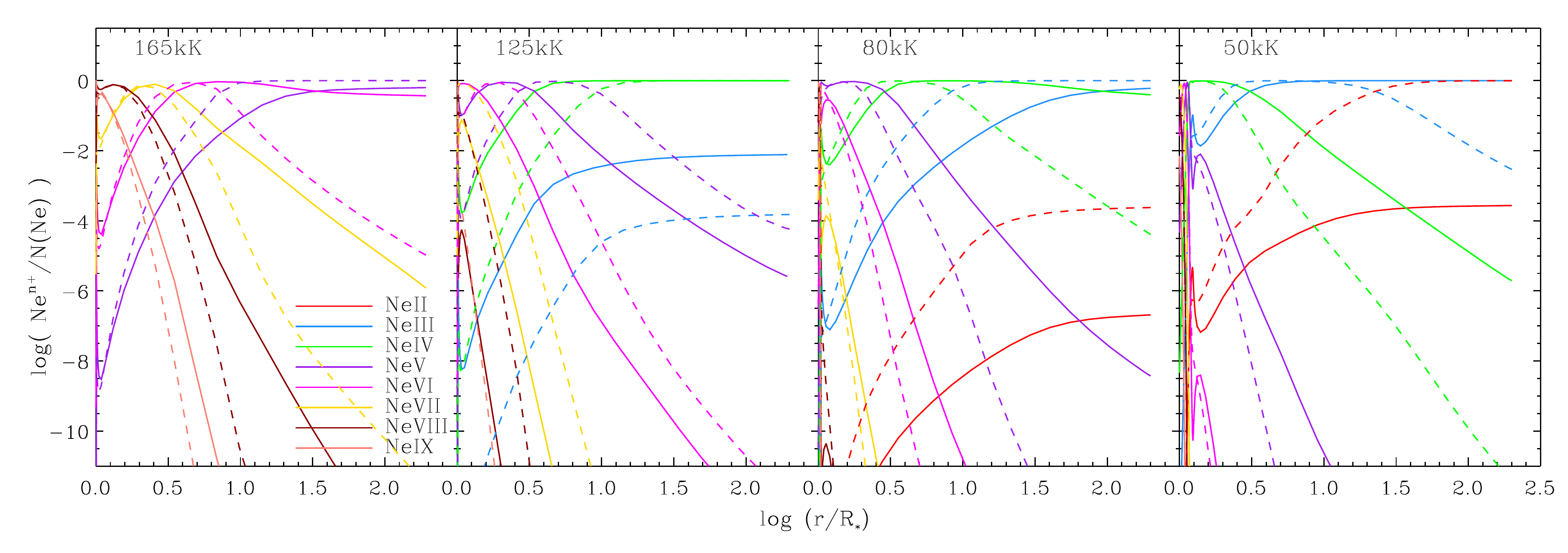}
\caption{Ne lines from different ionization stages and their variation with temperature (also radius), 
mass-loss rate and neon abundance. In the top-left panel, for clarity,
the models with $X_{\mathrm{Ne}}=0.01$ were omitted when the effect of Ne abundance is small.
The bottom panel shows the
ionization fractions of neon (for models with a neon mass fraction of $X_{\mathrm{Ne}}=0.02$); continuous lines indicate models with 
$\dot{M}=10^{-7}$ M$_{\odot}$ yr$^{-1}$, and dashed lines indicate
$\dot{M}=10^{-6}$ M$_{\odot}$ yr$^{-1}$ for models with $T_{\ast}\leq80$ kK, or
$\dot{M}=3\times10^{-7}$ M$_{\odot}$ yr$^{-1}$ for models with $T_{\ast}\geq100$ kK.
All the models shown are from track B ($M_{\ast}=0.6$ M$_{\odot}$) and have $v_{\infty}=2500$ km s$^{-1}$.}
\label{Nelines}
\end{figure}

Among all the spectral lines in the studied range, Ne VII $\lambda$ $973.3$ $\mathrm{\AA}$ shows 
the least sensitivity to mass loss, which makes it a good indicator of temperature within the range
of physical parameters studied here. Its behaviour is shown on the top-left panel
of Fig. \ref{Nelines}. \citet{2005ApJ...627..424H} and \citet{Bianchi2007} found,
however, that it can be sensitive to the neon abundance in PG1159 spectra, when it is not saturated. In section
\ref{subsec:Neabundance}, we discuss the effect of the neon abundance in the grid models.

\begin{figure}
\includegraphics[scale=0.435]{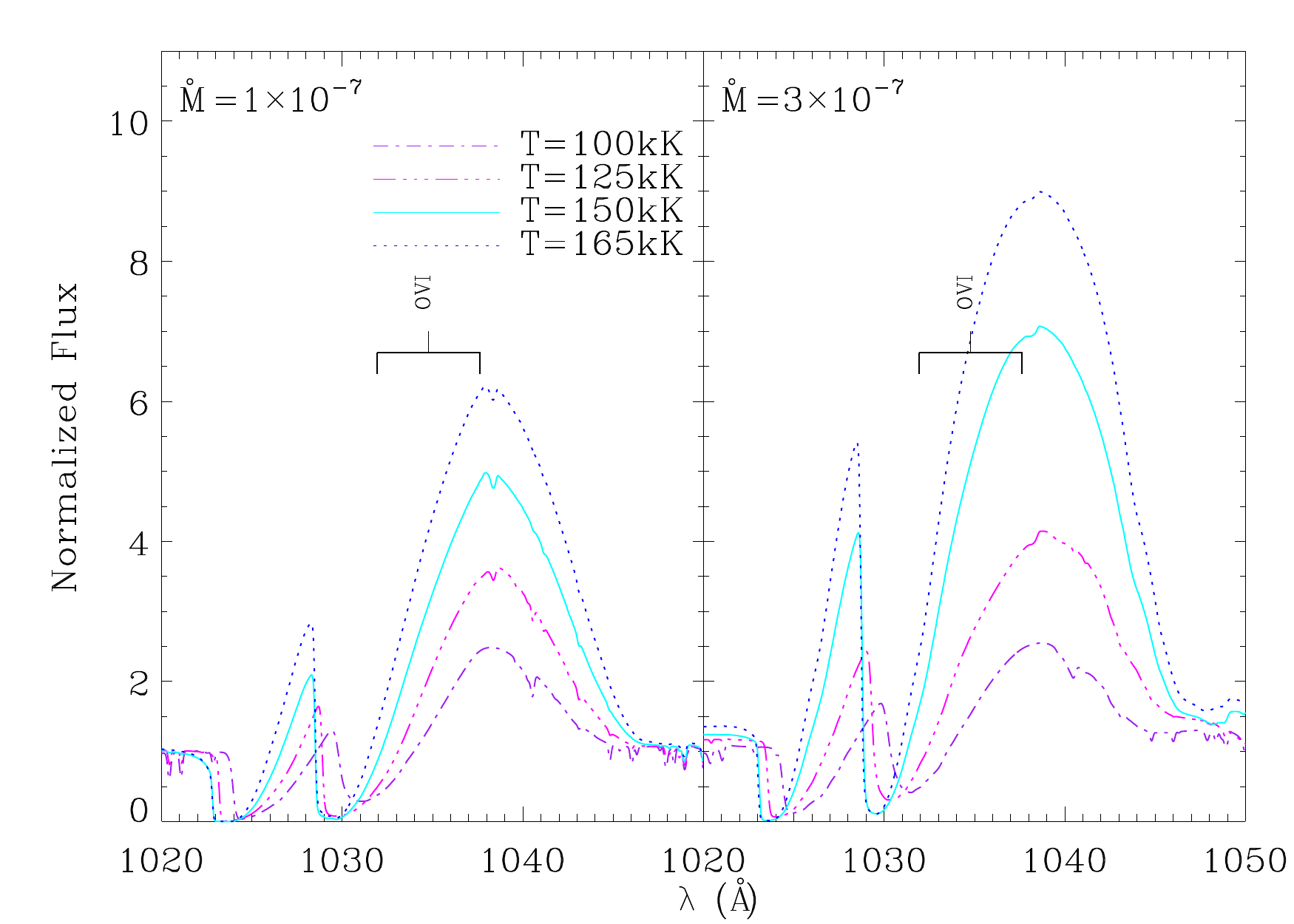}
\includegraphics[scale=0.435]{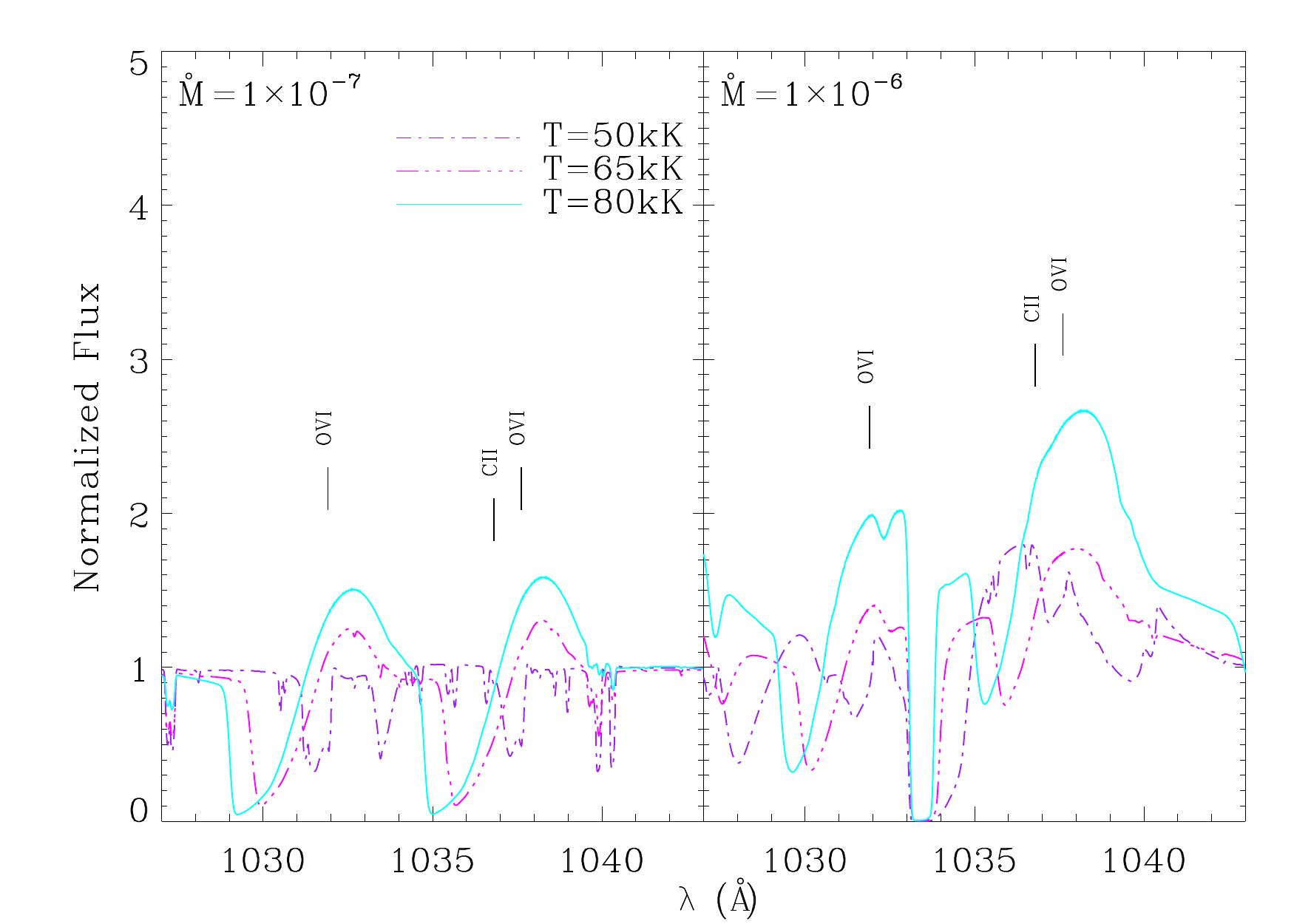}
\includegraphics[scale=0.435]{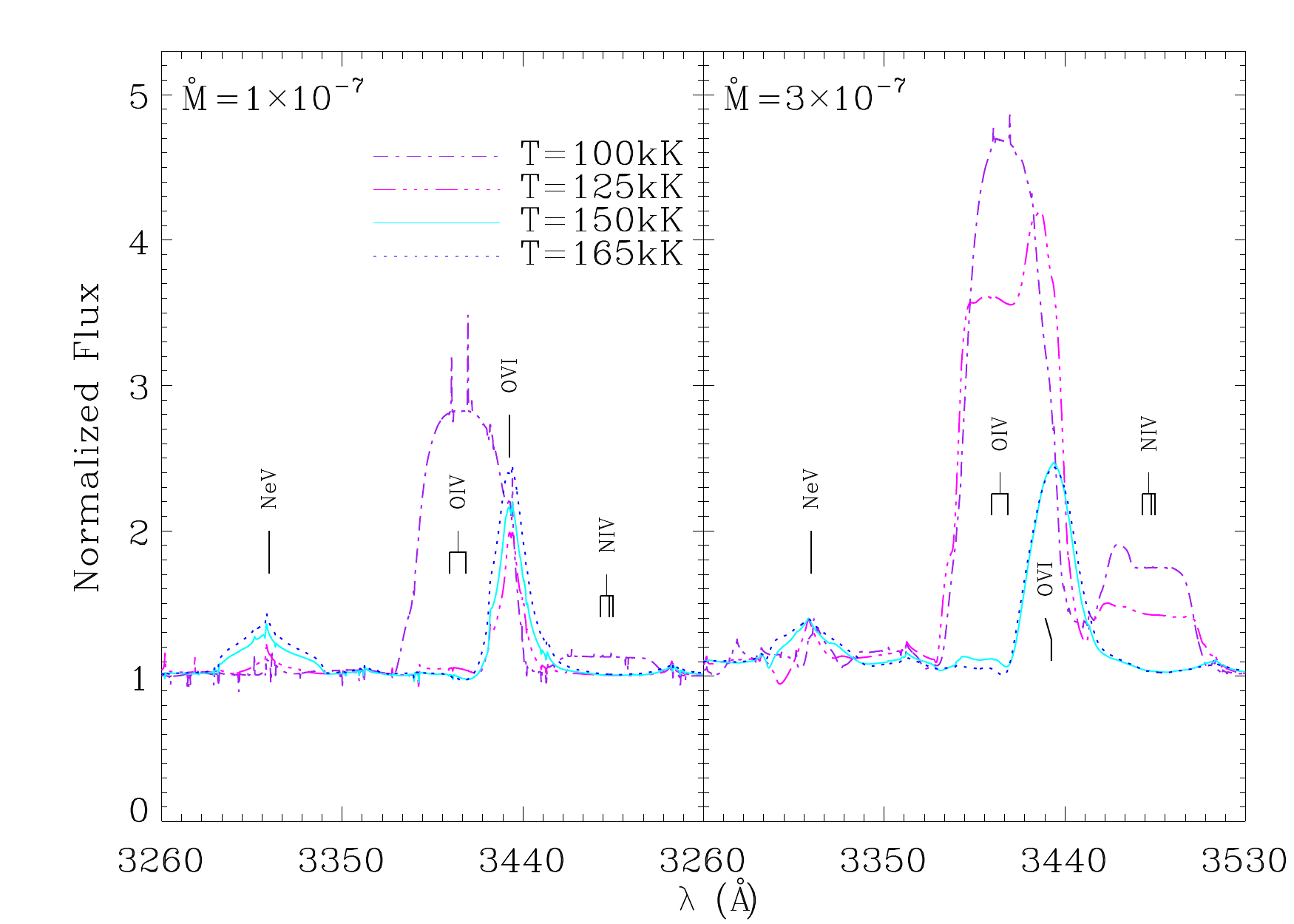}
\includegraphics[scale=0.435]{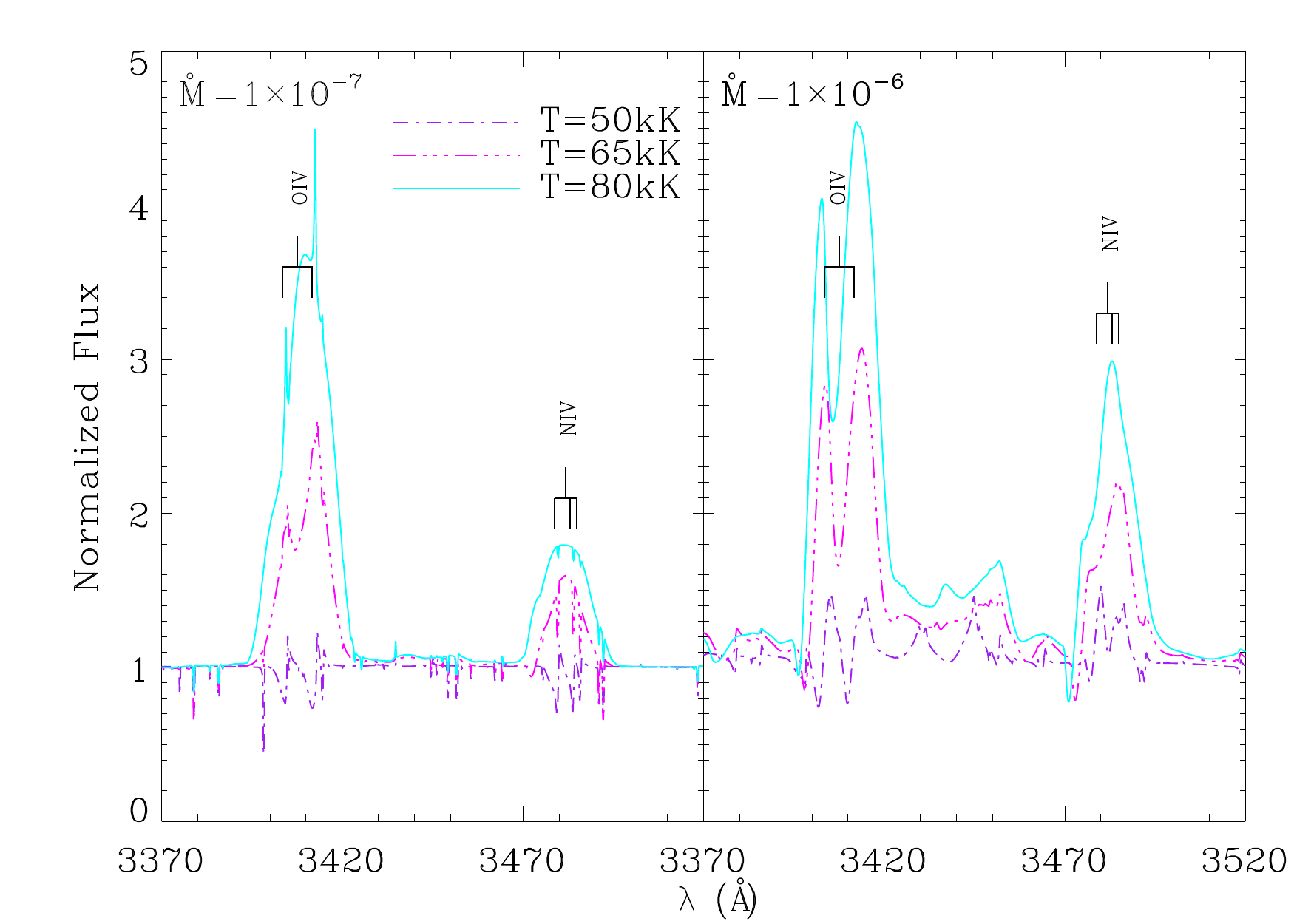}
\includegraphics[scale=0.435]{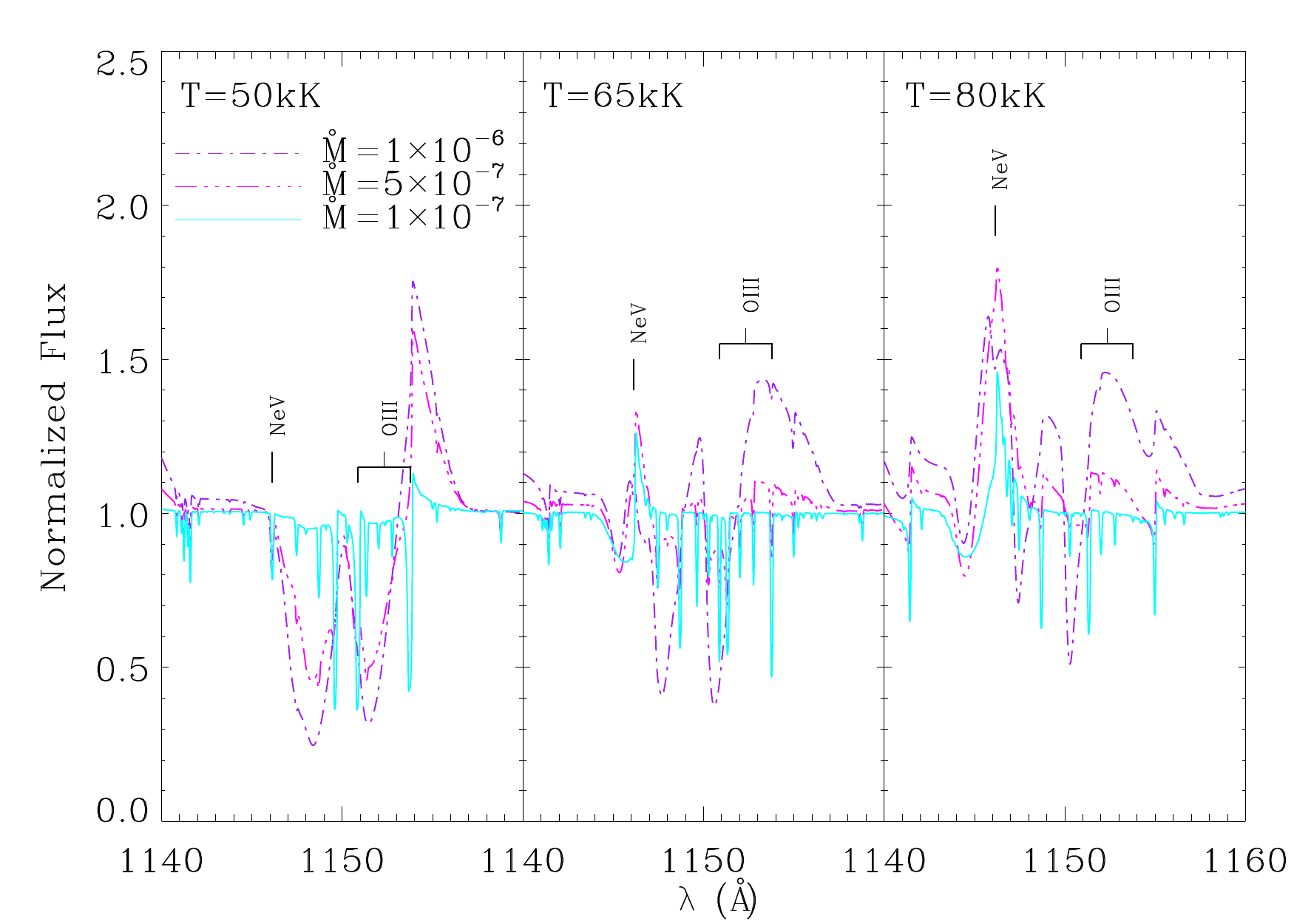}
\includegraphics[scale=0.435]{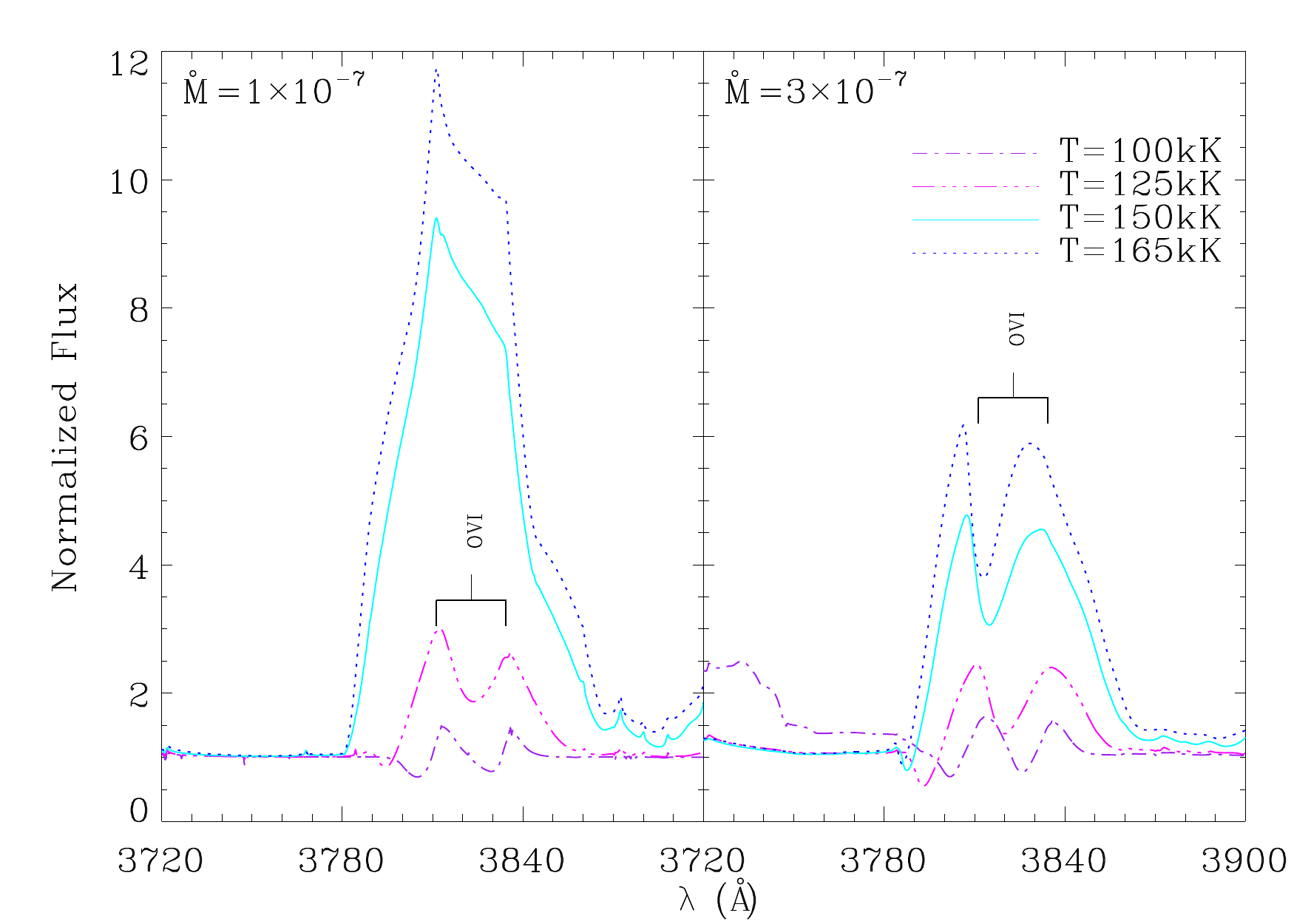}
\includegraphics[scale=0.43,angle=0]{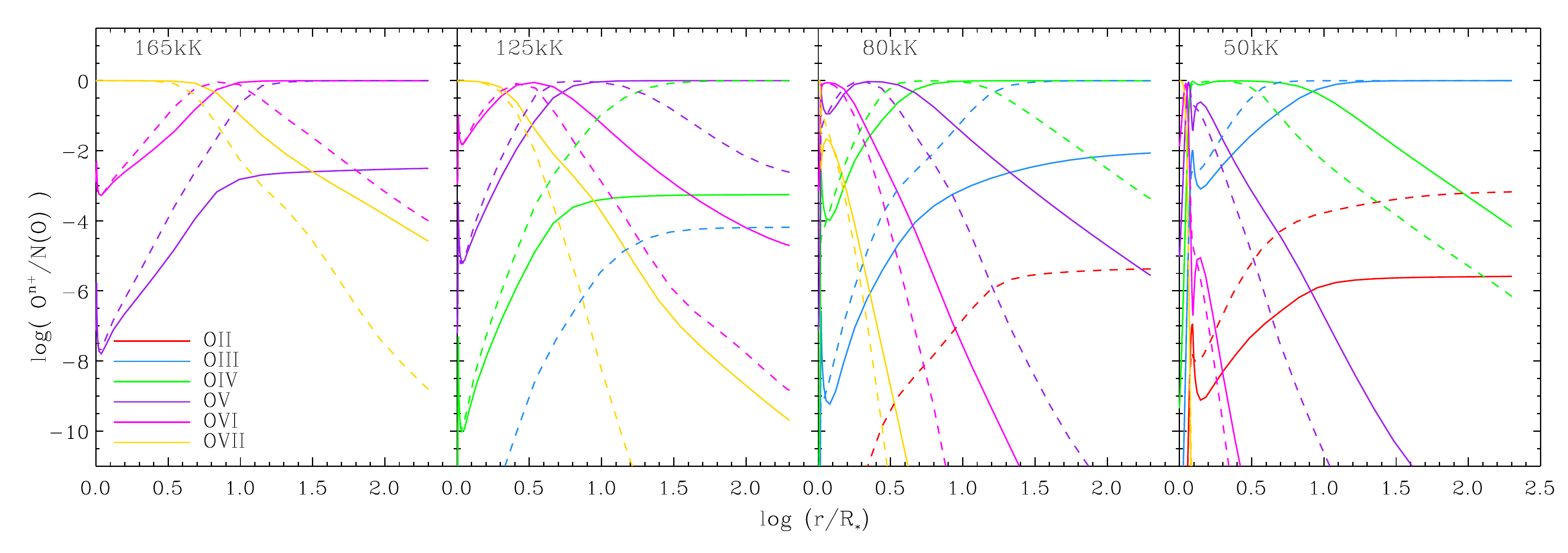}
\caption{Synthetic line profiles from different ionization stages of oxygen in various models.
The bottom panel shows the ionization fractions of
oxygen; continuous lines indicate models with $\dot{M}=10^{-7}$ M$_{\odot}$ yr$^{-1}$
and dashed lines indicate $\dot{M}=10^{-6}$ M$_{\odot}$ yr$^{-1}$ if $T_{\ast}\leq80$ kK, or 
$\dot{M}=3\times10^{-7}$ M$_{\odot}$ yr$^{-1}$ for models with $T_{\ast}\geq100$ kK. All the models shown are from track B 
($M_{\ast}=0.6$ M$_{\odot}$) and have $v_{\infty}=2500$ km s$^{-1}$.}\label{Olines}
\end{figure}

For the O VI $\lambda \lambda$ $1031.9$, $1037.6$ $\mathrm{\AA}$ doublet shown in Fig. 
\ref{Olines} (top-left panel), both temperature and mass loss alter significantly the
line profiles. The O VI $\lambda \lambda$ $3811.4$ and $3834.2$ $\mathrm{\AA}$ lines, seen in Fig. \ref{Olines}
(bottom-right panel), shows little variability with mass loss, within the range of values analysed,
in the two cooler models. In the
hotter models, on the other hand, it shows a behaviour opposite to that of all other strong lines in the range studied here, 
the line intensity decreasing with increasing mass loss. Also, the large variation of the line
profile between models with temperatures of $125000$ and $150000$ K can
help establish lower or higher limits for the temperature. In the bottom panels of
Fig. \ref{Olines}, we show oxygen ionization fraction plots for different
temperatures
and mass-loss rates. We see that the O VI ion gains importance as we progress
towards higher temperatures, which is reflected in the increasing intensity
of
the O VI lines in the hotter models. The mass-loss rate, on the other hand, has a
more complex effect on these lines: it alters not only the density of
the ion in the expanding atmosphere, but also changes its temperature structure,
altering the ionization fractions. Denser winds will be cooler, with lower
ionic species growing in importance. For massive stars, the O VI $\lambda \lambda$ $1031.9$, $1037.6$ $\mathrm{\AA}$ lines were found to strongly
depend on X-rays and to have a hard threshold with mass loss, while other lines strongly depend on clumping  \citep{Bianchi2002}. The effect of X-rays on the
far-UV O VI doublet will be discussed in section \ref{subsec:xrays}.

Other temperature indicators are the neon and oxygen features shown in Figs.
\ref{Nelines} (top-right panel) and \ref{Olines} (central-left panel), respectively.
These features are due to O IV (with lines at
$\lambda \lambda$ $3403.6$, $3411.7$, and $3413.6$ $\mathrm{\AA}$), O VI
(at $\lambda$ $3433.3$ $\mathrm{\AA}$) and Ne VI
(with contamination by Ne V lines). The Ne VI feature
can only be seen in hotter models ($T_{\ast}\geq150000$ K). The O IV feature
appears only in the two cooler ones (for $T_{\ast}=100000$ and $125000$ K), while it is
substituted by the O VI line as the temperature increases. For the right combination
of mass loss and temperature, these oxygen lines can also coexist.
Thus, these features can be used to constrain the
temperature regime of the object being studied. Also seen in the
central-left panel of Fig. \ref{Olines} are the Ne V $\lambda$ $3313.7$ $\mathrm{\AA}$ and N IV $\lambda \lambda$ $3478.7$,
$3483.0$, and $3484.9$ $\mathrm{\AA}$ lines, which can help constrain neon and nitrogen
abundances.

The strong C IV doublet at $\lambda \lambda$ $1548.2$, $1550.8$
$\mathrm{\AA}$ shows sensitivity to both mass loss and temperature, as can
be seen in the central-left panel of Fig. \ref{Clines}.
Nevertheless, once either mass loss
or temperature is established through other indicators, it can help
constraining the other parameter. Also, this line is an
important diagnostic of wind's terminal velocity, since this ion is abundant
also in the outer parts of the wind down to temperatures of $50000$ K.

\subsection{[WCL] types}
\label{subsec:WCL}

The winds of [WCL] CSPNe are characterized by much lower terminal velocities than the ones of [WCE] stars, which results in narrower spectral lines.
Also, these
cooler CSPNe seem to show a wider range in mass-loss, by comparing results of spectral analyses found in the literature. We took these facts
into consideration when calculating the grid.

These objects present very complex spectra, and a higher number of
spectral lines from elements with multiple ionization stages, thus making the stellar temperature easier to constrain.
Several strong C III and C IV lines appear in the grid models for these
stars, along with multiple ionization stages of neon, nitrogen,
oxygen, phosphorus, sulphur, and helium.

\begin{figure}
\includegraphics[scale=0.435]{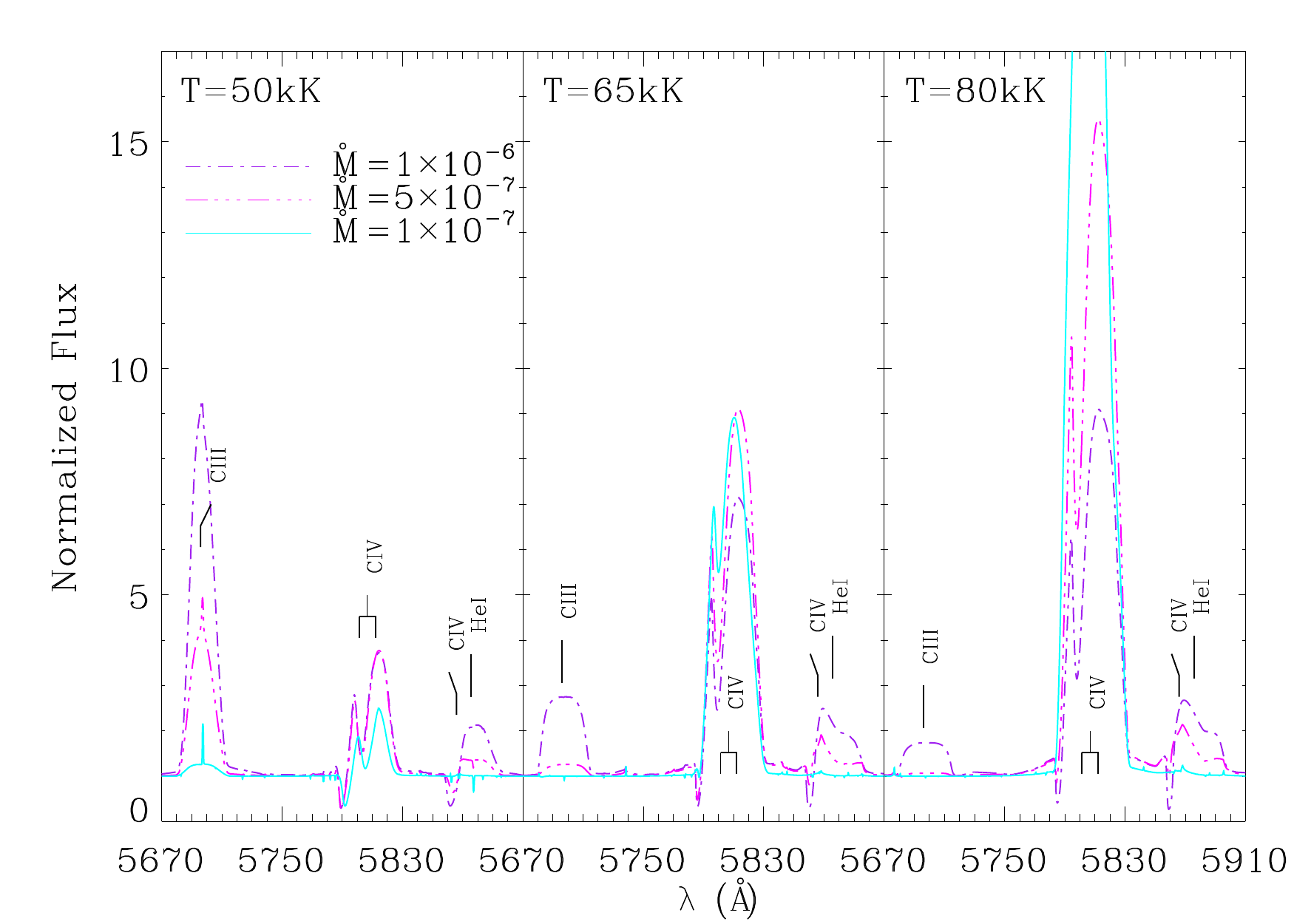}
\includegraphics[scale=0.435]{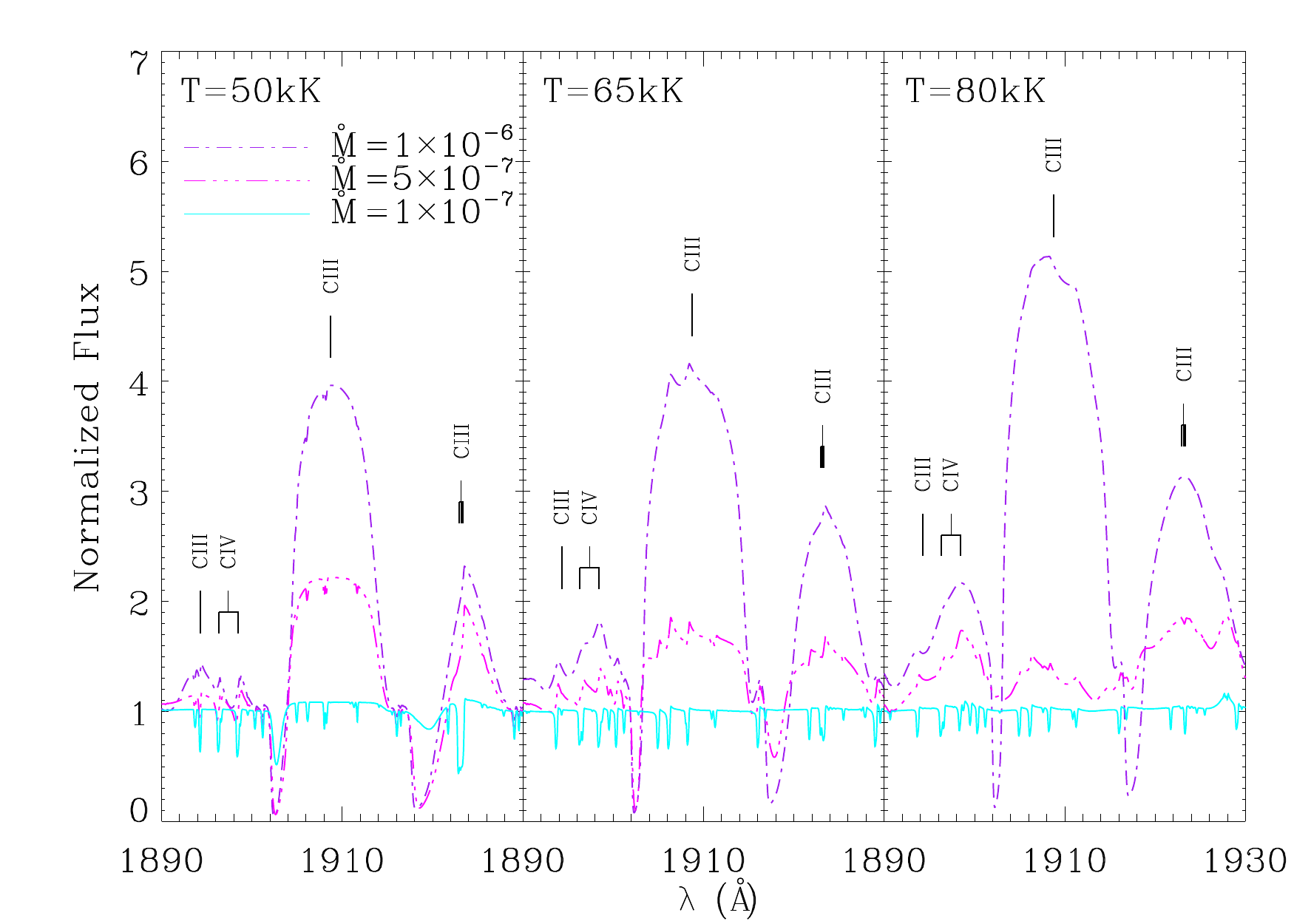}
\includegraphics[scale=0.435]{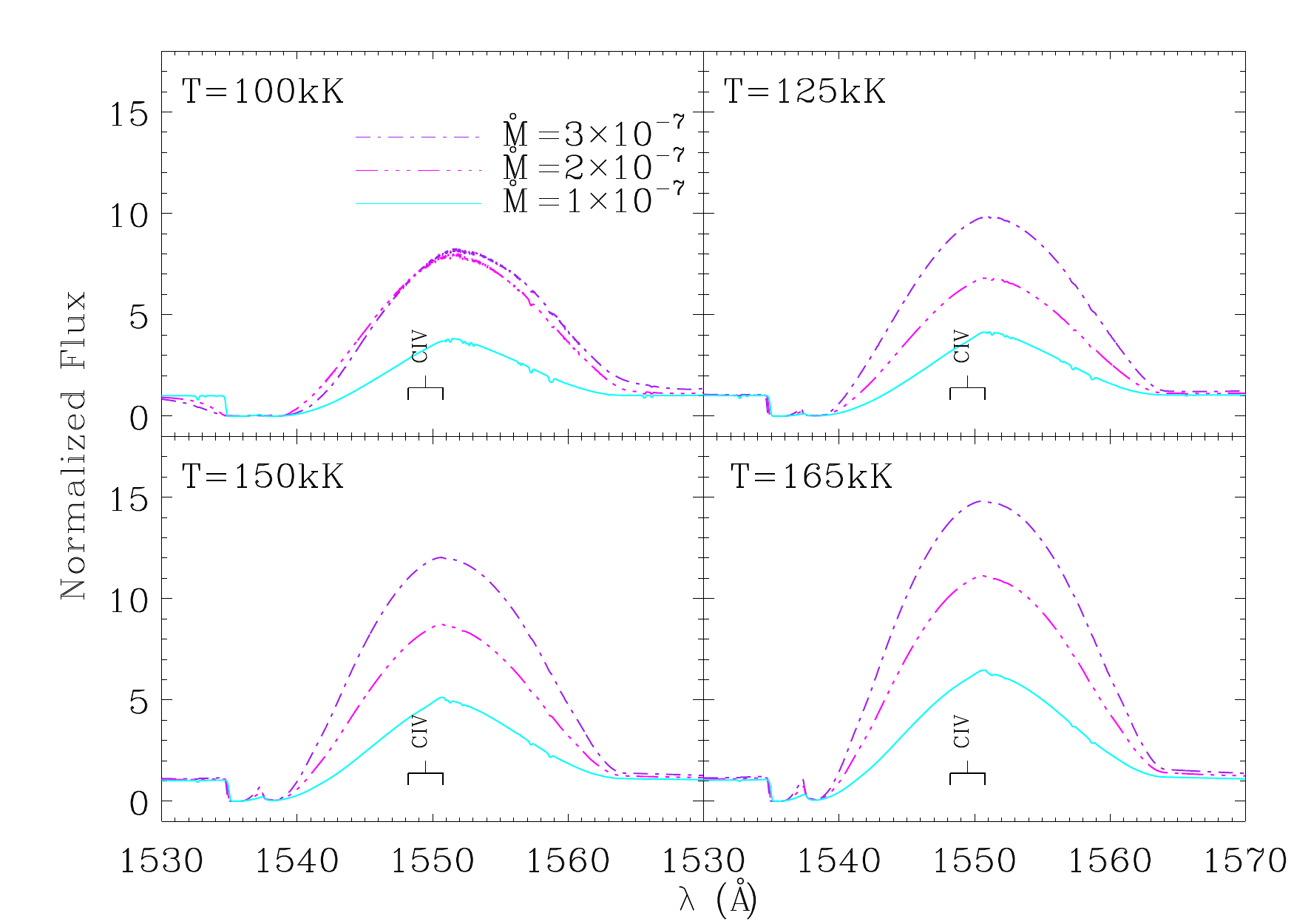}
\includegraphics[scale=0.435]{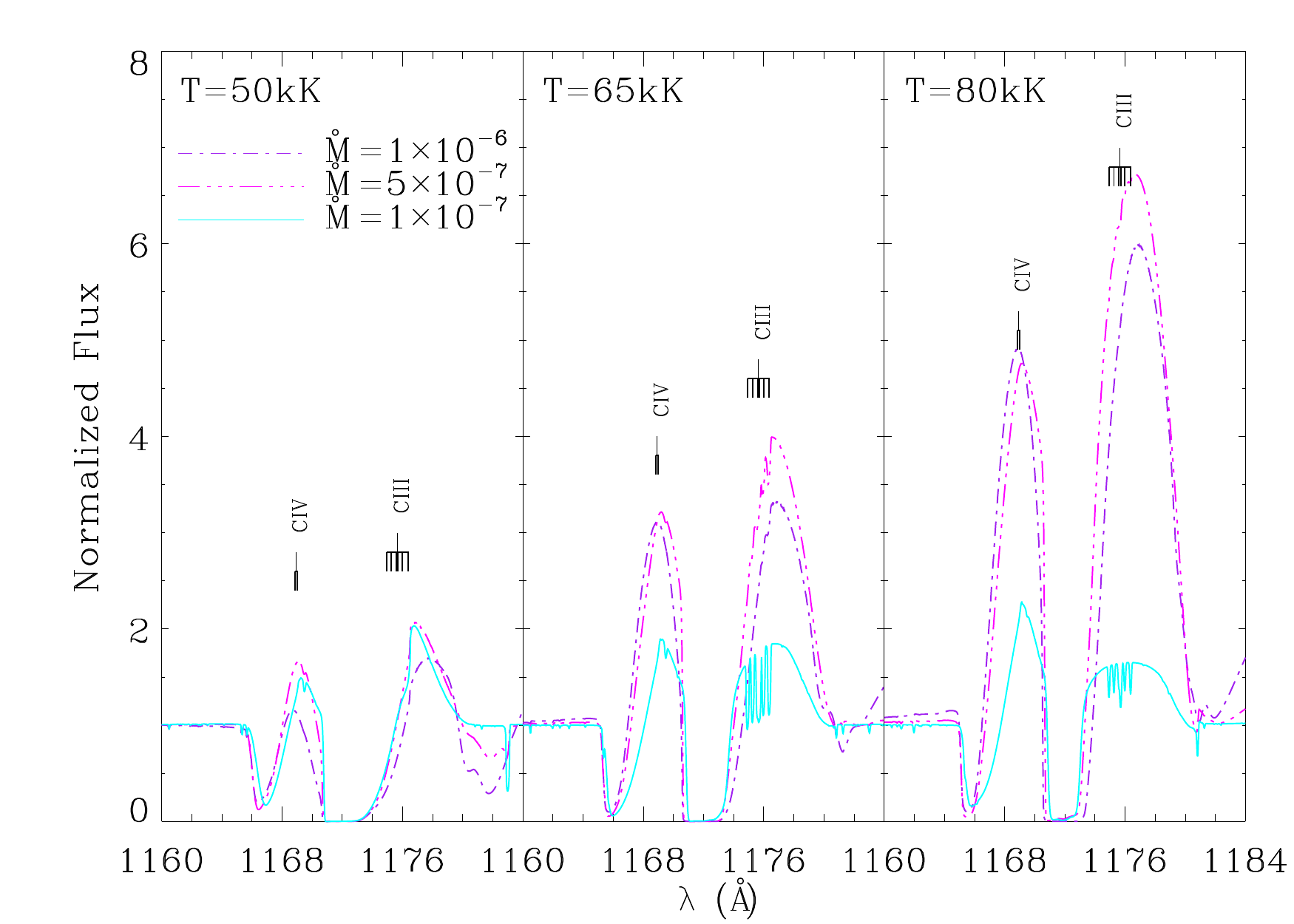}
\centering
\includegraphics[scale=0.435]{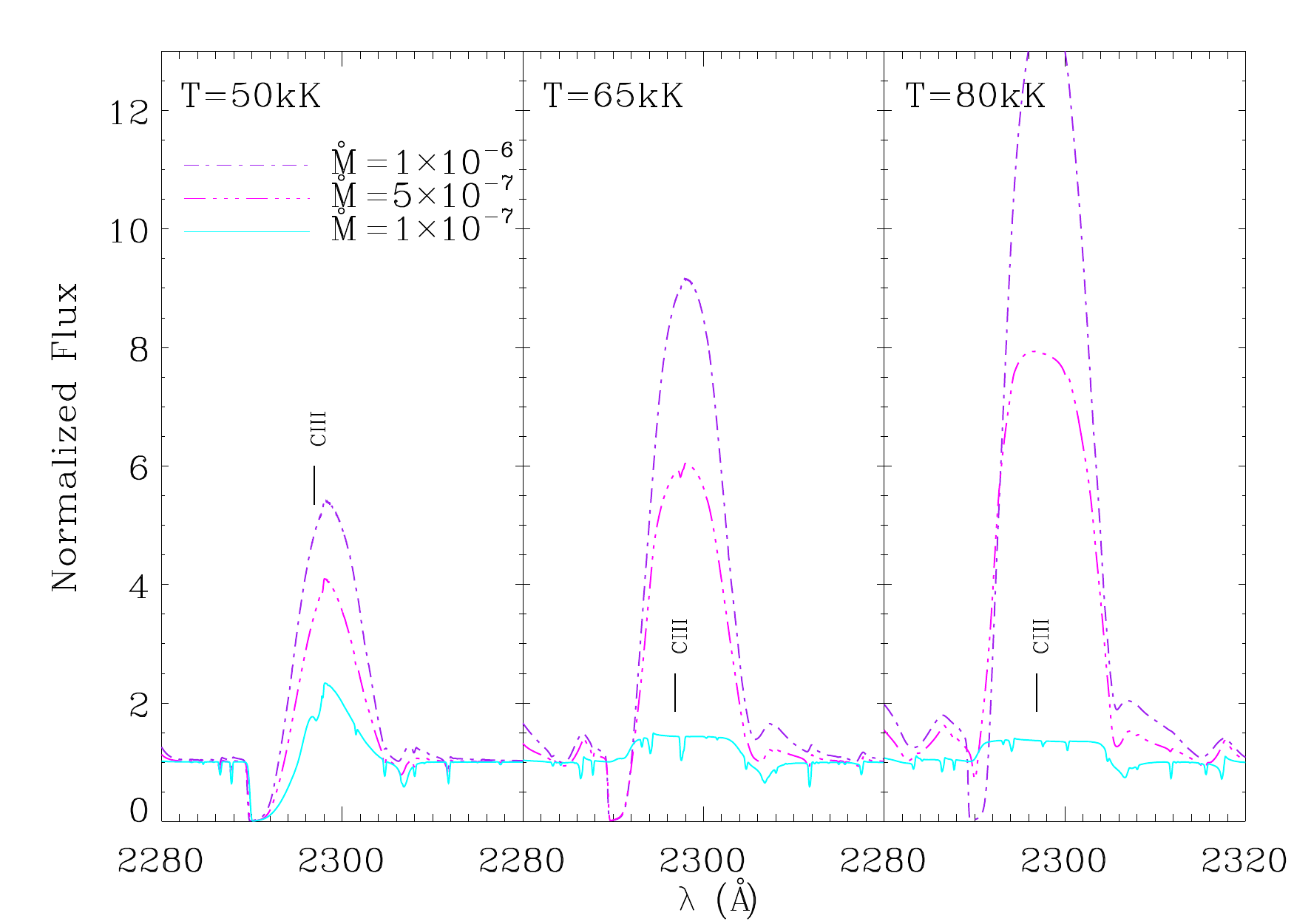}
\includegraphics[scale=0.43]{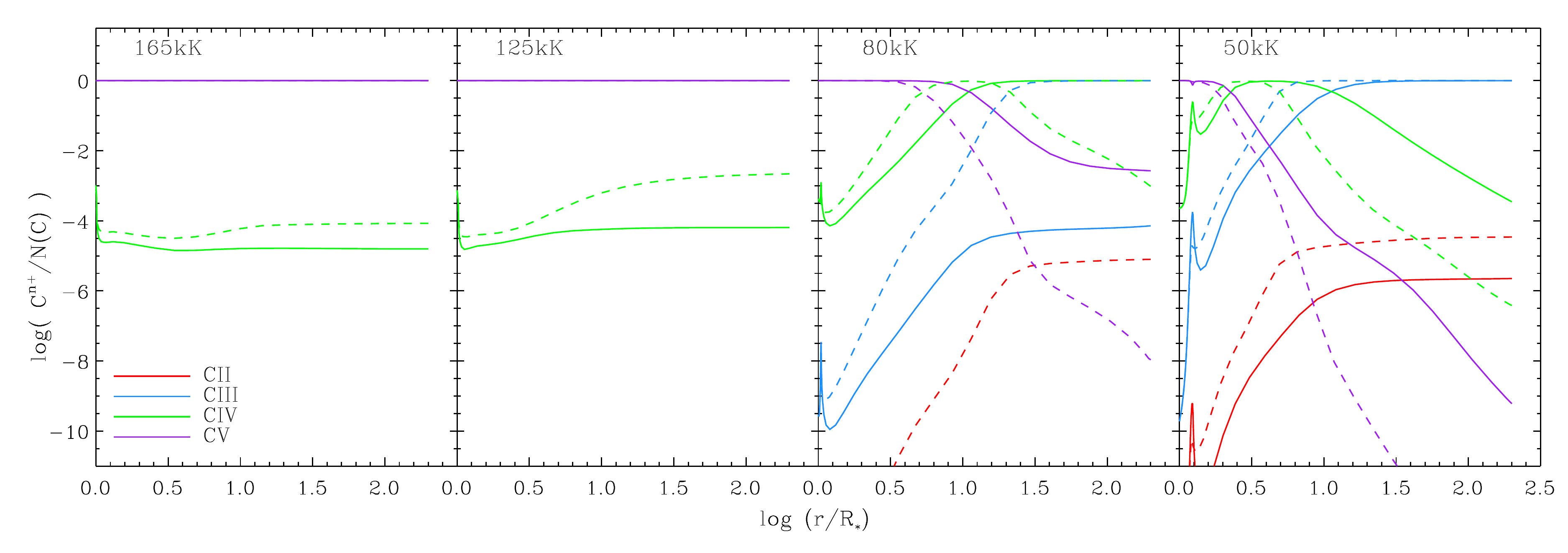}
\caption{Profiles of carbon lines of different ionization stages from models with different mass-loss rates 
and temperatures (also radii). The bottom panel shows the ionization fractions of
carbon; continuous lines indicate models with $\dot{M}=10^{-7}$ M$_{\odot}$ yr$^{-1}$
and dashed lines indicate $\dot{M}=10^{-6}$ M$_{\odot}$ yr$^{-1}$ if $T_{\ast}\leq80$ kK, 
or $\dot{M}=3\times10^{-7}$ M$_{\odot}$ yr$^{-1}$ for models 
with $T_{\ast}\geq100$ kK. All the models shown are from track B ($M_{\ast}=0.6$ M$_{\odot}$) and have $v_{\infty}=2500$ km s$^{-1}$.} \label{Clines}
\end{figure}

We selected some spectral features useful for distinguishing among
temperatures and mass-loss rate values, within the regime considered. The top-left panel of Fig. \ref{Clines} shows a
section of the optical spectra where strong lines from both C III
($\lambda$ $5695.9$ $\mathrm{\AA}$) and C IV ($\lambda \lambda$ $5801.3$,
$5812.0$, and $5865.9$ $\mathrm{\AA}$) can be seen, along with a
He I line ($\lambda$ $5875.7$ $\mathrm{\AA}$), making this
spectral region an important diagnostic of stellar temperature.
The O IV $\lambda \lambda$ $3403.6$, $3411.7$, and $3413.6$ $\mathrm{\AA}$ spectral features 
shown in the central-right panel of Fig. \ref{Olines} is likewise
an excellent indicator of
stellar temperature between T$_{\ast}=50000$ and $80000$ K,
showing very little sensitivity to mass-loss rate in the interval studied.
Also shown is the N IV $\lambda$ $3478.7$ $\mathrm{\AA}$ line.

In the late-[WC] type, the O VI doublet $\lambda \lambda$ $1031.9$, $1037.6$ $\mathrm{\AA}$, shown in Fig. 
\ref{Olines}, top-right panel, is still present,
albeit much weaker and
showing contamination by other spectral lines. Another useful diagnostic line for the
[WCL] subclass is the C III $\lambda$ $1908.7$ $\mathrm{\AA}$ line 
shown in the top-right panel of Fig. \ref{Clines}. It is much more sensitive to mass
loss than to temperature in the range shown, thus making it an important mass loss
discriminator.

\begin{figure}
\includegraphics[scale=0.435]{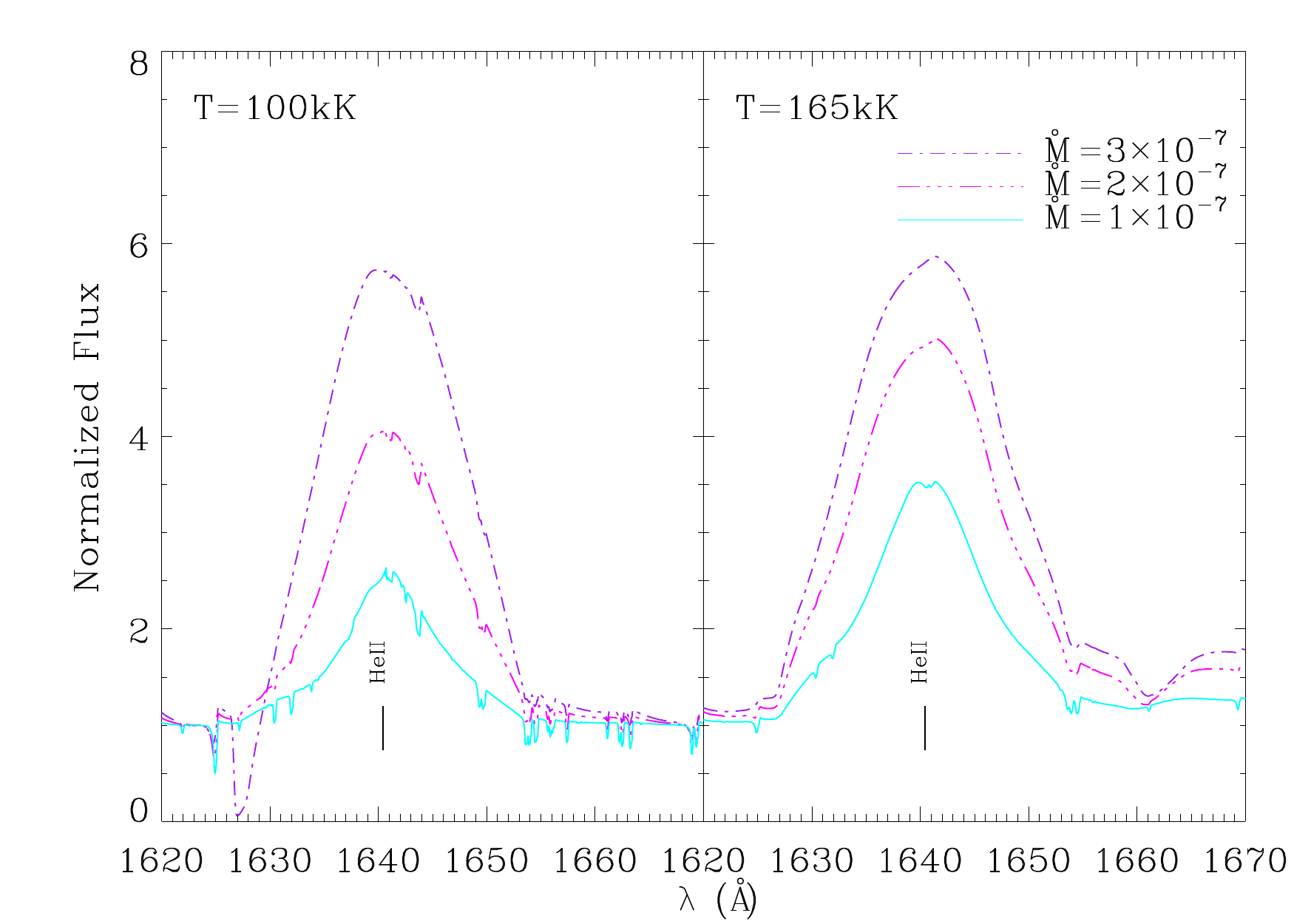}
\includegraphics[scale=0.435]{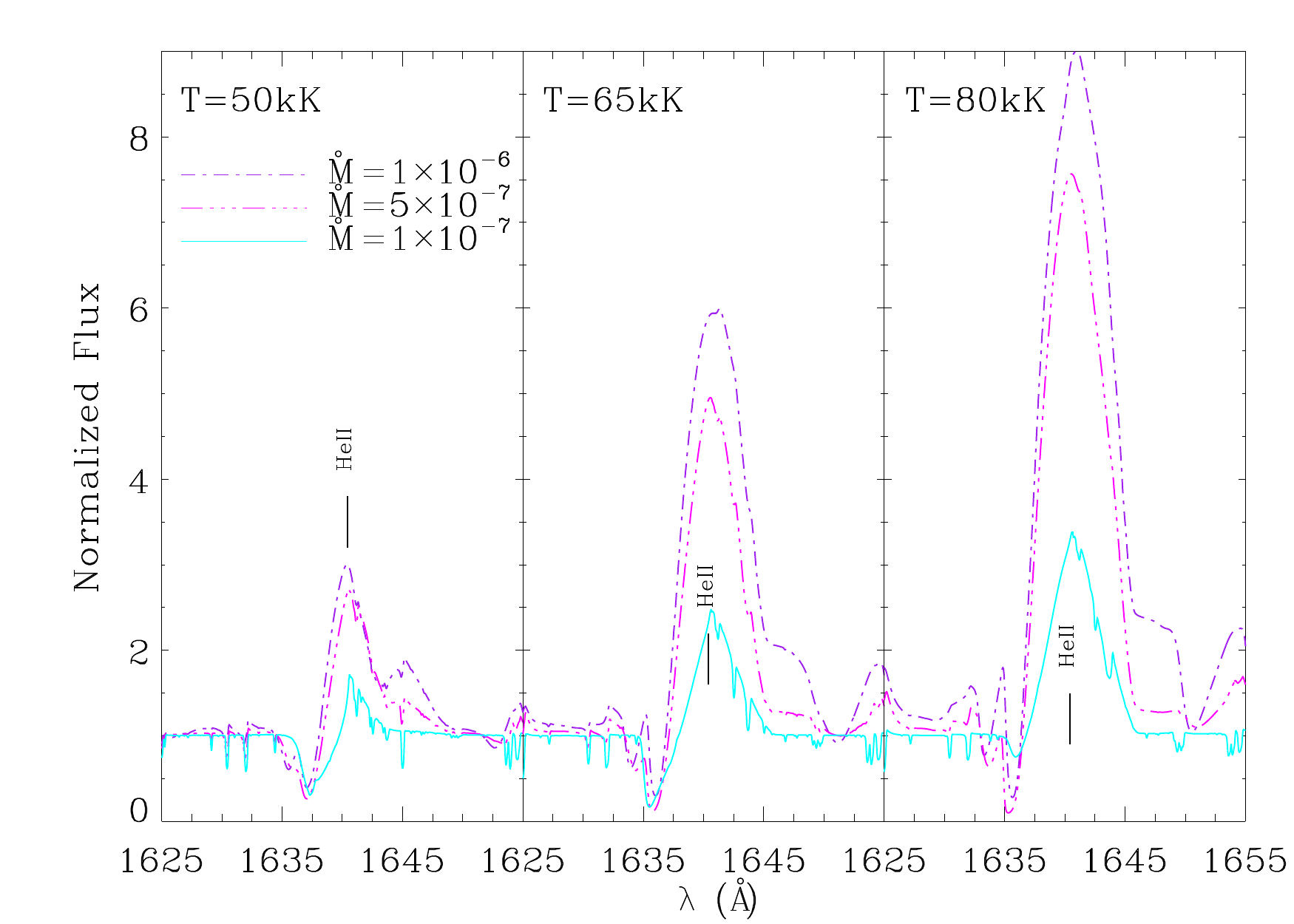}
\includegraphics[scale=0.435]{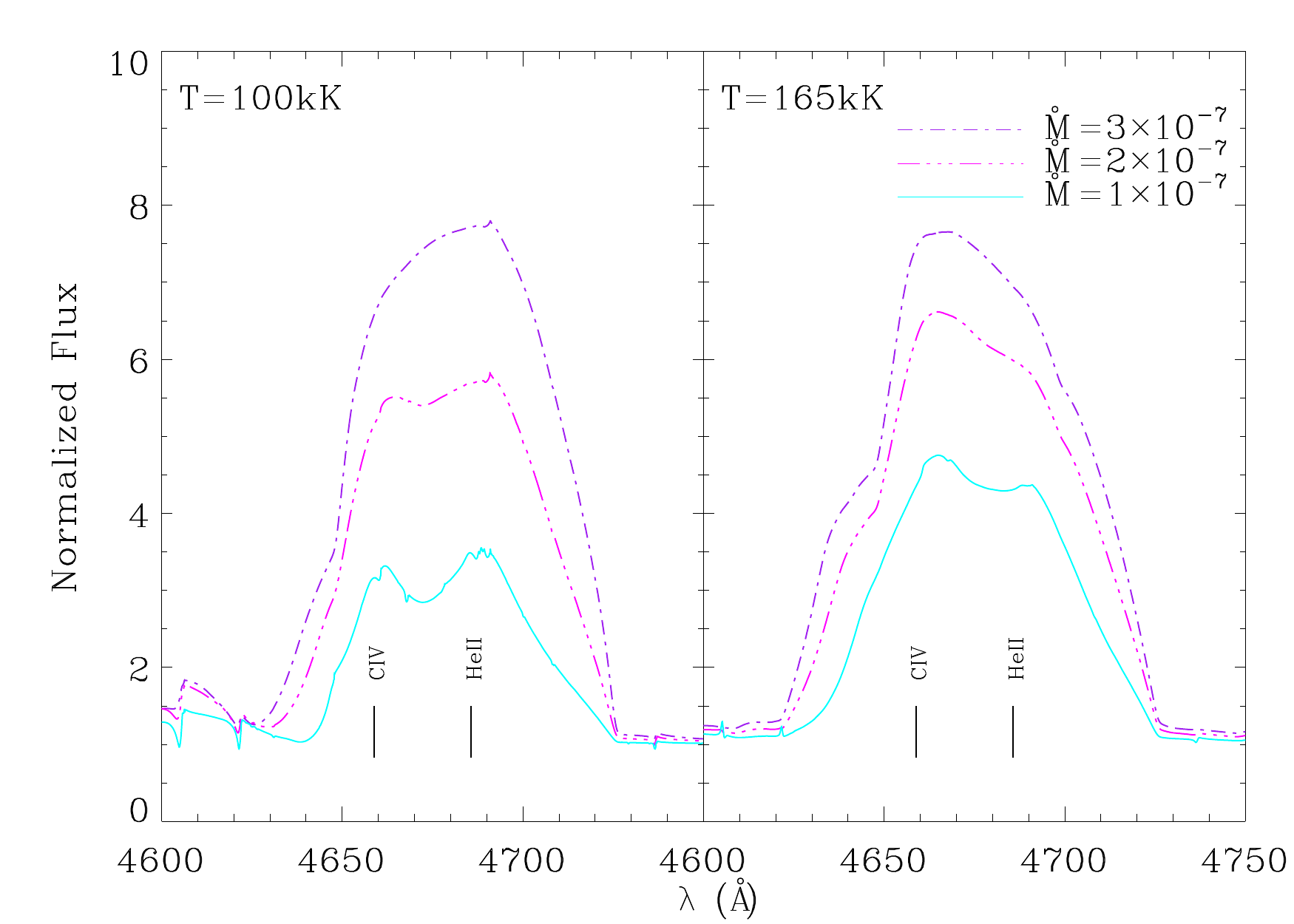}
\includegraphics[scale=0.435]{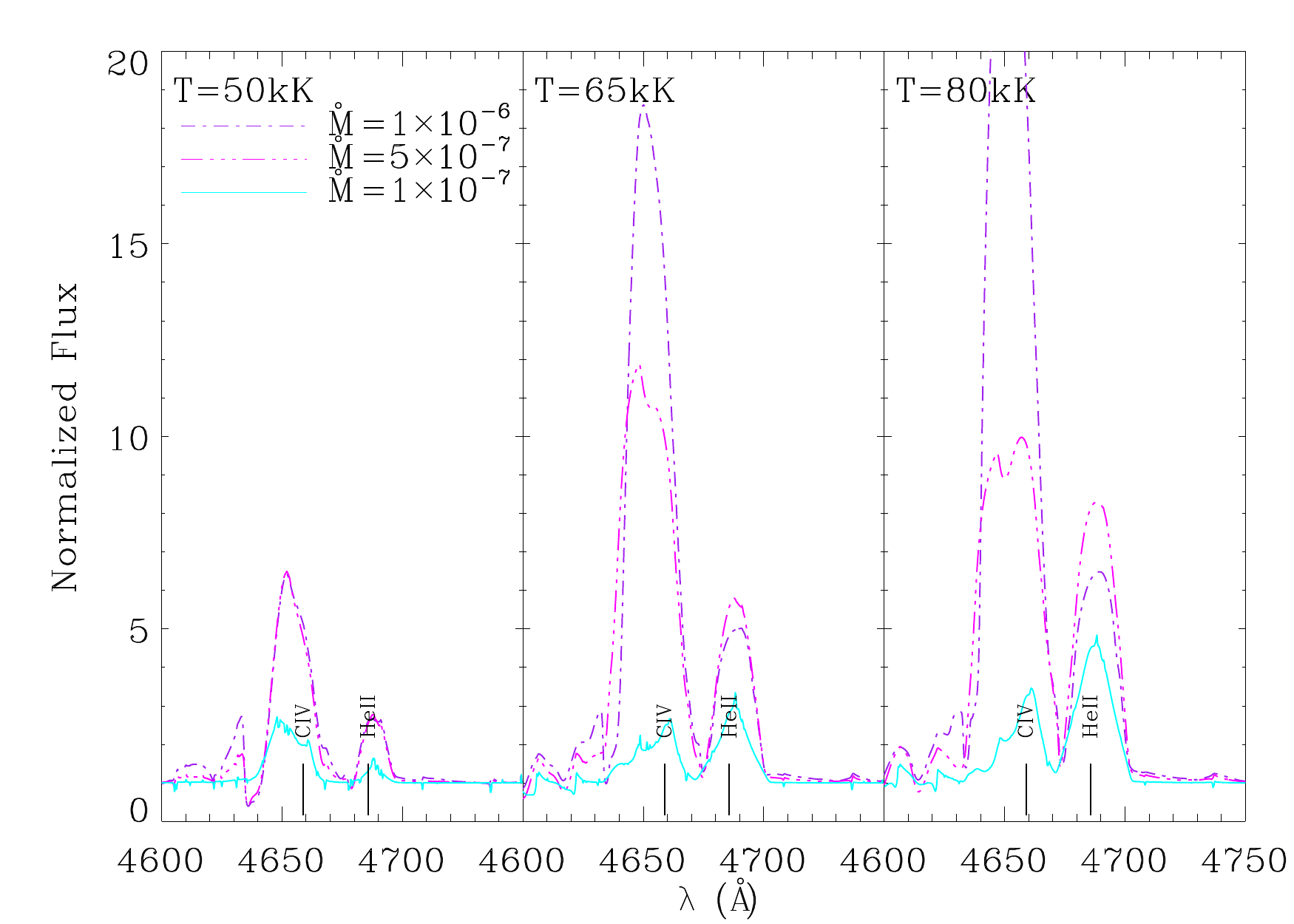}
\centering
\includegraphics[scale=0.43]{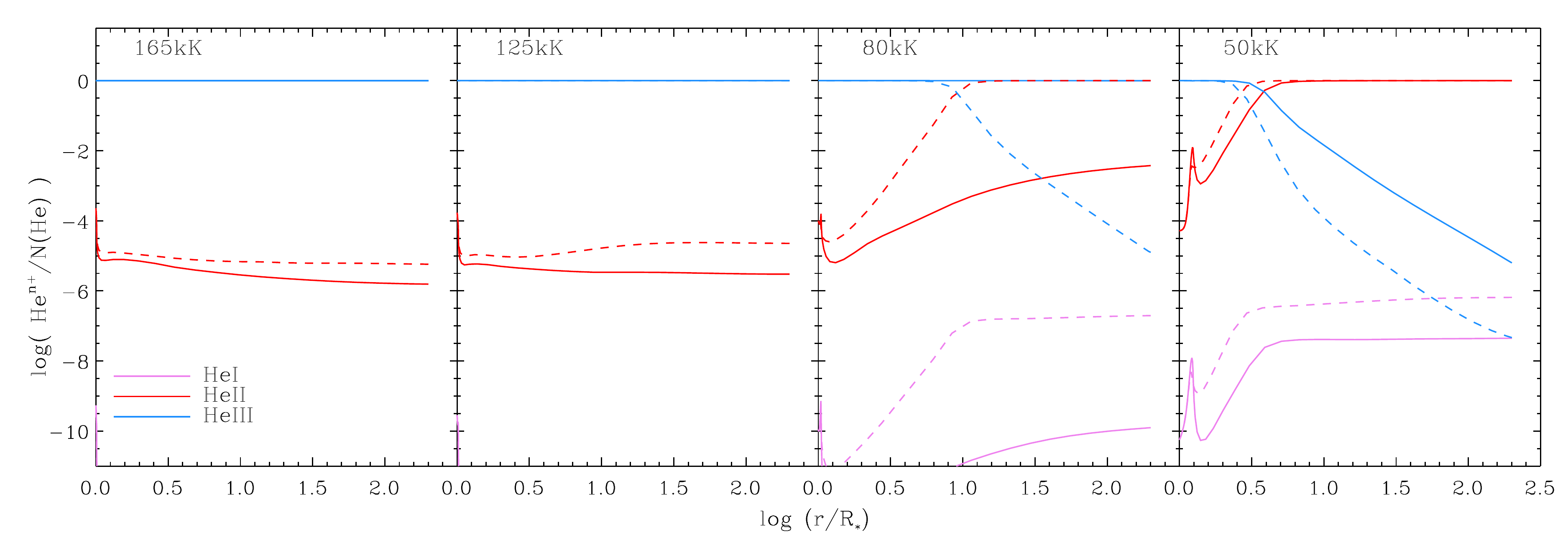}
\caption{Model profiles of the He II lines $\lambda$ $1640.4$ $\mathrm{\AA}$ 
(top panels) and $\lambda$ $4685.7$ $\mathrm{\AA}$ (central panels) at different temperatures (also radii) and mass-loss rates. The bottom panel 
displays the ionization fractions of helium, where continuous lines indicate models with 
$\dot{M}=10^{-7}$ M$_{\odot}$ yr$^{-1}$, and dashed lines indicate
$\dot{M}=10^{-6}$ M$_{\odot}$ yr$^{-1}$ if $T_{\ast}\leq80$ kK, or $\dot{M}=3\times10^{-7}$ M$_{\odot}$ yr$^{-1}$ for models with
$T_{\ast}\geq100$ kK. All the models shown belong to track B ($M_{\ast}=0.6$ M$_{\odot}$) and have $v_{\infty}=2500$ km s$^{-1}$.} \label{HeIIlines}
\end{figure}

Other lines such as He II $\lambda$ $1640.4$ $\mathrm{\AA}$ (shown on the
top-right panel of Fig. \ref{HeIIlines}), He II $\lambda$ $4685.7$ $\mathrm{\AA}$ and C IV 
$\lambda$ $4658.9$ $\mathrm{\AA}$ (central-right panel of Fig. \ref{HeIIlines}),
O III $\lambda \lambda$ $1150.9$ and $1153.8$ $\mathrm{\AA}$ (bottom-left panel of Fig. \ref{Olines}, also
showing a Ne V line at $\lambda$ $1146.1$ $\mathrm{\AA}$), C IV $\sim$ $\lambda$ $1169$ and C III $\sim$ $\lambda$ $1176$ $\mathrm{\AA}$ 
(both of them are blends of several lines) and C III $\lambda$ $2296.9$ $\mathrm{\AA}$ (Fig. \ref{Clines}, central-right and bottom panels) 
show high sensitivity to both mass loss and temperature.
Nevertheless, once temperature or mass loss are constrained by other indicators,
they can be an important diagnostic of the remaining parameter.

\begin{figure}
\includegraphics[scale=0.5]{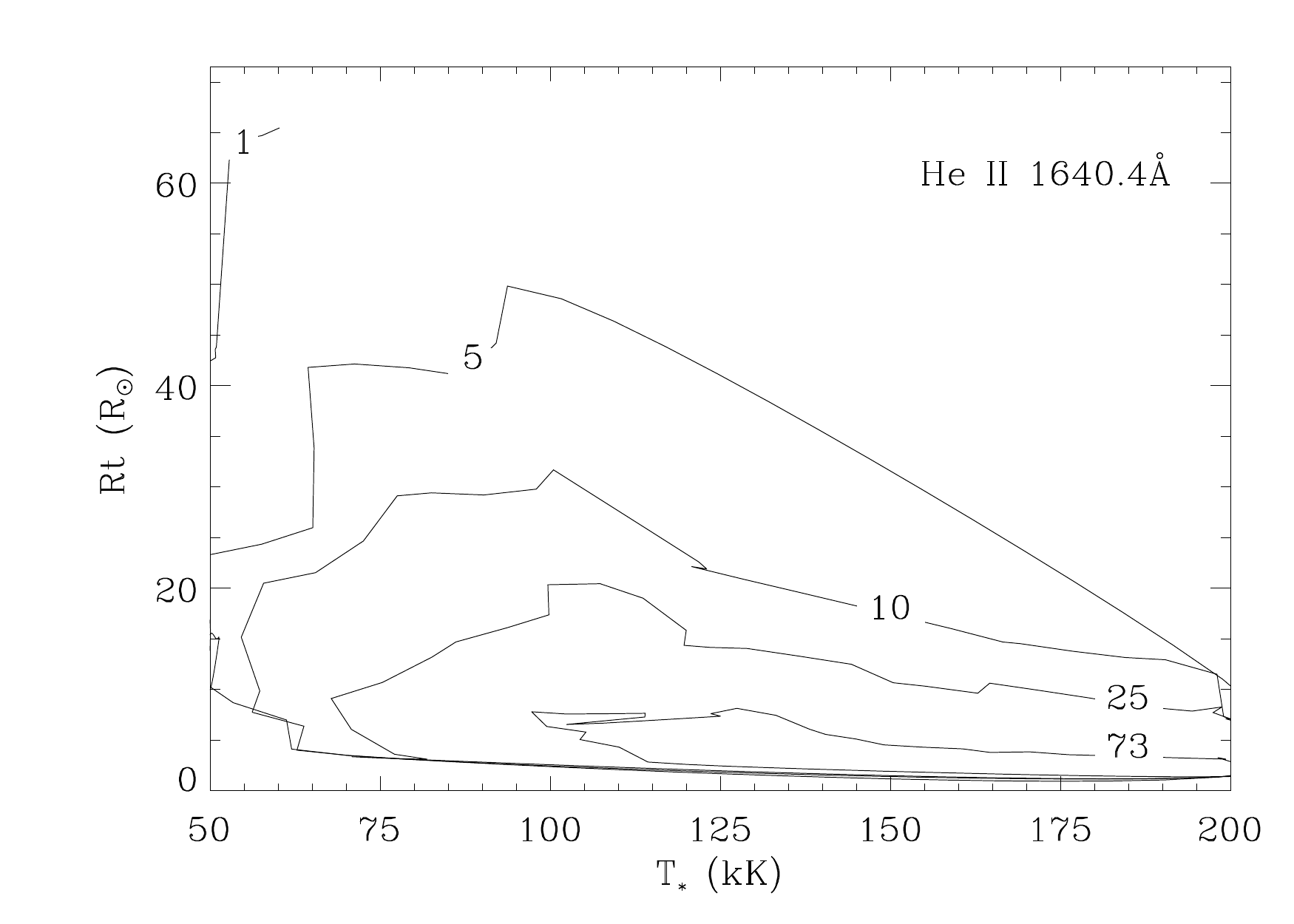}
\includegraphics[scale=0.5]{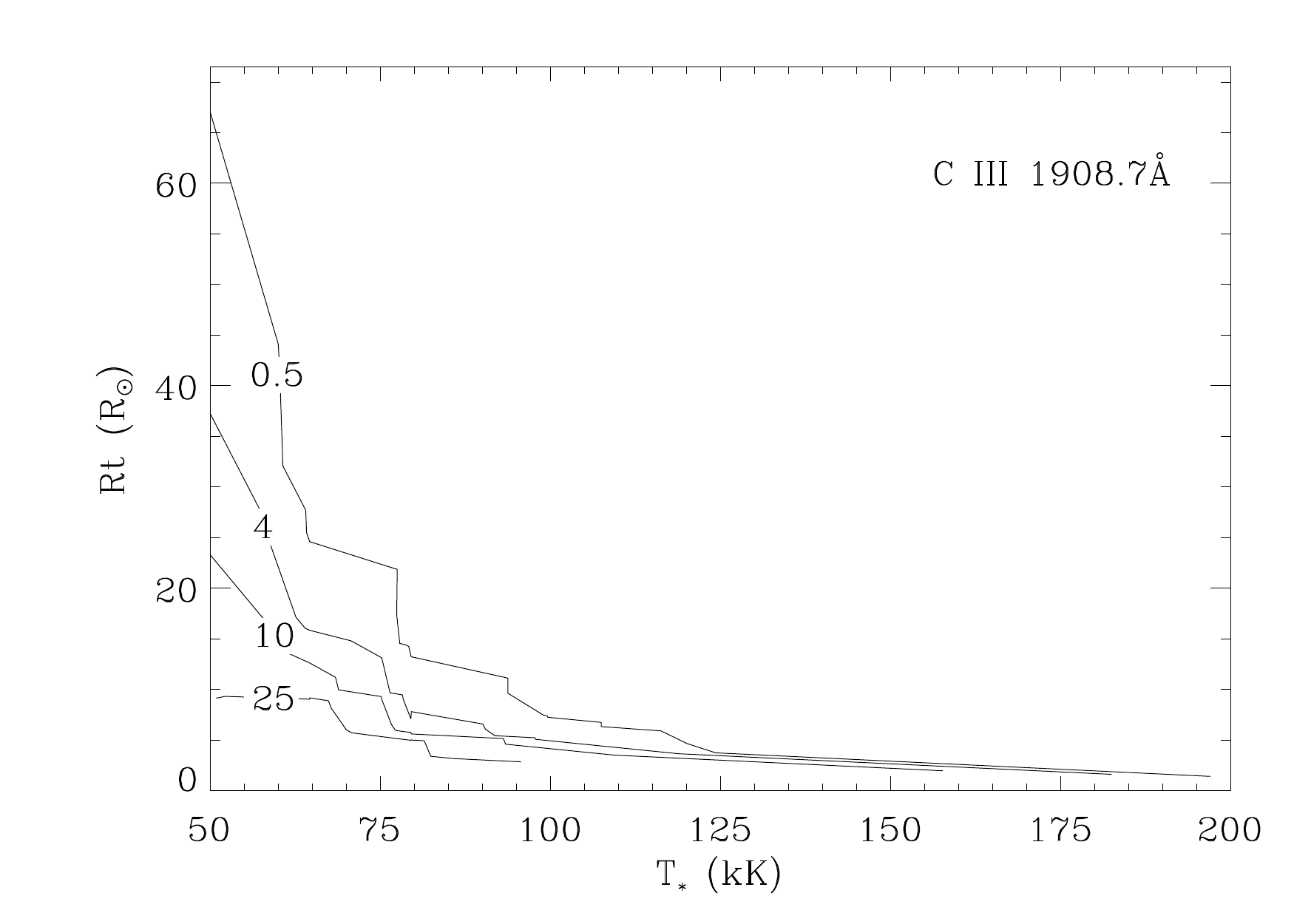}
\caption{Contours of constant equivalent widths for He II $\lambda$ $1640.4$ $\mathrm{\AA}$ and C III $\lambda$ $1908.7$ $\mathrm{\AA}$ emission lines. Labels are in
$\mathrm{\AA}$.} \label{contourplots}
\end{figure}

In order to further visualise the dependence of spectral diagnostics
on the photospheric and wind parameters, we made contour plots of the
equivalent widths of two spectral lines discussed in this section, having stellar
temperature and transformed radius on the axes. Contour plots of the He II $\lambda$ $1640.4$ $\mathrm{\AA}$ line, and the
C III $\lambda$ $1908.7$ $\mathrm{\AA}$ line are shown in Fig. \ref{contourplots},
where the contours indicate the equivalent widths of the lines. In the case of P-Cygni
profiles, CMFGEN calculates the equivalent widths of a line by subtracting that of the absorption
component from that of the emission. These two lines are very sensitive to mass loss
in the [WCE] and [WCL] parameter regimes, respectively, and thus can be used to
identify the interval of transformed radius and temperature that matches the data.

\subsection{Effect of Ne abundance}
\label{subsec:Neabundance}

A neon abundance of the order of 2 per cent by mass is expected from
evolutionary models for the [WC] stars, which present in their surfaces the
abundance pattern of the region between the hydrogen and helium burning shells
of the precursor AGB star. In AGB stars, neon is produced from nitrogen
through the $^{14}$N$(\alpha,\gamma)^{18}$F$(\beta^{+})^{18}$O$(\alpha,\gamma)
^{22}$Ne
 chain in the helium burning shell and then dredged up to the convective
intershell \citep[see, for example,][]{2006PASP..118..183W}.
Thus, the determination of neon abundances is important to constrain
evolutionary models, in particular, the dredge-up episodes. Also, the presence of several ionization stages of neon in
the winds of [WCE] stars is potentially a useful tool for the determination of stellar
temperature. In Fig. \ref{Nelines}, we show several neon features present in
the models in different temperature and mass loss regimes, for different
values of
neon abundance, ranging from solar to 11.5 times super-solar (which is adopted for the models in the grid). On the top-left panel,
the Ne VII $\lambda$ 973.3
$\mathrm{\AA}$ line seen in the grid models for [WCE] stars shows little sensitivity to
neon abundance in the models with $T_{\ast}=100000$ and $T_{\ast}=125000$ K. For higher temperatures,
this line starts to
 show a higher sensitivity to Ne abundances and can potentially help
 constrain it. Also, if used together with neon lines
 from different ionization stages, it can contribute to the determination of the
 stellar temperature, although it also depends on mass-loss rate.

In the central-right and left panels of Fig. \ref{Nelines}, we show the Ne III lines, which appear in the synthetic spectra for the [WCL] sub-class at
$\lambda$ 2553.4 $\mathrm{\AA}$ and at $\lambda \lambda$ 2677.9, 2678.6 $\mathrm{\AA}$, respectively.
Both lines present high sensitivity to neon abundance, temperature, and mass-loss rate,
and are, therefore, useful to constrain the neon abundance when the photospheric and wind parameters
have been determined through other indicators. They also help the determination of
stellar temperature when used in conjunction with other neon lines of different
ionization stages.

\subsection{Effect of X-rays}
\label{subsec:xrays}

X-rays emission of the order $L_{\mathrm{X}}\sim10^{-7}$ L$_{\ast}$ observed in massive O and
B-type stars \citep{Chlebowski1989,Chlebowski1991,Evans2003} is commonly believed to originate
from shocks in the stellar wind, which form due to the unstable nature of
the radiative force on spectral lines \citep{Lucy1980,Owocki1988}. Such a flux of
X-ray photons has been invoked in order to reproduce the strong O VI $\lambda \lambda$ $1031.9$, $1037.6$ $\mathrm{\AA}$ doublet 
in early O-type stars and to achieve 
consistency among observed N V, Si IV and C IV diagnostics \citep{Garcia2004a,Bianchi2009}.

Since the fast winds of CSPNe are also believed to be subjected to the radiation pressure in
spectral lines, they
are expected to be clumped and may exhibit X-ray fluxes. \cite{Guerrero2001} measured an 
X-ray luminosity of $\sim10^{-7}$ L$_{\ast}$ for the central star of NGC 6543, consistent with that predicted from 
shocks in the stellar wind. However,
the emission could also be explained by coronal activity of an undetected companion.

\begin{figure}
\centering
\includegraphics[scale=0.435]{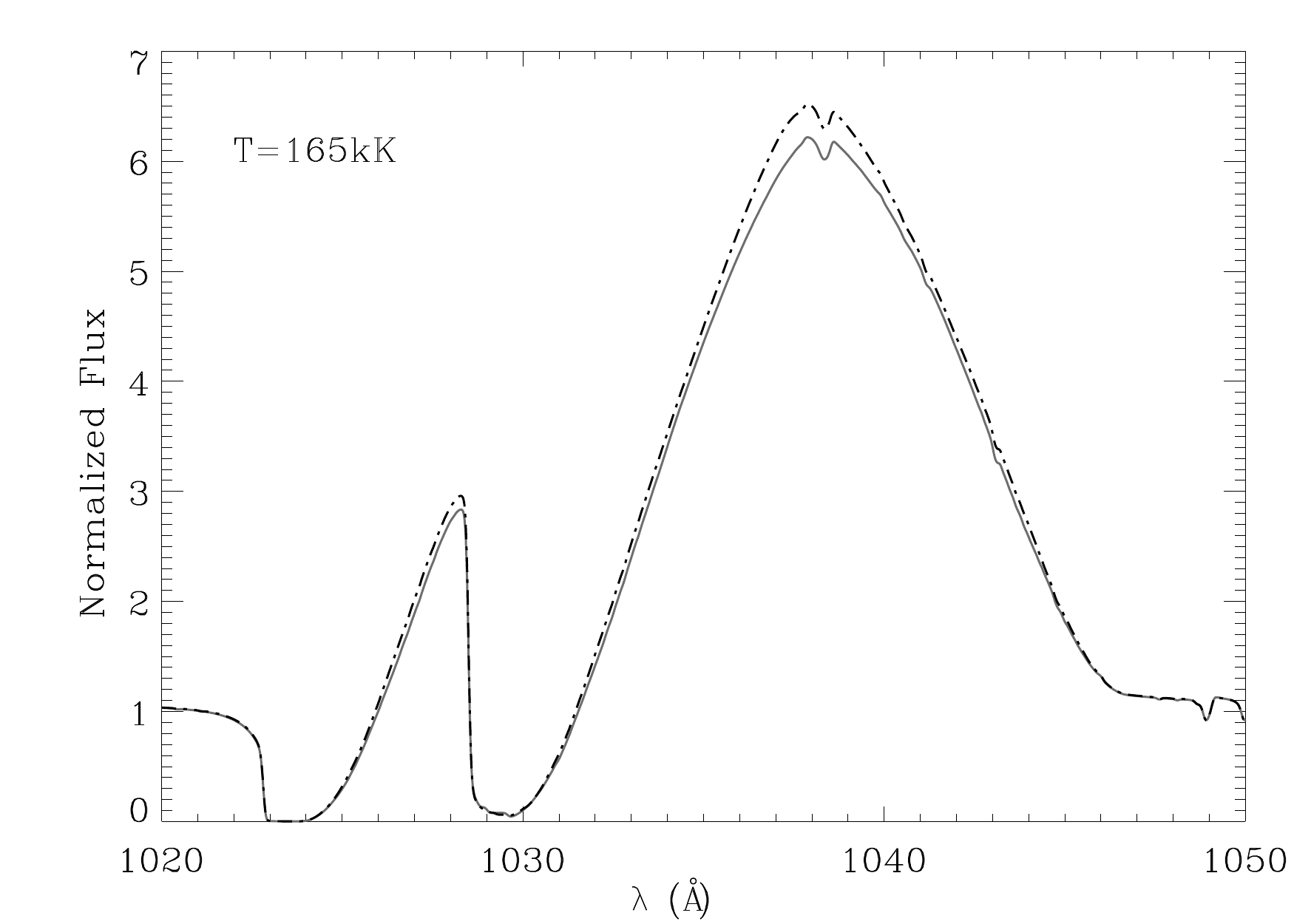}
\includegraphics[scale=0.435]{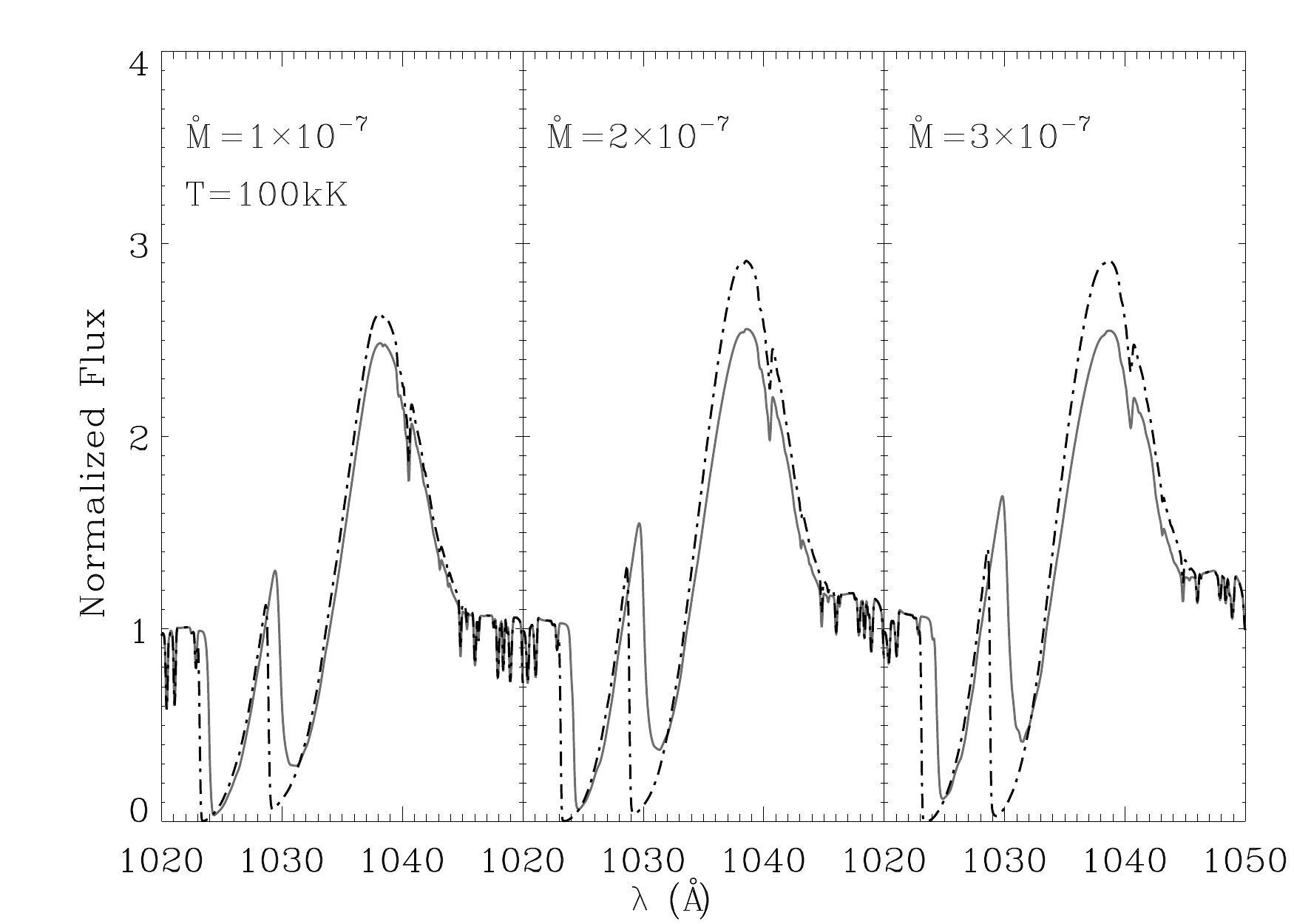}
\includegraphics[scale=0.435]{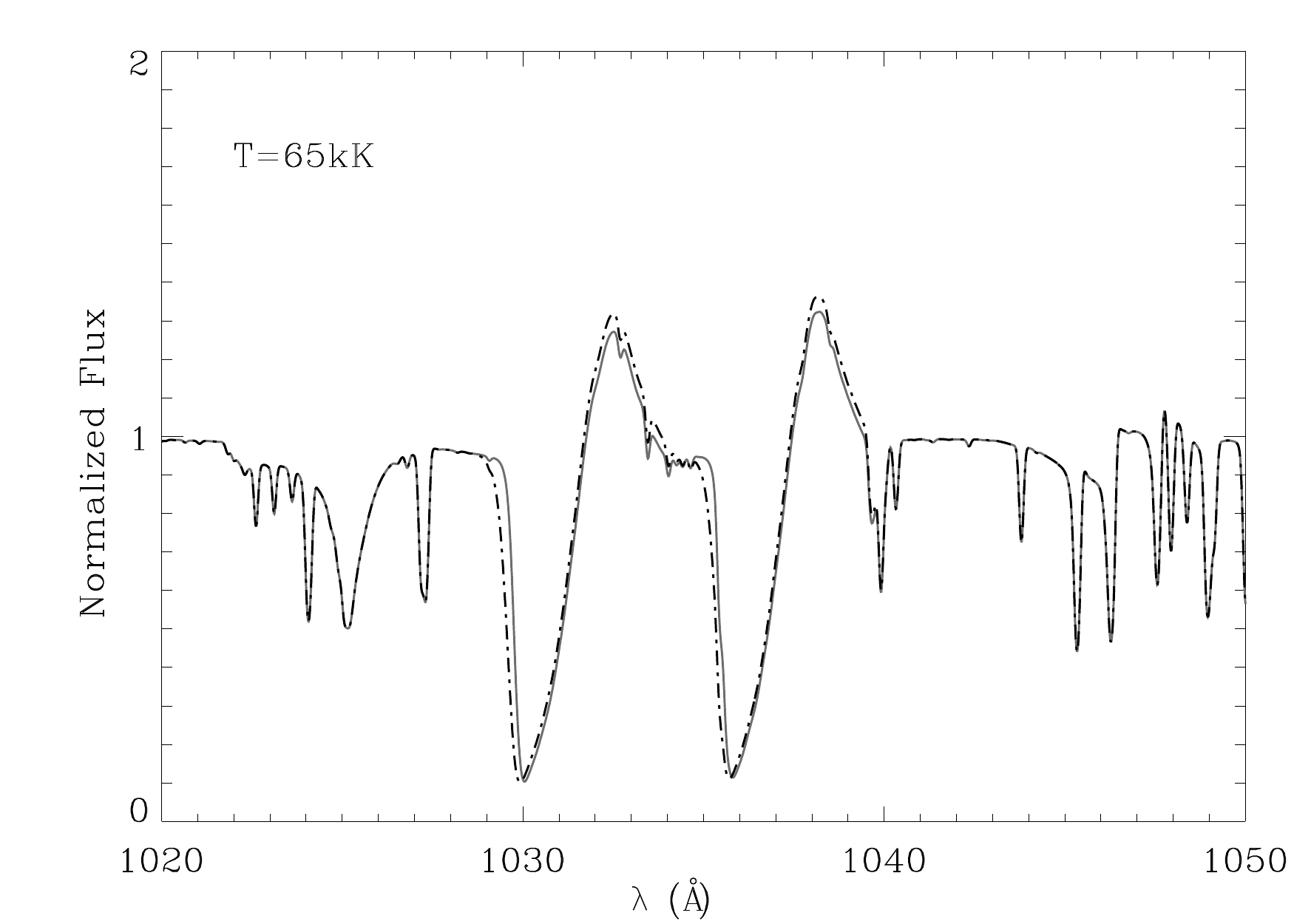}
\caption{Comparison between models with (dash-dotted line) and without
(continuous line) X-rays. The O VI $\lambda \lambda$ $1031.9$, $1037.6$ $\mathrm{\AA}$ doublet is shown due to its 
sensitivity to X-ray ionization (see text). For the $T_{\ast}=165$ kK model, an extreme  X-ray luminosity of
$\sim10^{-2}$ L$_{\ast}$ is shown, for the $T_{\ast}=100$ kK models, we show 
$L_{\mathrm{X}}\sim10^{-7}$ L$_{\ast}$, and for the $T_{\ast}=65$ kK model, $L_{\mathrm{X}}=4\times10^{-10}$ L$_{\ast}$}\label{xrays}
\end{figure}

We examined the effect of soft X-ray fluxes in the synthetic spectra computed for
our grid. We re-computed models with $T_{\ast}=65000$, $100000$, and $165000$ K including soft X-ray fluxes 
into the calculations. 
We assumed shock temperatures of $5 \times 10^{6}$ and $3 \times 10^{6}$ K, starting at velocities of
$300$ and $400$ km s$^{-1}$, respectively. In the model with $T_{\ast}=165000$ K,
only an extremely high X-ray flux yielding an
observed X-ray luminosity of $\sim10^{-2}$ L$_{\ast}$
(for a mass-loss rate of $1.0\times10^{-7}$ M$_{\odot}$ yr$^{-1}$),
caused a relevant
effect on the spectra, mainly in the O VI doublet $\lambda \lambda$ $1031.9$, $1037.6$ $\mathrm{\AA}$,
as shown in Fig. \ref{xrays}. In models with progressively lower temperatures,
we found that lower X-ray luminosities modify the O VI doublet profile. For the $T_{\ast}=100000$ K model, a reasonable
X-ray luminosity of $\sim10^{-7}$ L$_{\ast}$ (for $\dot{M}=1.0\times10^{-7}$, $\dot{M}=2.0\times10^{-7}$,
and $\dot{M}=3.0\times10^{-7}$ M$_{\odot}$ yr$^{-1}$) significantly alters the
O VI doublet (but not other spectral features), as shown in Fig. \ref{xrays}. At a temperature of $T_{\ast}=65000$ K 
the convergence of models with X-ray fluxes proved to be very hard to reach. Nevertheless, we
were able to obtain a converged model for a mass-loss rate of $\dot{M}=5.0\times10^{-8}$ M$_{\odot}$ yr$^{-1}$
with an X-ray luminosity of $4\times10^{-10}$ L$_{\ast}$. Again, the only spectral feature
affected was the far-UV O VI doublet, which is also shown in Fig. \ref{xrays}.

\section{Spectral Analysis of the central star of NGC 6905}
\label{sec:ObservedSpectra}

\begin{figure}
\includegraphics[scale=0.435]{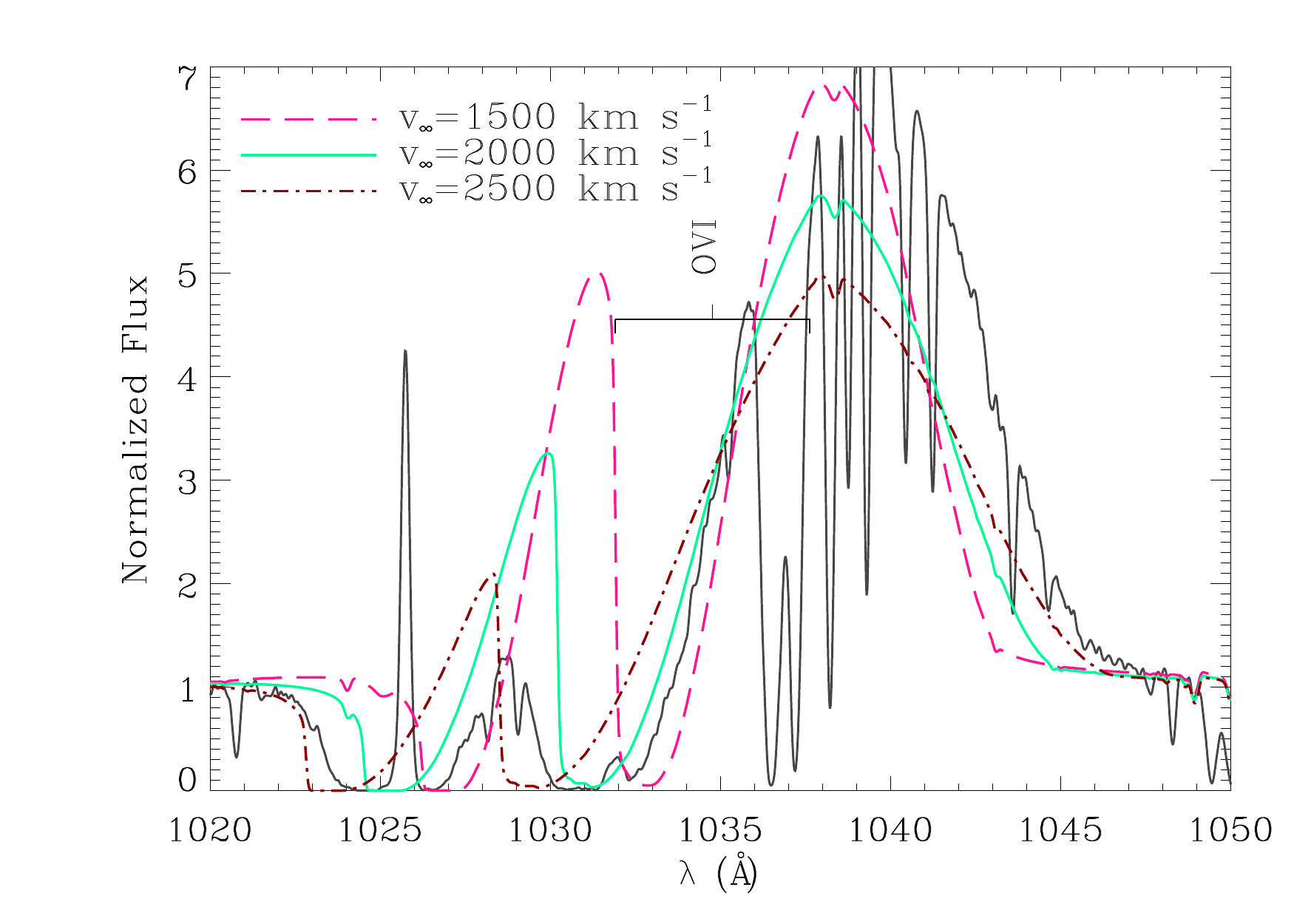}
\includegraphics[scale=0.435]{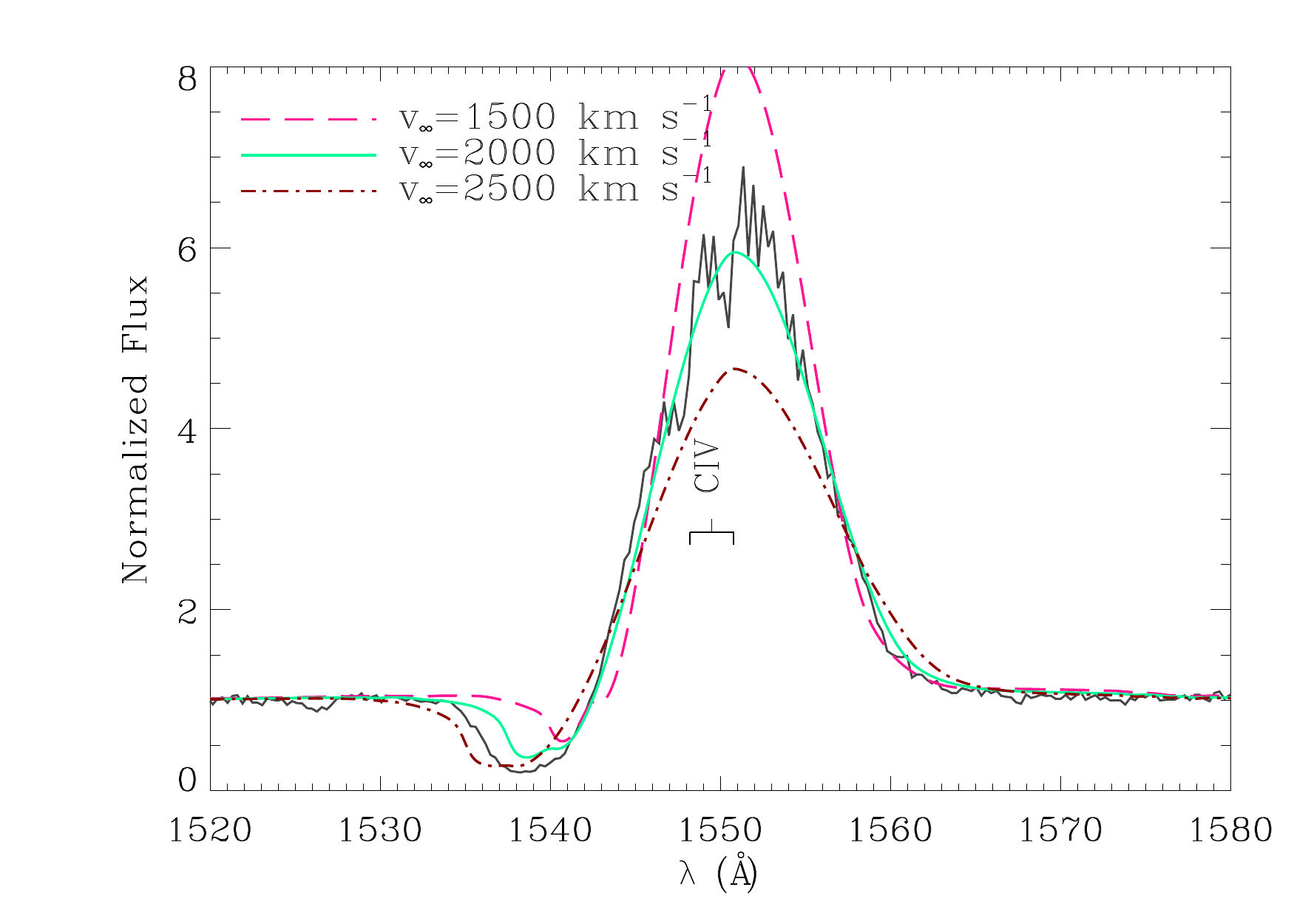}
\caption{ The two lines showing the strongest observed (continuous black line) P-Cygni profiles in the spectra of NGC 6905, 
O VI $\lambda \lambda$ $1031.9$, $1037.6$ $\mathrm{\AA}$ in the FUSE range and C IV $\lambda \lambda$ $1548.2$, $1550.8$ $\mathrm{\AA}$ 
in the STIS G140L range, are shown, as well as models computed with three values of $v_{\infty}$ from our model grid. 
All other parameters in the models are identical and correspond to our derived values.}
\label{terminalvelocities}
\end{figure}

\begin{figure}
\centering
\includegraphics[scale=0.435]{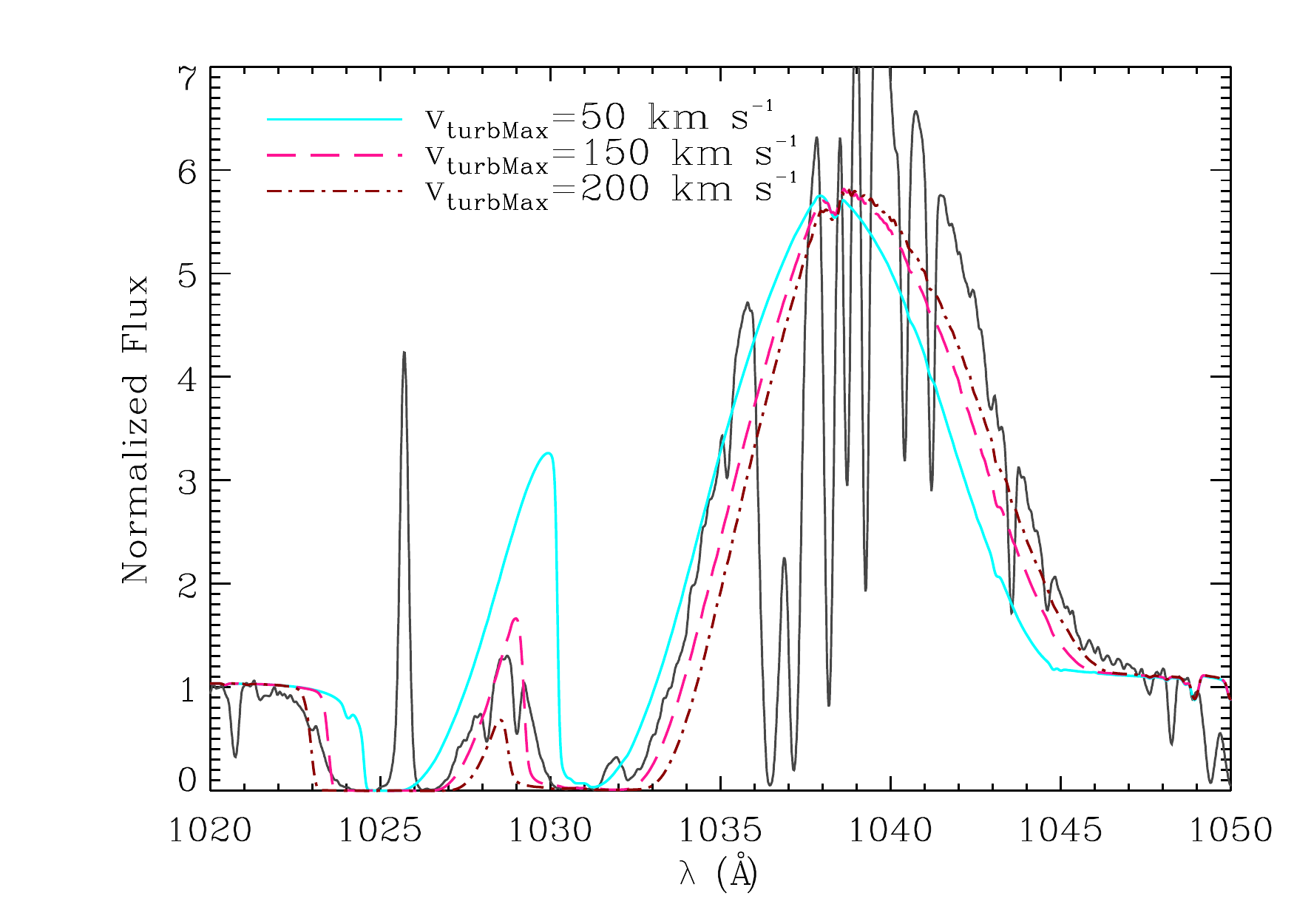}
\includegraphics[scale=0.435]{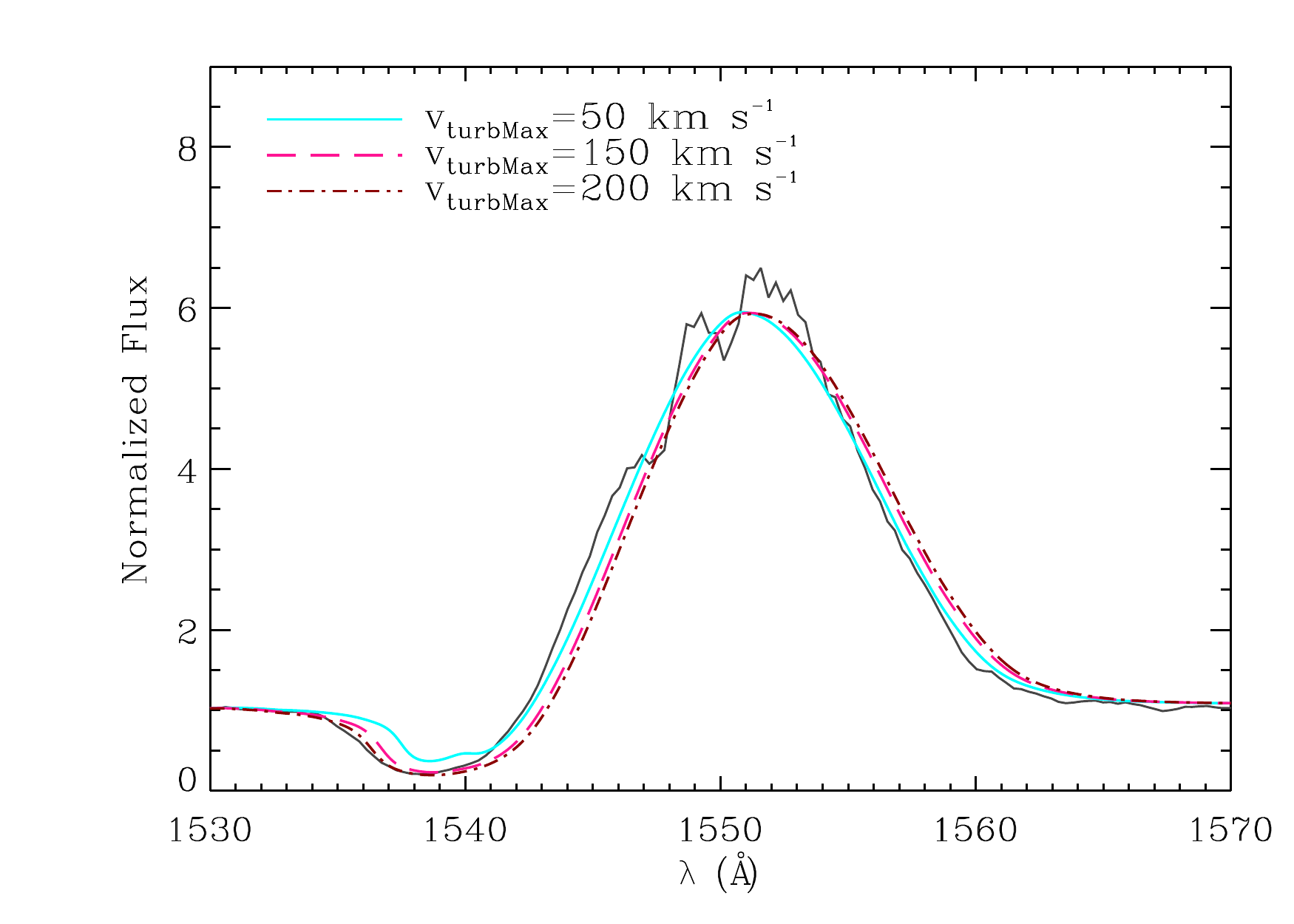}
\caption{ The two strongest P-Cygni line profiles present in the observed spectra of NGC 6905 (continuous black line), the O VI 
$\lambda \lambda$ $1031.9$, $1037.6$ $\mathrm{\AA}$ and the C IV $\lambda \lambda$ $1548.2$, $1550.8$ $\mathrm{\AA}$, are compared with models with 
$T_{\ast}=150$ kK,
$\dot{M}=10^{-7}$ M$_{\odot}$ yr$^{-1}$, $v_{\infty}=2000$ km s$^{-1}$ and
different turbulence velocities.}\label{vturb}
\end{figure}

\begin{figure}
\centering
\includegraphics[scale=0.435]{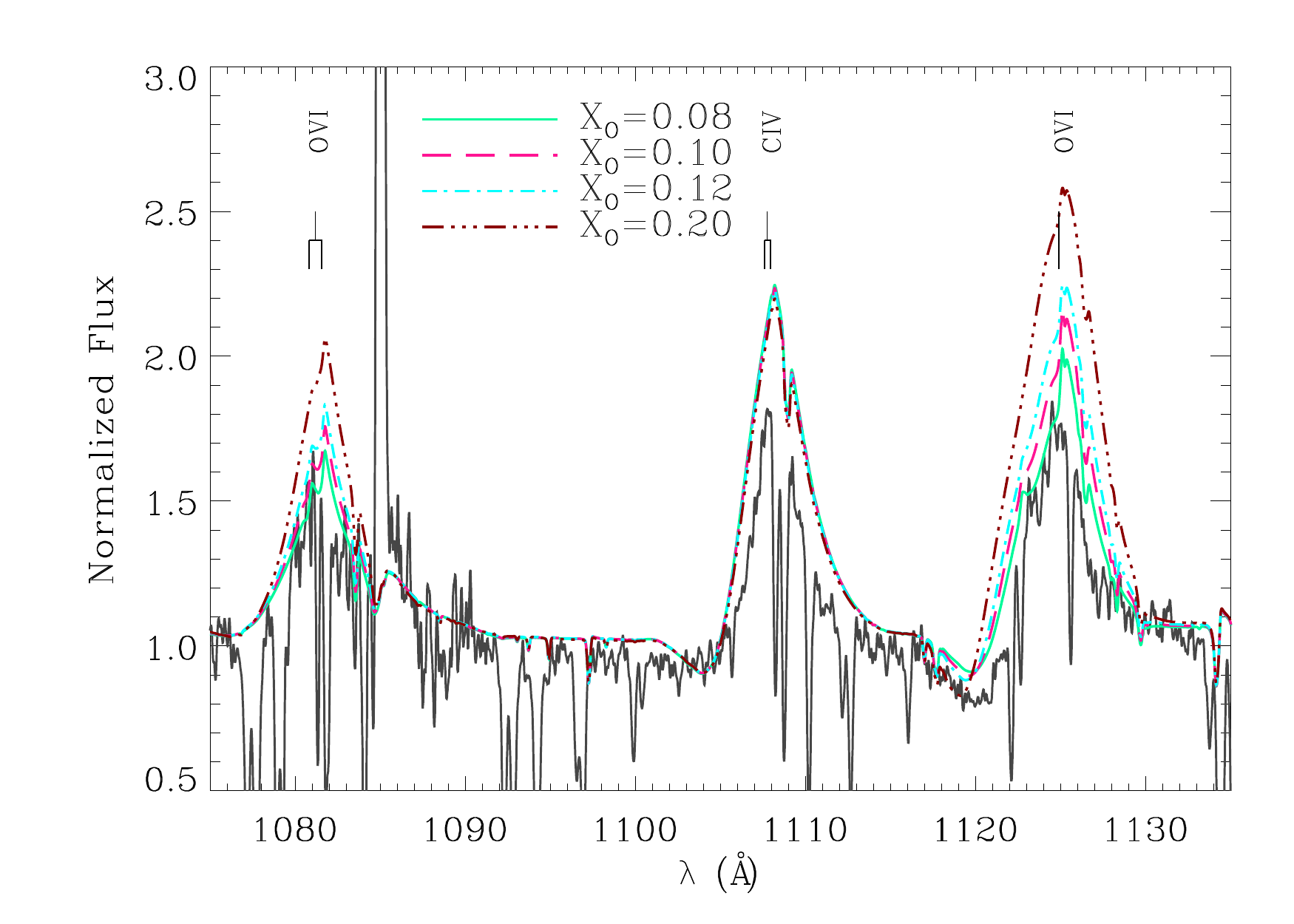}
\includegraphics[scale=0.435]{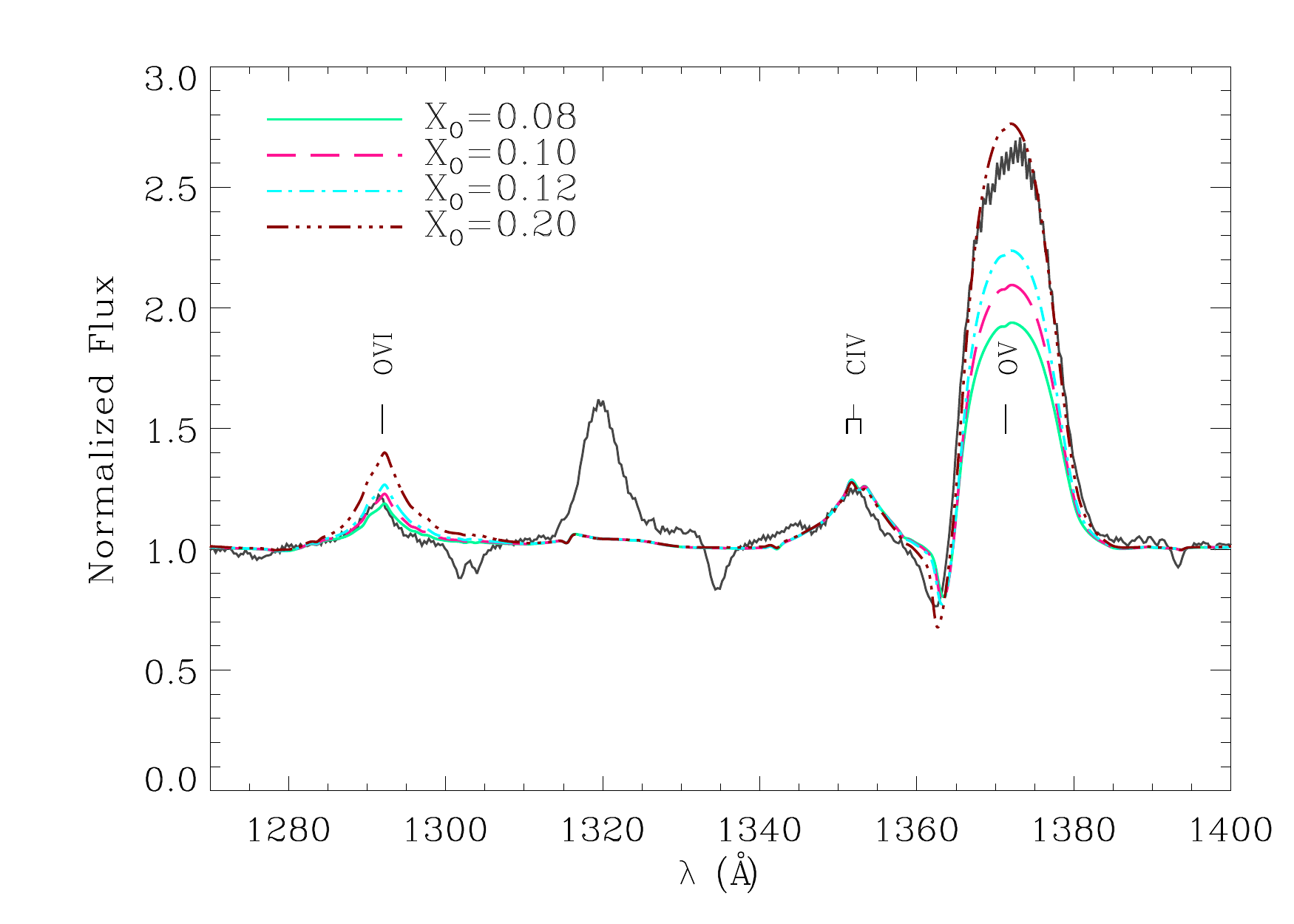}
\includegraphics[scale=0.435]{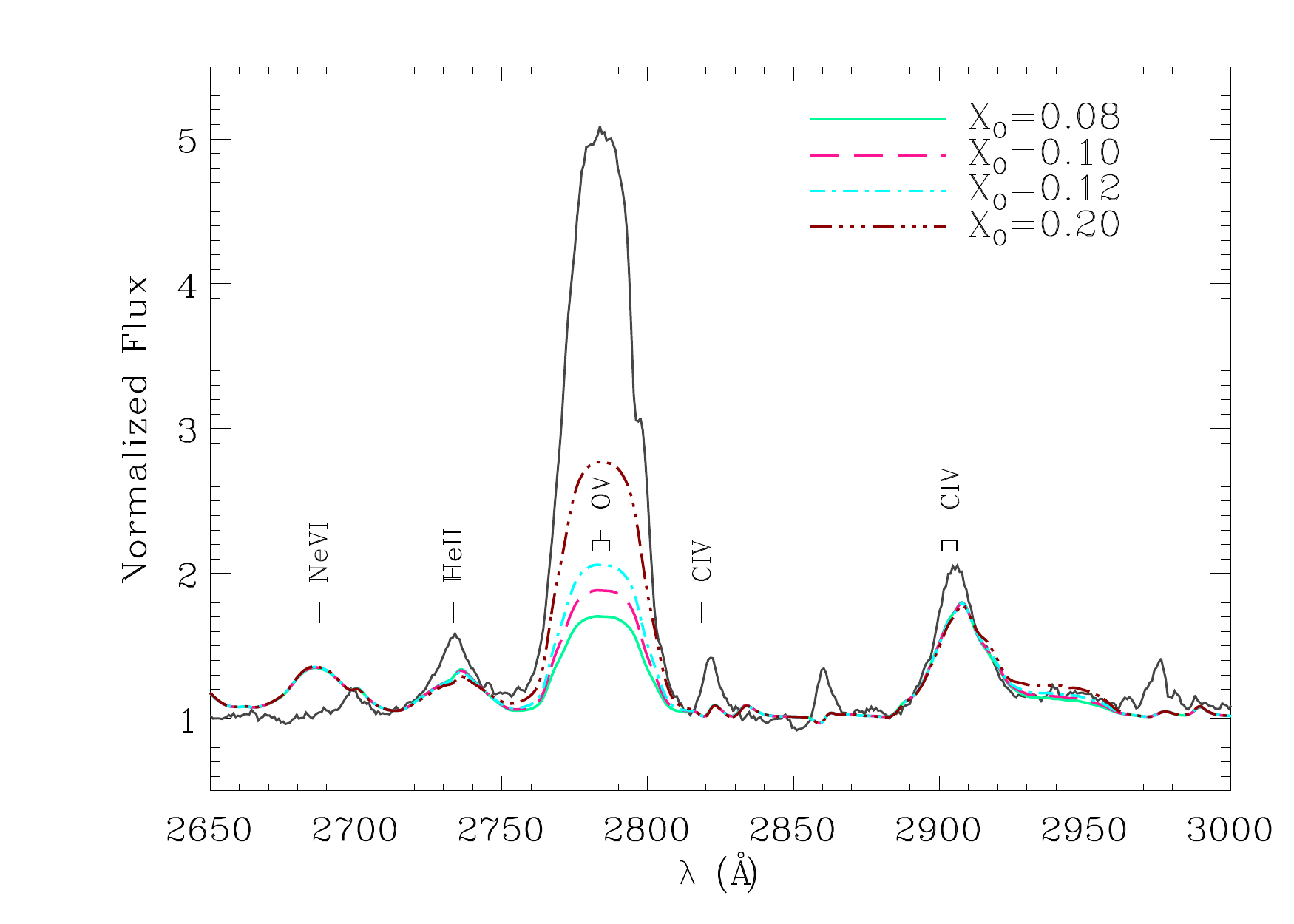}
\caption{ Observed spectra of the central star of NGC 6905 (continuous black line) and models with $T_{\ast}=150$ kK,
$\dot{M}=10^{-7}$ M$_{\odot}$ yr$^{-1}$, $v_{\infty}=2000$ km s$^{-1}$ and different
values of oxygen abundance. The oxygen mass fraction adopted for the grid models is $X_{\mathrm{O}}=0.08$.}\label{Oabundance}
\end{figure}

\begin{figure}
\centering
\includegraphics[scale=0.435]{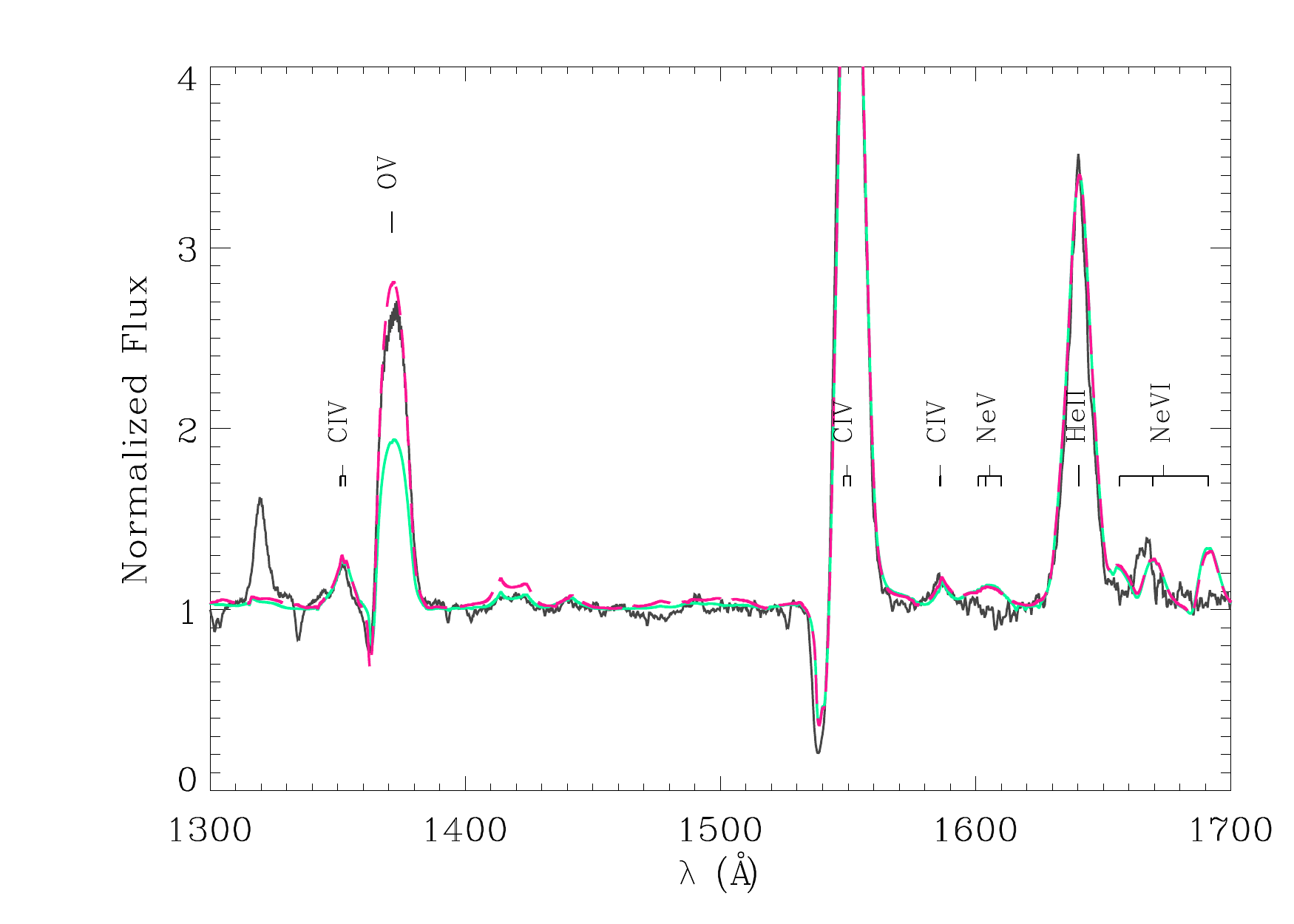}
\includegraphics[scale=0.435]{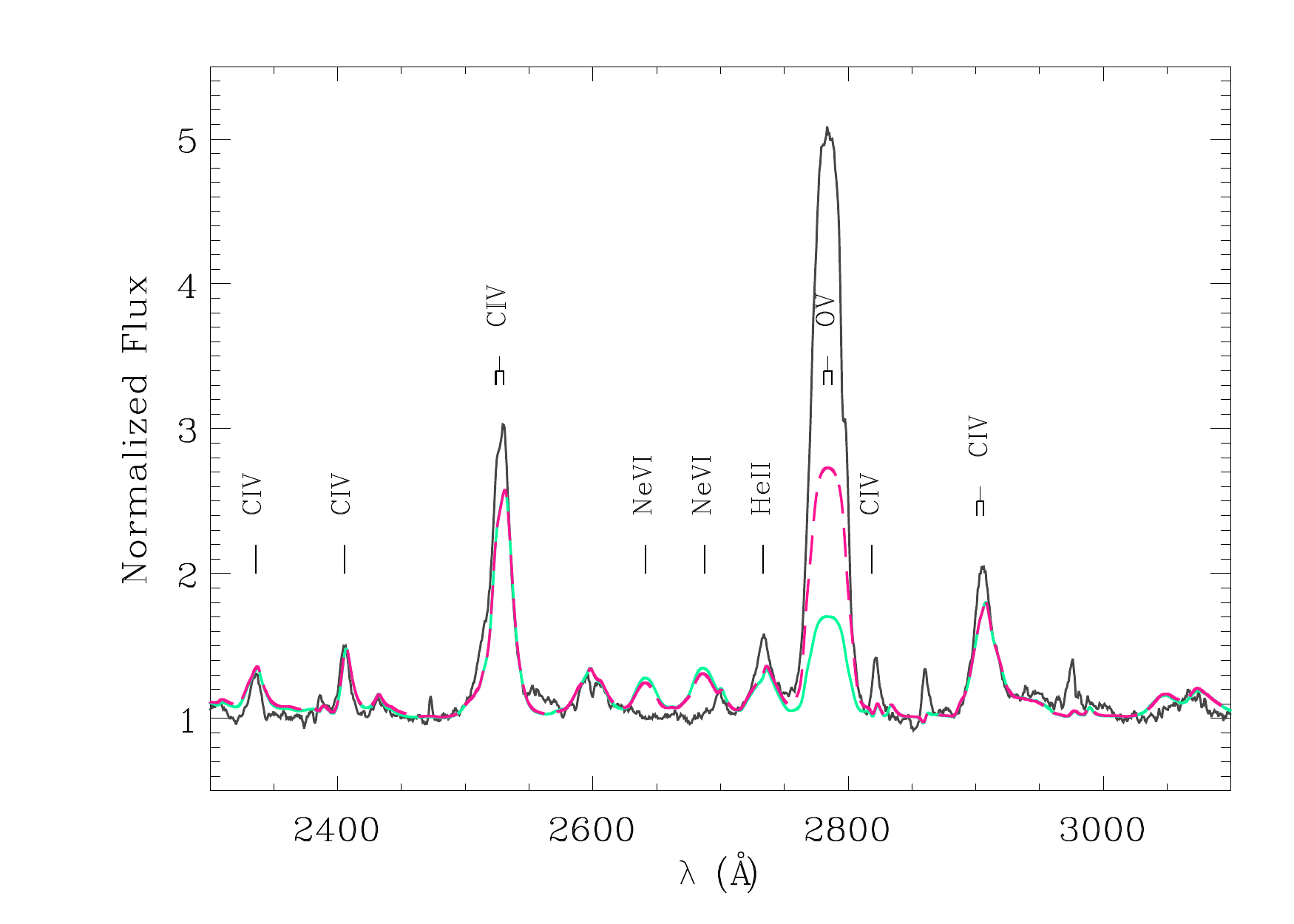}
\caption{ The effect on the synthetic spectra of including new ions in the calculation is seen almost exclusively in the profiles of 
the O V lines. The observed spectra are shown as a continuous black line. The pink/dashed and green/continuous lines, 
show models with and without the inclusion of new ions, respectively. The 
models have $T_{\ast}=150$ kK, $\dot{M}=10^{-7}$ M$_{\odot}$ yr$^{-1}$, $v_{\infty}=2000$ km s$^{-1}$.}\label{newions}
\end{figure}

\begin{figure}
\centering
\includegraphics[angle=0,scale=0.435]{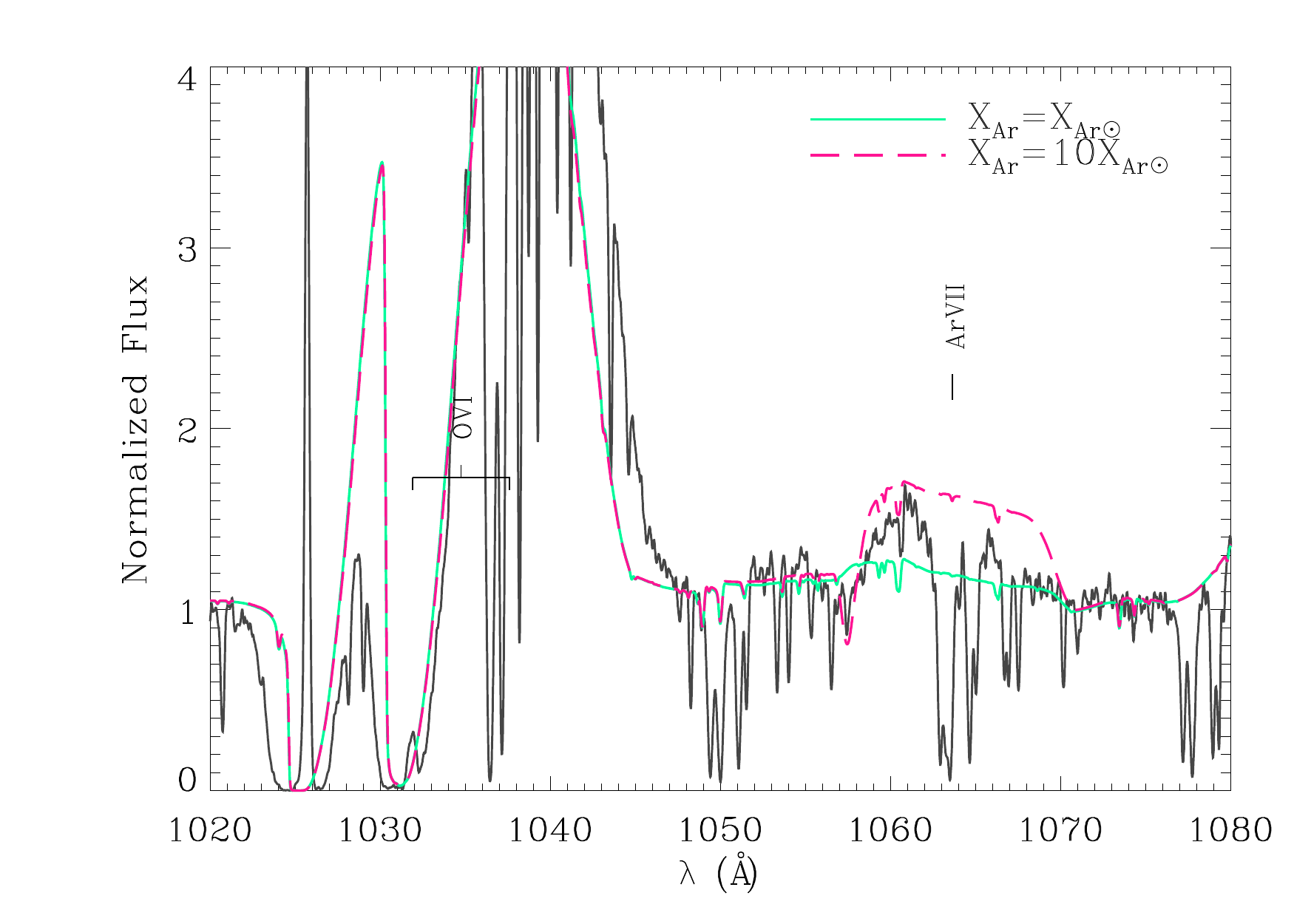}
\caption{ Ar VII $\lambda$ $1063.55$ $\mathrm{\AA}$ in models with different argon abundances. 
Both models shown have $T_{\ast}=150$ kK, $\dot{M}=10^{-7}$ M$_{\odot}$ yr$^{-1}$, 
$v_{\infty}=2000$ km s$^{-1}$. The observed spectrum of NGC 6905 is shown as a continuous black line.}
\label{Argon}
\end{figure}

\begin{figure}
\centering
\includegraphics[angle=0,scale=0.435]{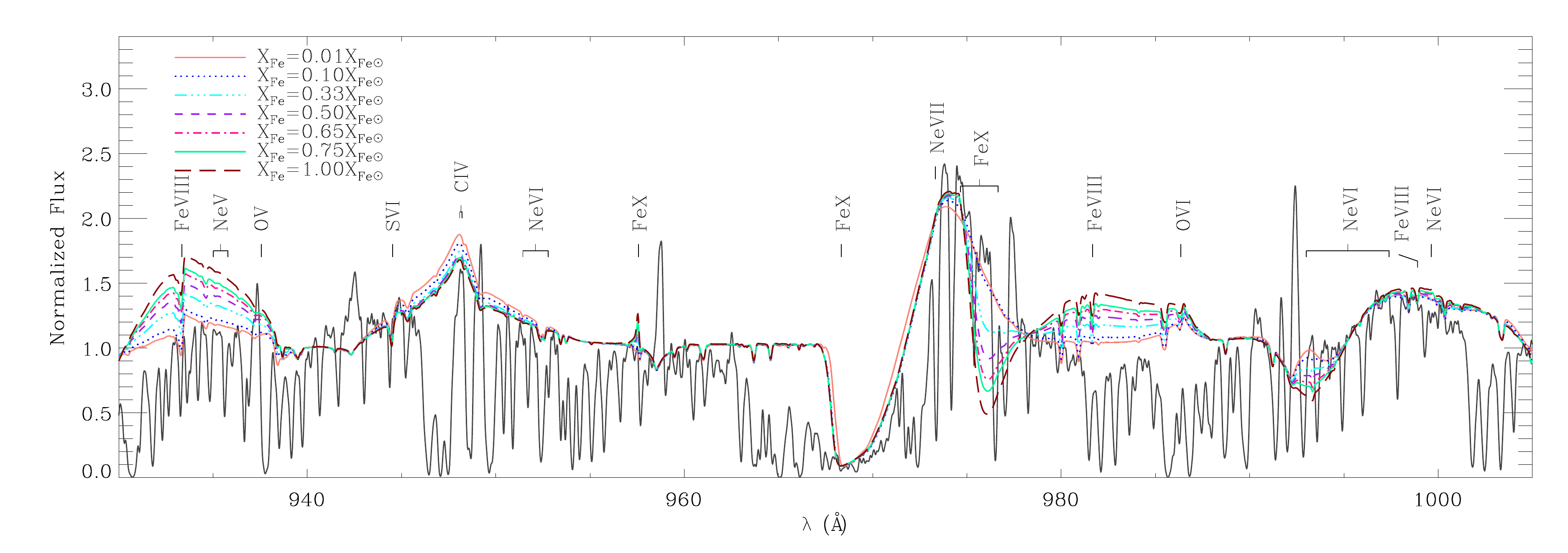}
\caption{ The observed spectrum of the central star of NGC 6905 (continuous black line) is compared with synthetic spectra 
for different iron abundances. The models have $T_{\ast}=150$ kK, $\dot{M}=10^{-7}$ M$_{\odot}$ yr$^{-1}$, 
$v_{\infty}=2000$ km s$^{-1}$. The numerous narrow absorptions are from interstellar H$_{2}$.}
\label{Iron}
\end{figure}

To illustrate the usefulness of the model grid, we apply it here to derive
the physical parameters of the [WCE] central star of NGC 6905.
We compare our grid models with data (Table \ref{spectra}) available at the Multimission
Archive at STScI (MAST) from FUSE and HST spectrographs. FUSE covers the
wavelength range 905-1187 $\mathrm{\AA}$ with a resolving power of $\la$
20000. We also used HST-STIS spectra obtained with G140L (R $\sim$ 1200) and
G230L (R $\sim$ 750) gratings with wavelength coverage of
1150-1736 $\mathrm{\AA}$ and 1570-3180 $\mathrm{\AA}$, respectively. In all figures showing the observed spectra of NGC 6905, 
the resolution of the models, in the STIS wavelength range, is matched to the resolution of the observations by convolving the 
synthetic spectra with the instrument line spread functions, while for the spectral region 
covered by FUSE, which is full of interstellar hydrogen absorptions, the models
and observations were convolved with a Gaussian of $FWHM=0.1$ $\mathrm{\AA}$ to facilitate
the visualization. In Figs. \ref{terminalvelocities}-\ref{Iron}, the interstellar absorption was omitted from the synthetic spectra to 
facilitate visualization and will be further discussed below. The line identifications in these figures only mark the strongest lines in each model.

We obtained an initial estimate of the wind's terminal velocity by measuring the blue edge velocity, $v_{\mathrm{edge}}$, from the strong P-Cygni
profile of the C IV $\lambda$ $1548.2$ $\mathrm{\AA}$ line. We derived $v_{\infty}=2170$ km s$^{-1}$, correcting $v_{\mathrm{edge}}$ for an assumed 
turbulence velocity, $v_{\mathrm{turb}}$, of 10 per cent $v_{\infty}$ \citep{1999isw..book.....L}.
\begin{equation}
v_{\infty} \simeq v_{\mathrm{edge}}-2 \times v_{\mathrm{turb}}.
\end{equation}
This estimate
is confirmed by comparing the observed spectra with grid models computed for wind's terminal velocities $v_{\infty}=2500$, $2000$ and $1500$ km s$^{-1}$. 
An example is shown in Fig. \ref{terminalvelocities}.

We then used the grid of models with the appropriate wind's terminal velocity
to constrain the temperature and the transformed radius values of the central star of NGC 6905. Based on these
results, we then perform a more detailed analysis, varying additional parameters
not explored by the grid.

\subsection{Using the grid to constrain temperature and transformed radius}

The central star of NGC 6905 is a [WCE], therefore its spectra are poor in lines of 
multiple ions of the same element. Among the features identified in the spectra,
only oxygen shows strong lines of multiple ionization stages. Several O VI
lines are present ($\lambda \lambda$ $1031.9$, $1037.6$
$\mathrm{\AA}$, $\lambda \lambda$ $1080.8$, $1081.6$ $\mathrm{\AA}$,
$\lambda \lambda$ $1122.5$, $1124.8$, $1124.9$ $\mathrm{\AA}$, $\lambda$ $1291.9$ $\mathrm{\AA}$,
$\lambda$ $1996.1$ $\mathrm{\AA}$, $\lambda$ $2070.4$ $\mathrm{\AA}$, $\lambda$ $2082.04$ $\mathrm{\AA}$
and $\lambda$ $2431.409$ $\mathrm{\AA}$), along with two strong O V lines at $\lambda$
$1371.3$ $\mathrm{\AA}$ and $\lambda \lambda$ $2781.0$,
$2787.0$ $\mathrm{\AA}$. However, as
will be discussed below, these two O V lines are only matched by grid models incompatible
with other spectral diagnostics and both show sensitivity to the inclusion of 
heavy elements in the calculations, while other lines do not. Therefore, they are more useful as indirect indicator of abundance and not to uniquely constrain temperature.
Several C IV lines are also observed and used as temperature diagnostic.

\begin{figure}
\centering
\includegraphics[scale=0.435]{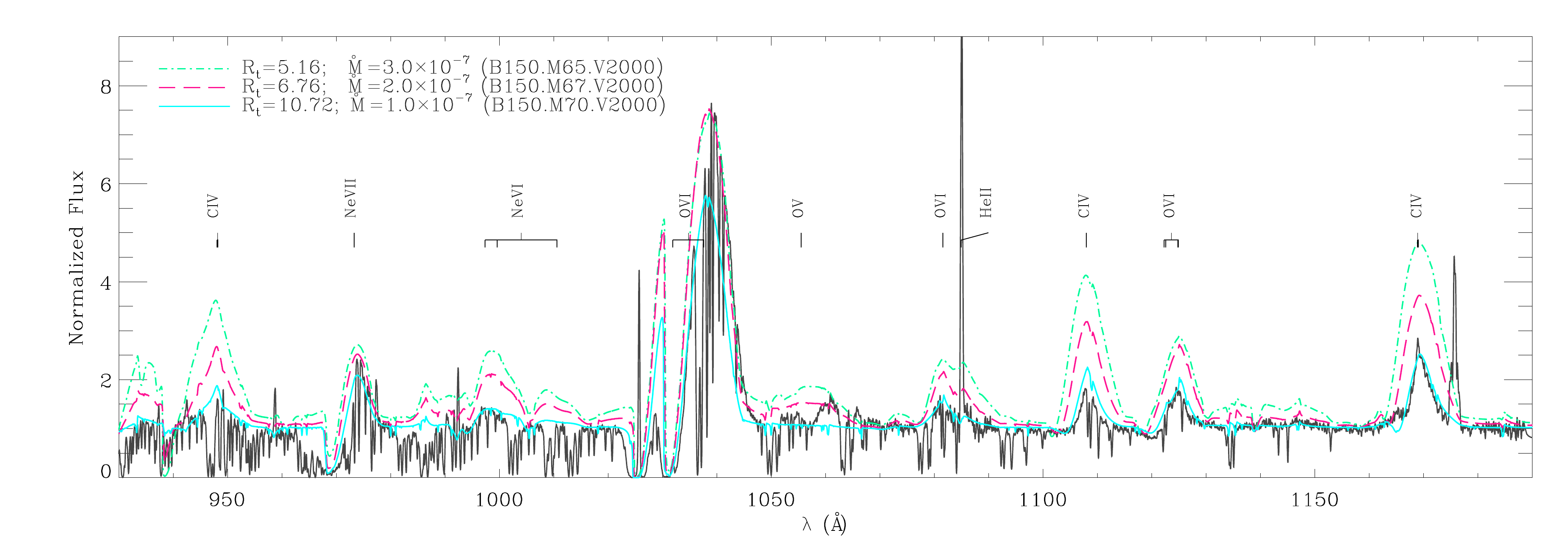}
\includegraphics[scale=0.435]{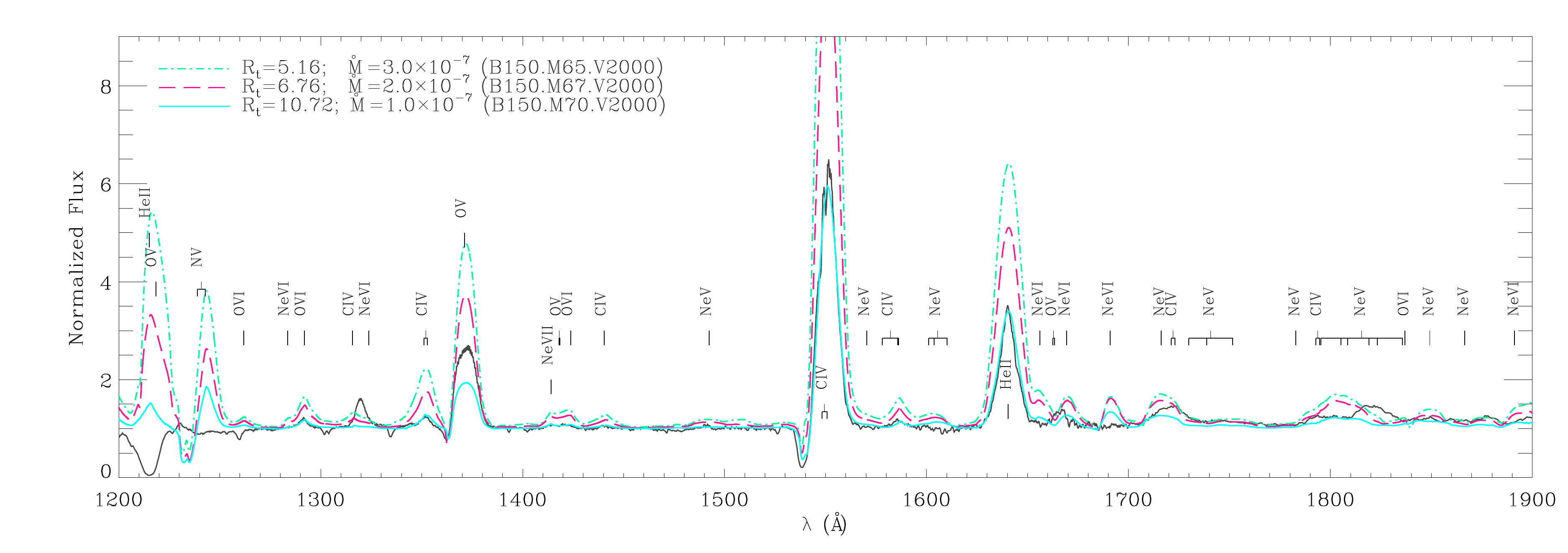}
\includegraphics[scale=0.435]{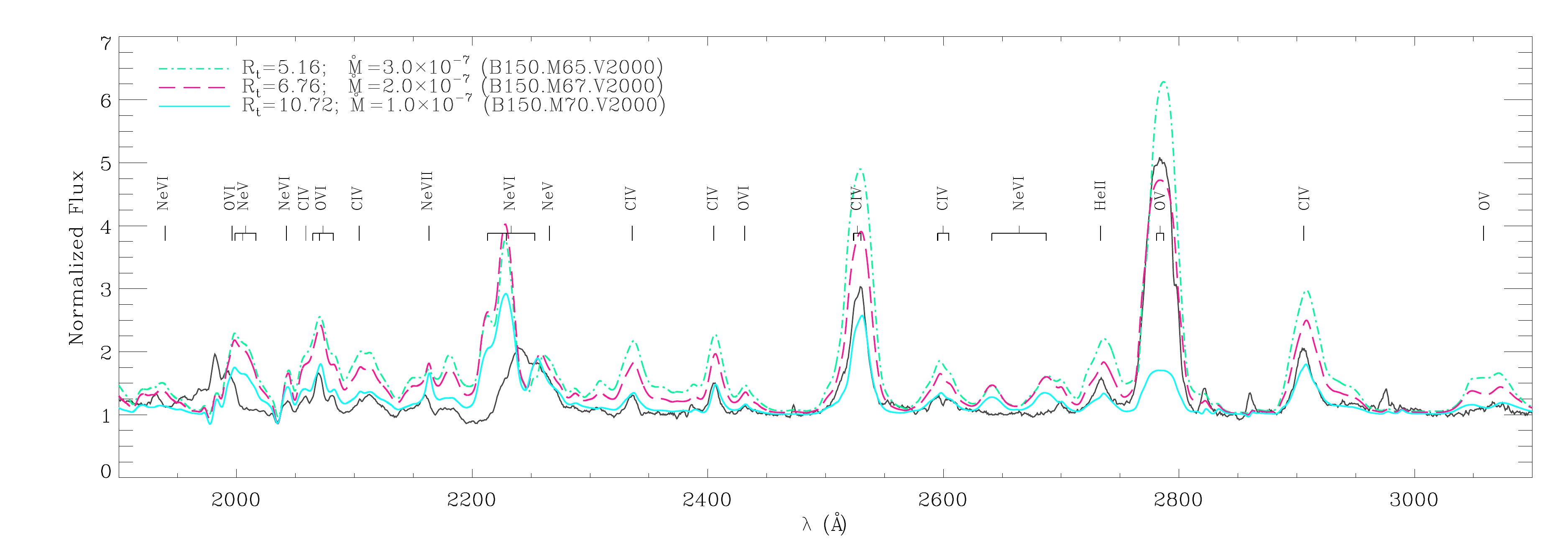}
\caption{ UV and far-UV observed spectra (continuous black line) of the central star of NGC 6905 and three
models with $T_{\ast}=150$ kK and different values of transformed radius (transformed radius and mass-loss rate are given in 
units of solar radii and M$_{\odot}$ yr$^{-1}$, respectively). Among the grid models, the light blue continuous line is 
the one that fits best the observed diagnostics, except for the two O V lines which are better fit by higher mass-loss rates and some weak neon 
features (discussed in the text).}\label{NGC6905T150}
\end{figure}

\begin{figure}
\centering
\includegraphics[scale=0.435]{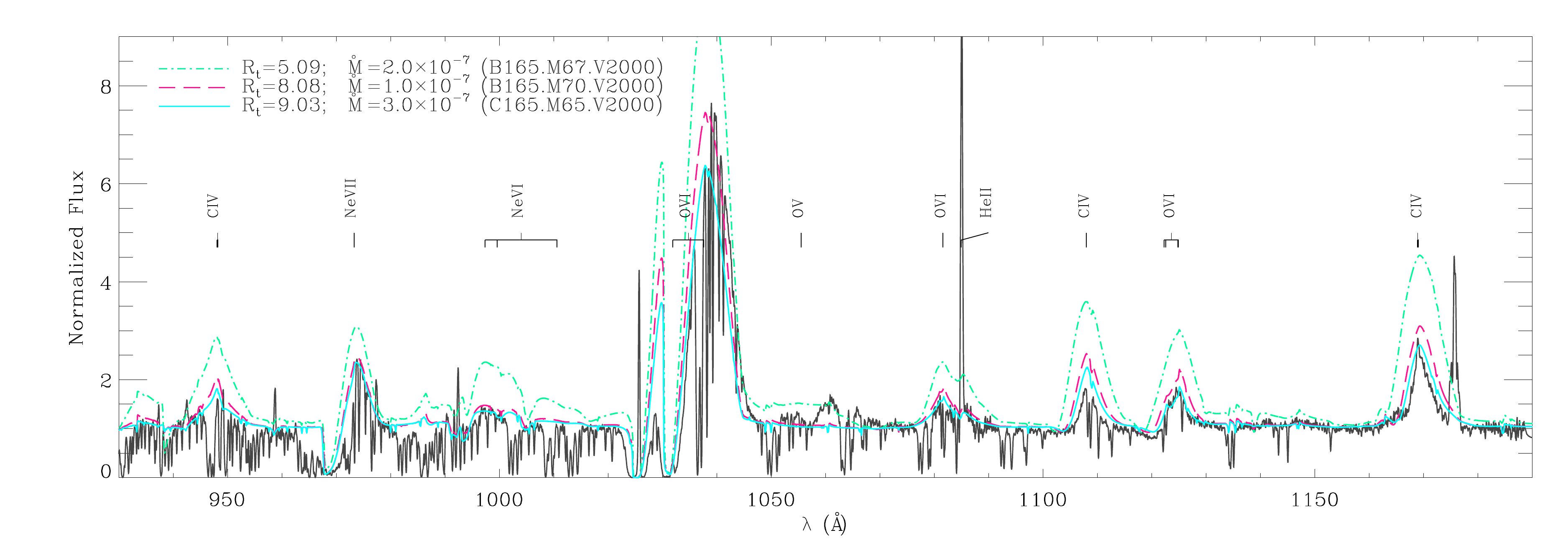}
\includegraphics[scale=0.435]{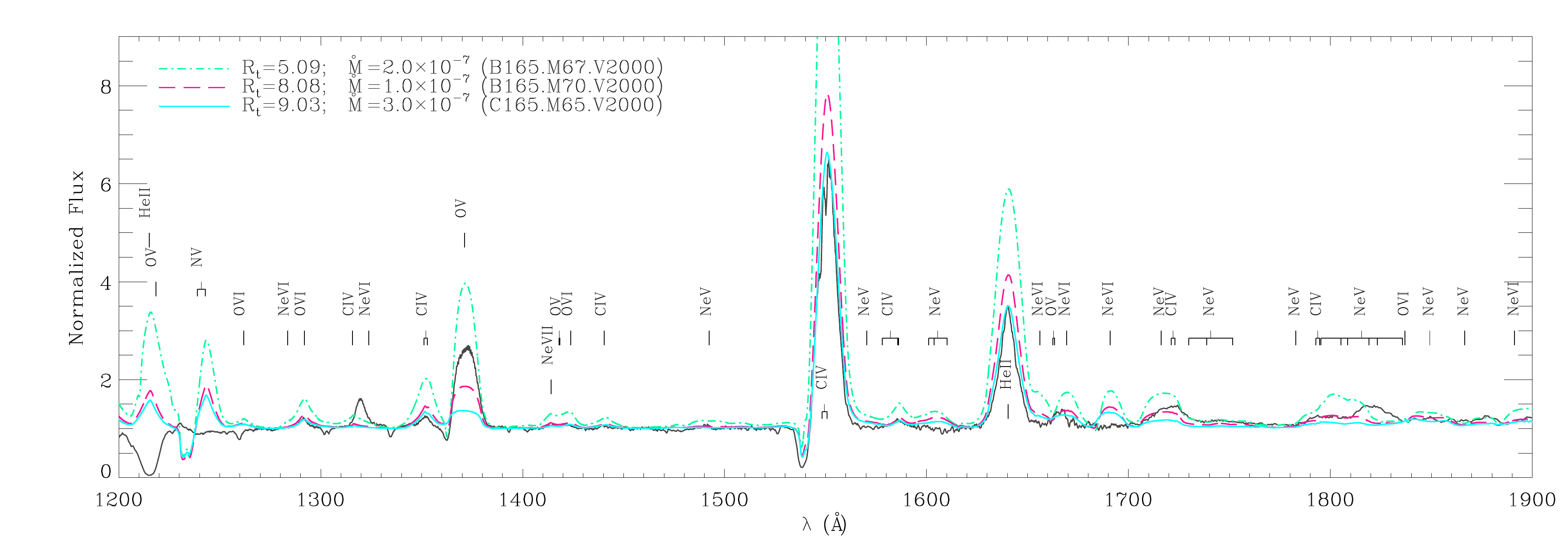}
\includegraphics[scale=0.435]{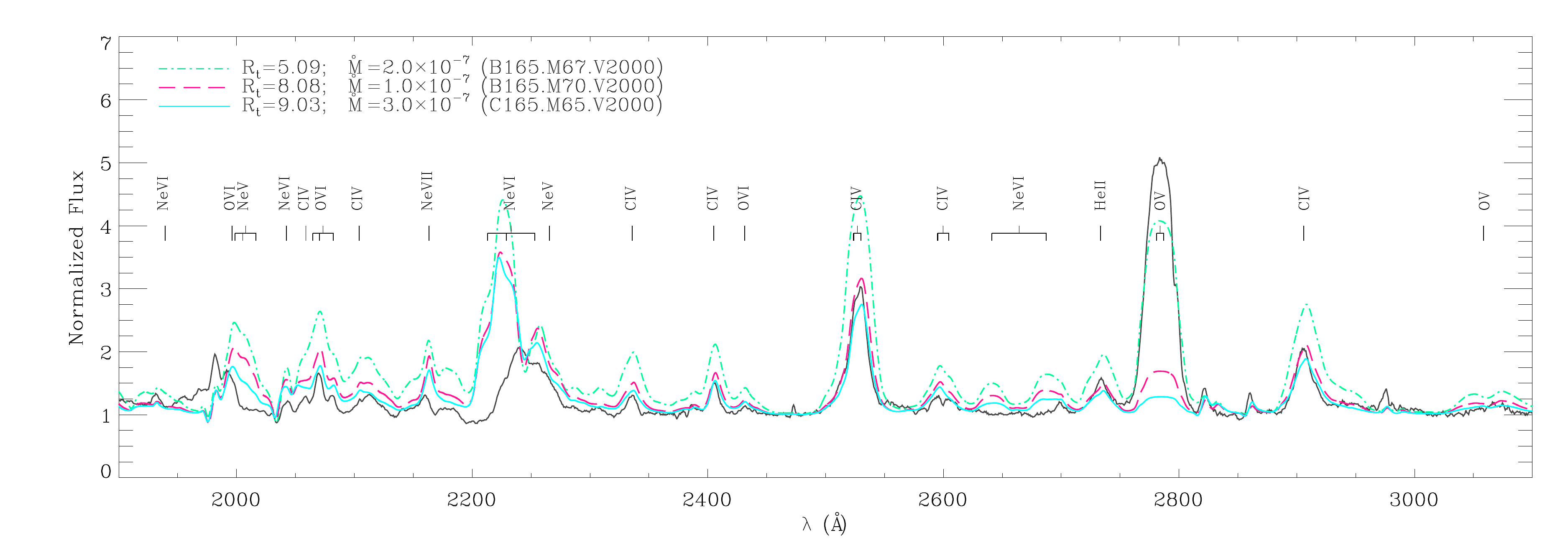}
\caption{ The UV and far-UV observed spectra of the CSPN NGC 6905 (continuous black line, as in Fig.\ref{NGC6905T150}) are shown along with three grid 
models with $T_{\ast}=165$ kK and different values of transformed radius. The discrepancy between the O V lines and all other 
diagnostics is even larger than for $T_{\ast}=150$ kK models (Fig. \ref{NGC6905T150}).}\label{NGC6905T165}
\end{figure}

As a first step, we compare the observed spectra of the central star of NGC 6905 with our grid models with a wind's terminal velocity of $2000$ km s$^{-1}$, the closest value available
in the grid to the one inferred from the C IV $\lambda$ $1548.2$ $\mathrm\AA$ 
P-Cygni absorption edge. The comparison rules out models with $T_{\ast}=125000$ K, which show very weak O VI $\lambda \lambda$ $1031.9$, $1037.6$
$\mathrm{\AA}$, of about half the observed intensity, for all the mass-loss rate values in our grid, besides a too weak Ne VII
$\lambda$ $973.3$ $\mathrm{\AA}$ line, which, as shown in Fig. \ref{Nelines},
at this temperature, shows little sensitivity to mass loss and neon abundance. The carbon, helium and other O VI lines can be 
simultaneously matched by models of this temperature. The O V $\lambda$ $1371.3$ $\mathrm{\AA}$ is too strong in $T_{\ast}=125000$ K models and 
shows a deep absorption component which is not observed. As we will show below, this line is sensitive to the inclusion of heavier ions in 
the models and their abundances and tends to become stronger as new ions are added to the models.
Models with a temperature of $200000$ K can also be
ruled out since, for the mass-loss rate values that allow reasonable fits of the carbon and helium lines, both O V lines are absent and very 
strong unobserved Ne VIII features appear in the synthetic spectra, which also show weak O VI lines. The models with temperatures
of $T_{\ast}=150000$ and $165000$ K better agree with the observed spectra and are shown in Figs. \ref{NGC6905T150} and \ref{NGC6905T165}, 
respectively, for different values of
transformed radius. Among the $T_{\ast}=150000$ K models, the B150.M70.V2000 one with $\dot{M}=10^{-7}$ M$_{\odot}$ yr$^{-1}$, and 
$R_{\mathrm{t}}=10.72$ R$_{\odot}$ fits best all the observed
diagnostics, except for the O V $\lambda \lambda$ $2781.0$, $2787.0$ $\mathrm{\AA}$
and $\lambda$ $1371.3$ $\mathrm{\AA}$ lines, which are only fitted by models of this $T_{\ast}$, by adopting a mass-loss rate twice and 1.3 times higher, 
respectively. Such a higher mass loss,
however, worsens the
fit of all the C IV, He II and O VI lines, except for the O VI
$\lambda \lambda$ $1031.9$, $1037.6$ $\mathrm{\AA}$ doublet, where the intensity of
the second component is improved, as shown in Fig. \ref{NGC6905T150}. The relative strength of this line, however, 
is also very sensitive to clumping, since emission lines are affected by it, but not resonance ones \citep{2007A&A...476.1331O}, as well as to soft X-rays, 
and other factors. Among models with temperature of $T_{\ast}=165000$ K, the
C165.M65.V2000 one, with a mass-loss rate of $\dot{M}=10^{-7}$ M$_{\odot}$ yr$^{-1}$, and a transformed radius of $R_{\mathrm{t}}=9.03$ R$_{\odot}$ 
fits well the observed spectra, again except for the O V lines which are
even more discrepant than in the $T_{\ast}=150000$ K models, as can be seen in Fig. \ref{NGC6905T165}. At any of these two temperatures, 
the fits of the O V lines require lower values of transformed radius, which are incompatible with the other diagnostic lines. This effect is 
worse in the $T_{\ast}=165000$ K models than it is in 
the $T_{\ast}=150000$ K ones, from what we conclude that, among the grid models, the B150.M70.V2000 one gives the best fit.

Below, we extend the analysis beyond the model grid comparison and address in
depth not only the O V lines discrepancy but also other disagreements between the observations
and the best-fitting model, and refine the parameters.
We compare the observations with synthetic spectra calculated assuming different
values of turbulence velocity, show the effect of different neon, oxygen and argon 
abundances, put constraints on the nitrogen and iron abundances, and study the impact of adding heavier
ions in the synthetic spectra, which we find to affect the ionization fractions of
oxygen and to improve the O V lines fit.

\subsection{Extending the analysis}

In Fig. \ref{terminalvelocities}, the O VI $\lambda \lambda$ $1031.9$, $1037.6$ $\mathrm{\AA}$
doublet and the absorption profile of C IV $\lambda \lambda$ $1548.2$, $1550.8$
$\mathrm{\AA}$ are narrower in the $v_{\infty}=2000$ km s$^{-1}$ models than in the observations, suggesting that 
the turbulence velocity in NGC 6905 is higher than that adopted in the calculation of the model grid. Fig. \ref{vturb}
shows that higher values of turbulence velocity, in fact, improve the match of the O VI and C IV profiles.

\begin{figure}
\centering
\includegraphics[scale=0.435]{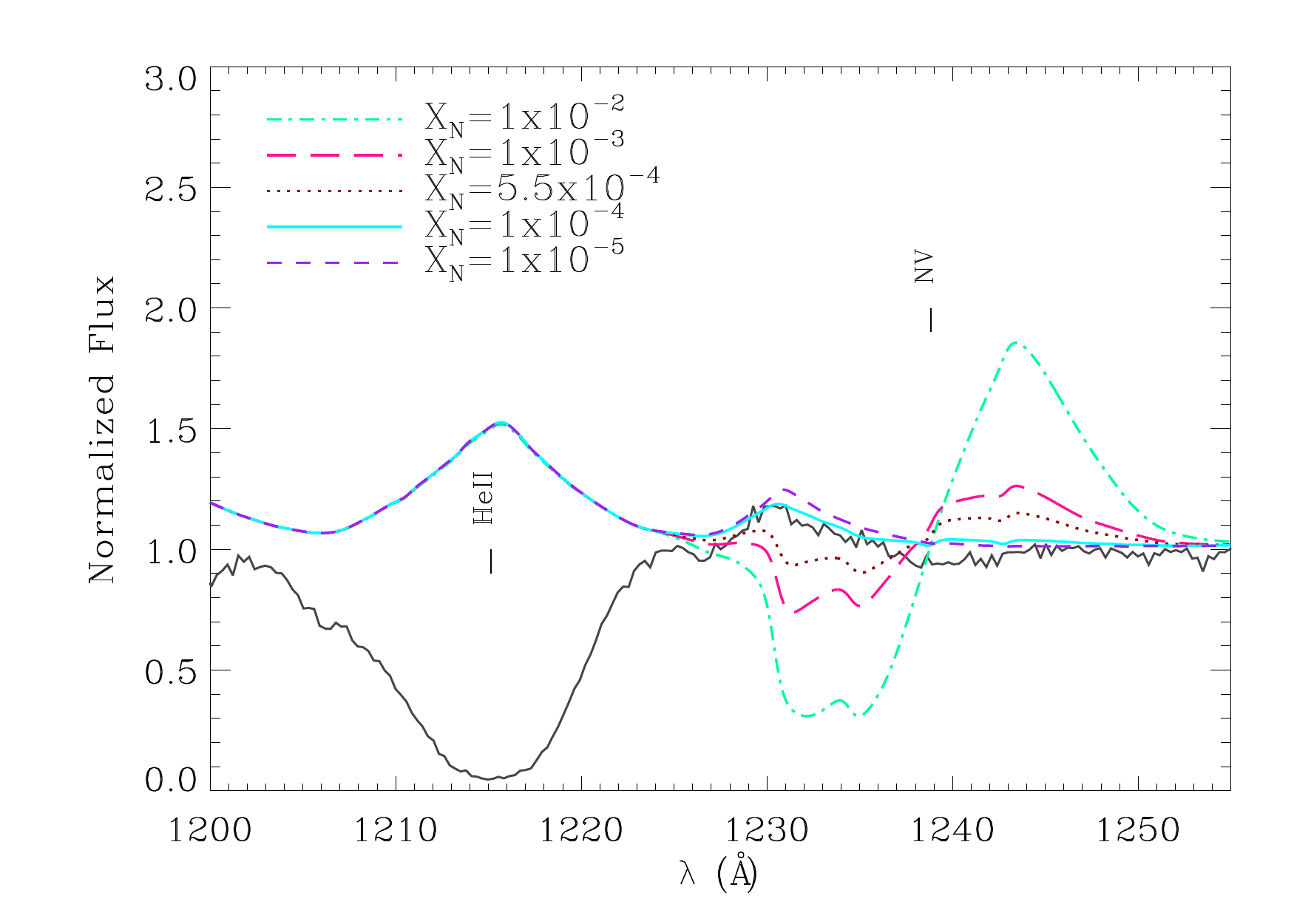}
\caption{ The observed (continuous black line) Ly$_{\alpha}$ interstellar absorption and N V $\lambda \lambda$ $1238.8$, $1242.8$ 
$\mathrm{\AA}$ doublet compared with models with $T_{\ast}=150$ kK,
$\dot{M}=10^{-7}$ M$_{\odot}$ yr$^{-1}$, $v_{\infty}=2000$ km s$^{-1}$ and different
values of nitrogen abundance.}\label{Nitrogen}
\end{figure}

Another difference between the best-fitting grid model and observed spectra is the absence,
in the observed spectra, of N V $\lambda \lambda$ $1238.8$, $1242.8$ $\mathrm{\AA}$ seen in all the models. This is due to a lower nitrogen 
abundance in this star than we assumed for the grid. Fig. \ref{Nitrogen} shows models with different nitrogen abundances and $T_{\ast}=150000$ K,
$\dot{M}=10^{-7}$ M$_{\odot}$ yr$^{-1}$, and $v_{\infty}=2000$ km s$^{-1}$. Nitrogen abundances of $X_{\mathrm{N}}<5.5\times10^{-4}$ better match the 
spectral region shown in Fig. \ref{Nitrogen}, while other spectral lines show no significant differences.

\begin{figure}
\centering
\includegraphics[scale=0.435]{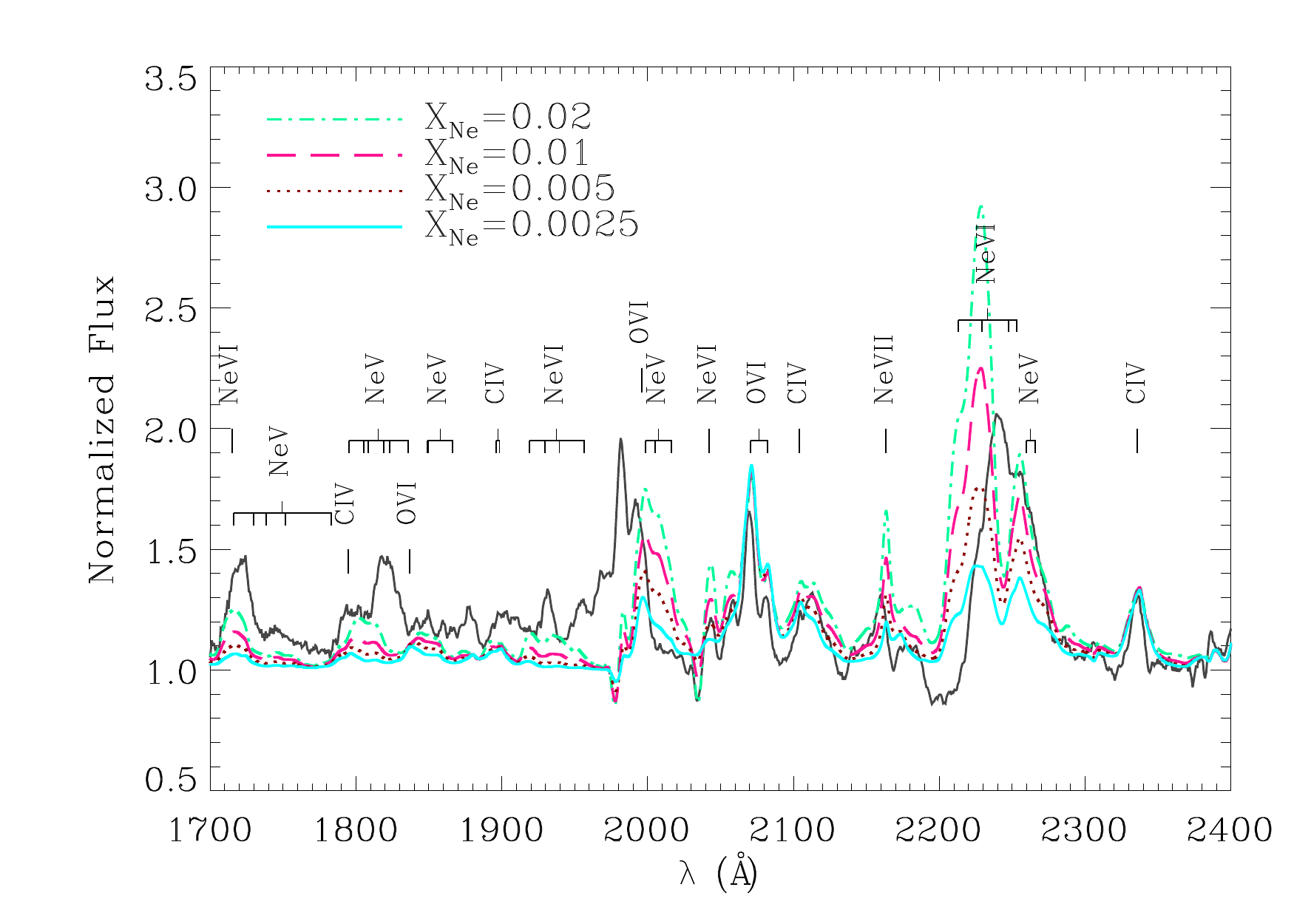}
\includegraphics[scale=0.435]{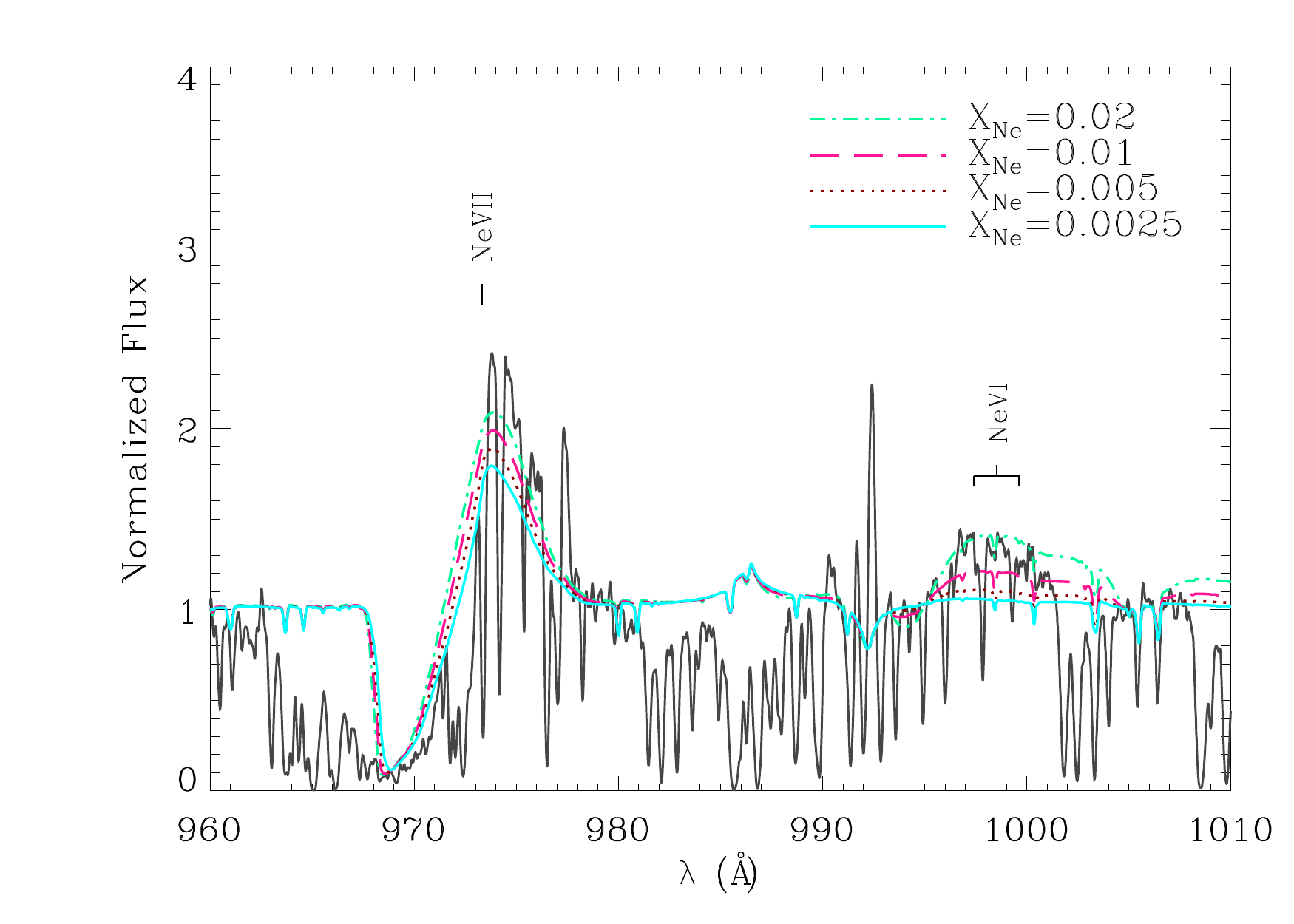}
\includegraphics[scale=0.435]{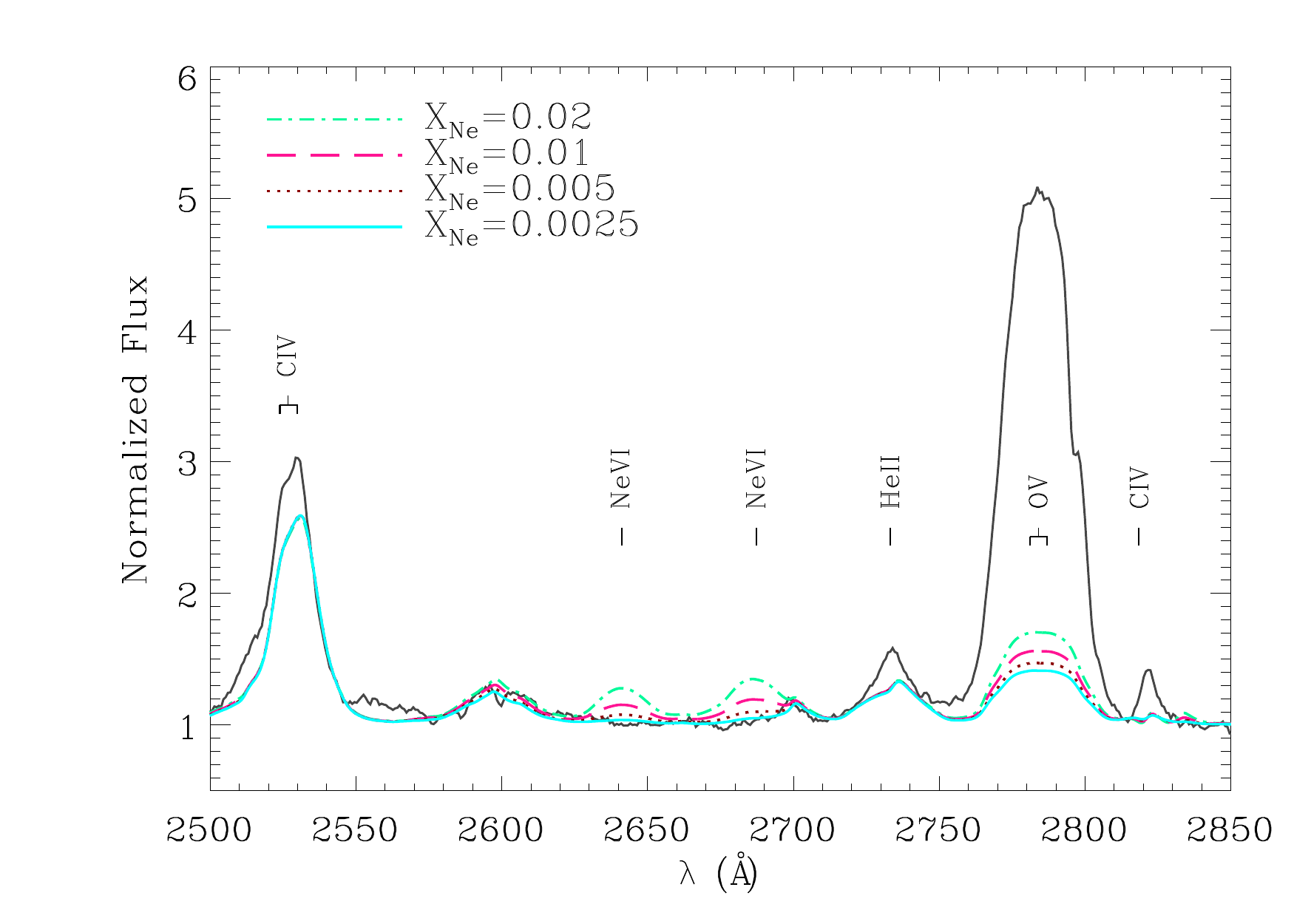}
\includegraphics[scale=0.435]{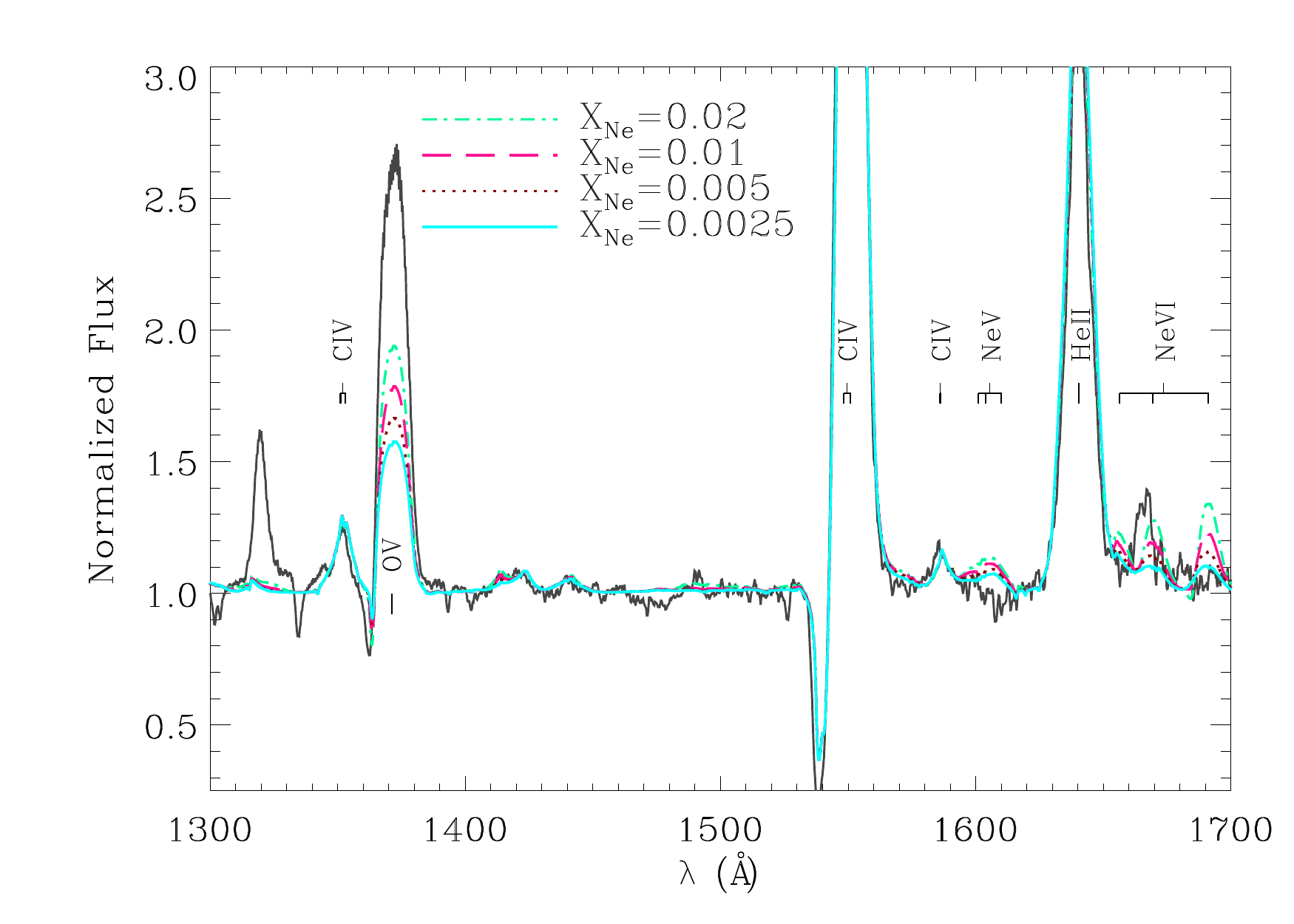}
\includegraphics[scale=0.435]{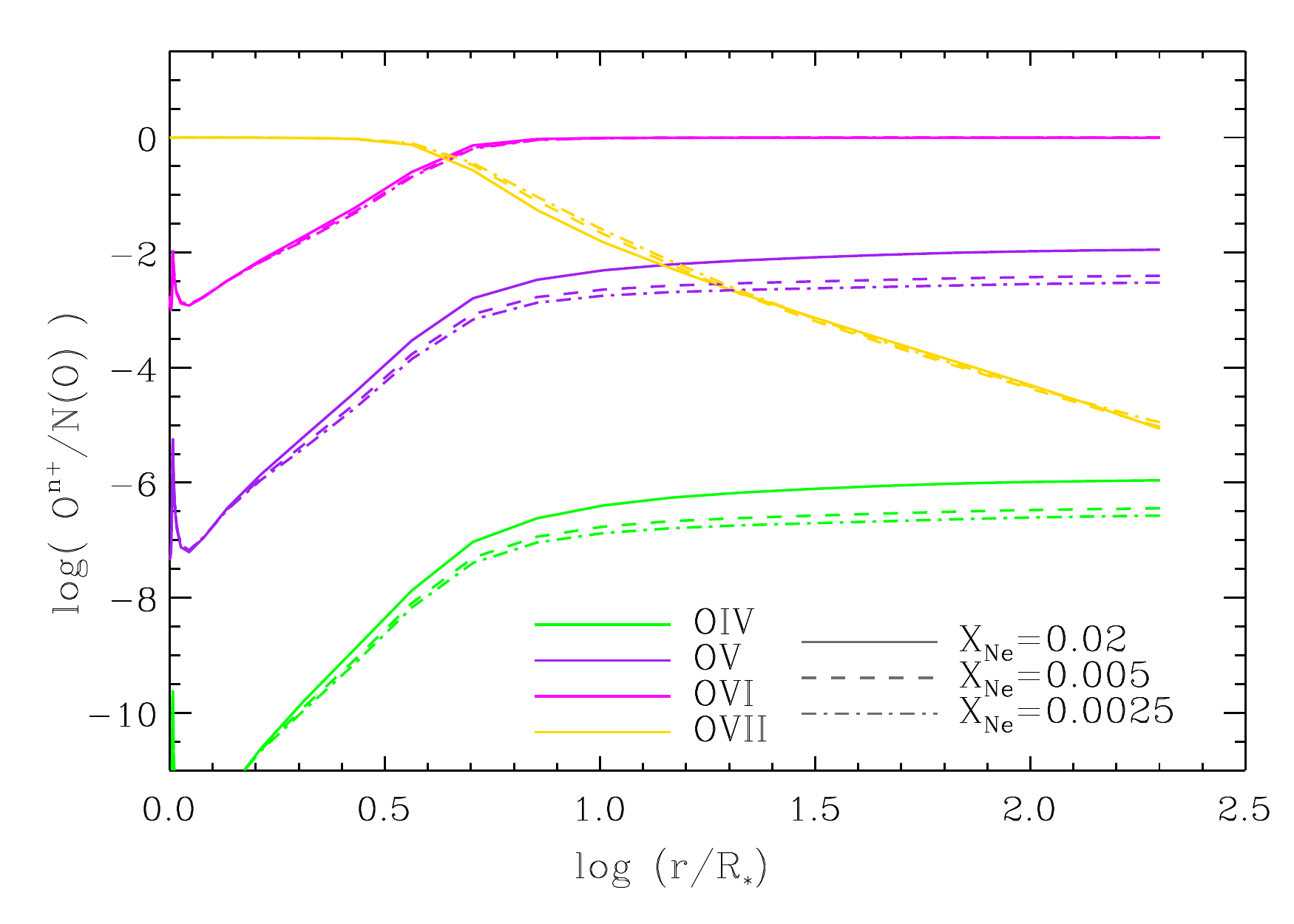}
\caption{ Comparison between NGC 6905 UV and far-UV observed spectra
(continuous black line) and models with $T_{\ast}=150$ kK,
$\dot{M}=10^{-7}$ M$_{\odot}$ yr$^{-1}$, $v_{\infty}=2000$ km s$^{-1}$ and different
values of neon abundance. The bottom panel shows the oxygen ions fractions for the different neon abundances.}\label{NeonAbundance}
\end{figure}

Several neon features which appear in the synthetic spectra are not observed, 
indicating that this object may have a lower Ne abundance. Fig. \ref{NeonAbundance} 
shows models with different neon abundances and $T_{\ast}=150000$ K, 
$\dot{M}=10^{-7}$ M$_{\odot}$ yr$^{-1}$, and $v_{\infty}=2000$ km s$^{-1}$. The neon
abundance in the spectra of the central star of NGC 6905 is difficult
to constrain due to the lack of strong neon lines other than Ne VII $\lambda$ $973.3$ $\mathrm{\AA}$, 
and due to the mismatch between Ne VI and Ne V features present in the models in the region between 1700-2300
$\mathrm{\AA}$ and the observations. In Fig. \ref{NeonAbundance}, we see that
none of the neon abundances adopted improves
the fit of the Ne VI complex $\lambda \lambda$ $2213.1$, $2229.1$ $\mathrm{\AA}$. Also, models with
higher neon abundances produce some features which are not seen in the observed
spectra such as a Ne VII line at
$\lambda$ $2163.4$ $\mathrm{\AA}$ and Ne VI lines at $\lambda \lambda$ $2641.1$, $2687.4$
$\mathrm{\AA}$. On the other hand, the lower neon abundances produce very weak Ne VII
$\lambda$ $973.3$ $\mathrm{\AA}$ and Ne VI $\lambda \lambda$ $997.4$, $999.6$ $\mathrm{\AA}$ lines.
The fact that some diagnostics are improved by a higher neon content
while others are worsened seems to point to an incorrect temperature. We did not use the neon lines 
as a temperature diagnostic as they are, also in the $T_{\ast}=165000$ K models, incompatible with 
other diagnostics. However, we have observed (not shown here) that models with temperatures of $T_{\ast}=200000$ K better reproduce several 
of the neon structures that are incompatible in cooler models. We consider the present analysis inconclusive regarding the neon abundance.
An interesting outcome of the comparison among models with different neon abundances
is a slight enhancement of the O V $\lambda$ $1371.3$ $\mathrm{\AA}$ and
$\lambda \lambda$ $2781.0$, $2787.0$ $\mathrm{\AA}$ lines as neon content in the models increases,
which is the result of an increase in the relative fraction of the O V ion, as
can be seen in the bottom panel of Fig. \ref{NeonAbundance}.   

In an attempt to reproduce the strong O V lines simultaneously with the 
other diagnostics, we investigated the effect of
higher oxygen abundances, which is shown in Fig. \ref{Oabundance}.
For the grid models we assumed an oxygen mass fraction of $X_{\mathrm{O}}=0.08$. 
A higher oxygen abundance increases the intensity of the O V lines in the synthetic spectra, 
but worsens the fit of most of the O VI lines. Besides, even an oxygen mass fraction of 
$X_{\mathrm{O}}=0.20$ is not enough to reproduce the strong observed O V 
$\lambda \lambda$ $2781.0$, $2787.0$ $\mathrm{\AA}$ line. The O VI $\lambda \lambda$ 
$1031.9$, $1037.6$ $\mathrm{\AA}$ doublet does not show
any significant change from the increase in the oxygen abundance.

Because of the effect seen in the O V lines due to an increase in neon abundance,
we also experimented with the inclusion of new ions in the models in an attempt to
reproduce the O V lines observed intensities. A very complete
model including Ni VII, Ni VIII, Ni IX, Co VII, Co VIII, Co IX, Ca VI, Ca VII,
Ca VIII, Ca IX, Ca X, Ar VI, Ar VII, Ar VIII, Si V, Si VI, Mg V, Mg VI, Mg VII,
Na VI, Na VII, Na VIII, Na IX in addition to the ions
already present in the previous model was calculated, assuming solar abundances
for all the additional elements. As a result, the intensity of the O V $\lambda$ $1371.3$ $\mathrm{\AA}$ and 
$\lambda \lambda$ $2781.0$, $2787.0$ $\mathrm{\AA}$ lines in the
synthetic spectra increased as shown in Fig. \ref{newions}, while other major
features which we used to constrain stellar parameters were unaffected. This is a result of the increase of the O V ion fraction. 
In particular, the magnesium ions followed by the sodium ones produced most of the effect, 
while the cobalt and nickel ions cause smaller, but noticeable changes of the O V line strengths.
The rest of the spectra show minor differences, mainly due to calcium and argon lines.

We further explore this model by altering the abundances considered for some elements and show 
the effect on the spectra of different abundances of the elements 
argon and iron. Fig. \ref{Argon} shows the Ar VII $\lambda$ $1063.55$ $\mathrm{\AA}$ line. 
This line was first identified as a 
photospheric absorption by \citet{Werner2007} in CSPNe and WDs, for which they found 
roughly solar abundances, in line with nucleosynthesis calculations for AGB 
stars. We find a model with ten times the solar abundance to better reproduce 
the observed feature in this spectral region, confirming the results of \citet{Herald2009} 
who identified this same line in the FUSE spectra of some of the hottest [WC] CSPNe.
Fig. \ref{Iron} shows the effects of different iron abundances in the synthetic 
spectra. In a recent work, \citet{Werner2010} identified Fe X 
$\lambda \lambda$ $979.3$, $1022.9$ $\mathrm{\AA}$ lines in the FUSE spectra of the hot PG1159 stars RX J2117.1+3412, 
K 1?16, Longmore 4, NGC 246, H1504+65, finding a solar iron abundance for these stars. In NGC 6905, we find that iron abundances higher than 
$0.3$ times the solar one produce some unobserved features, especially an absorption on the Ne VII $\lambda$ $973.3$ $\mathrm{\AA}$ 
P-Cygni profile, and can be ruled out.

\subsubsection{Interstellar absorption}

\begin{figure}
\centering
\includegraphics[angle=0,scale=0.435]{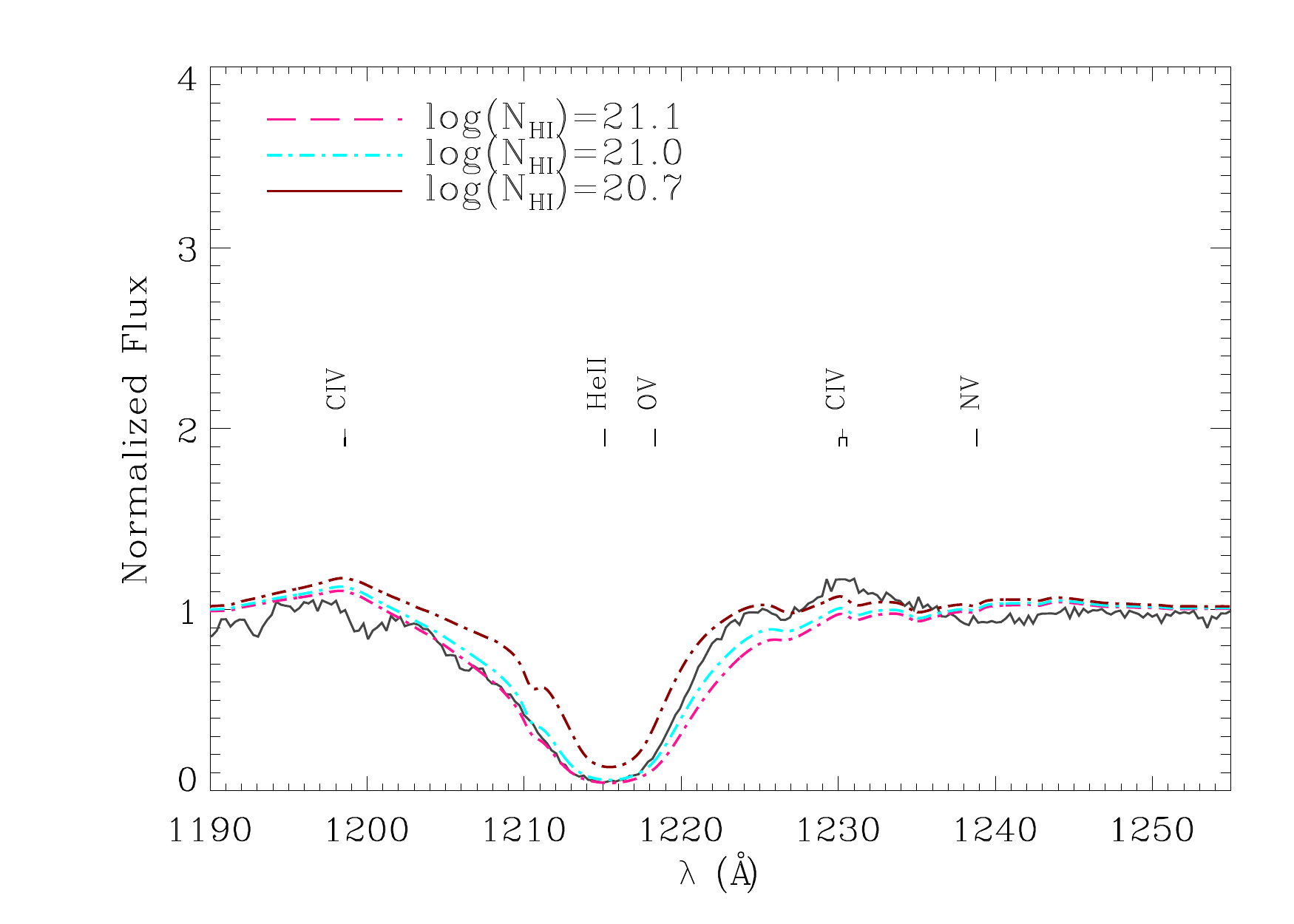}
\caption{The Lyman-$\alpha$ interstellar absorption in the spectrum of NGC 6905 (continuous black line) is compared 
with our best-fitting model shown with the effect of different values of the neutral hydrogen column density.}
\label{Halphaabs}
\end{figure}

\begin{figure}
\centering
\includegraphics[angle=0,scale=0.435]{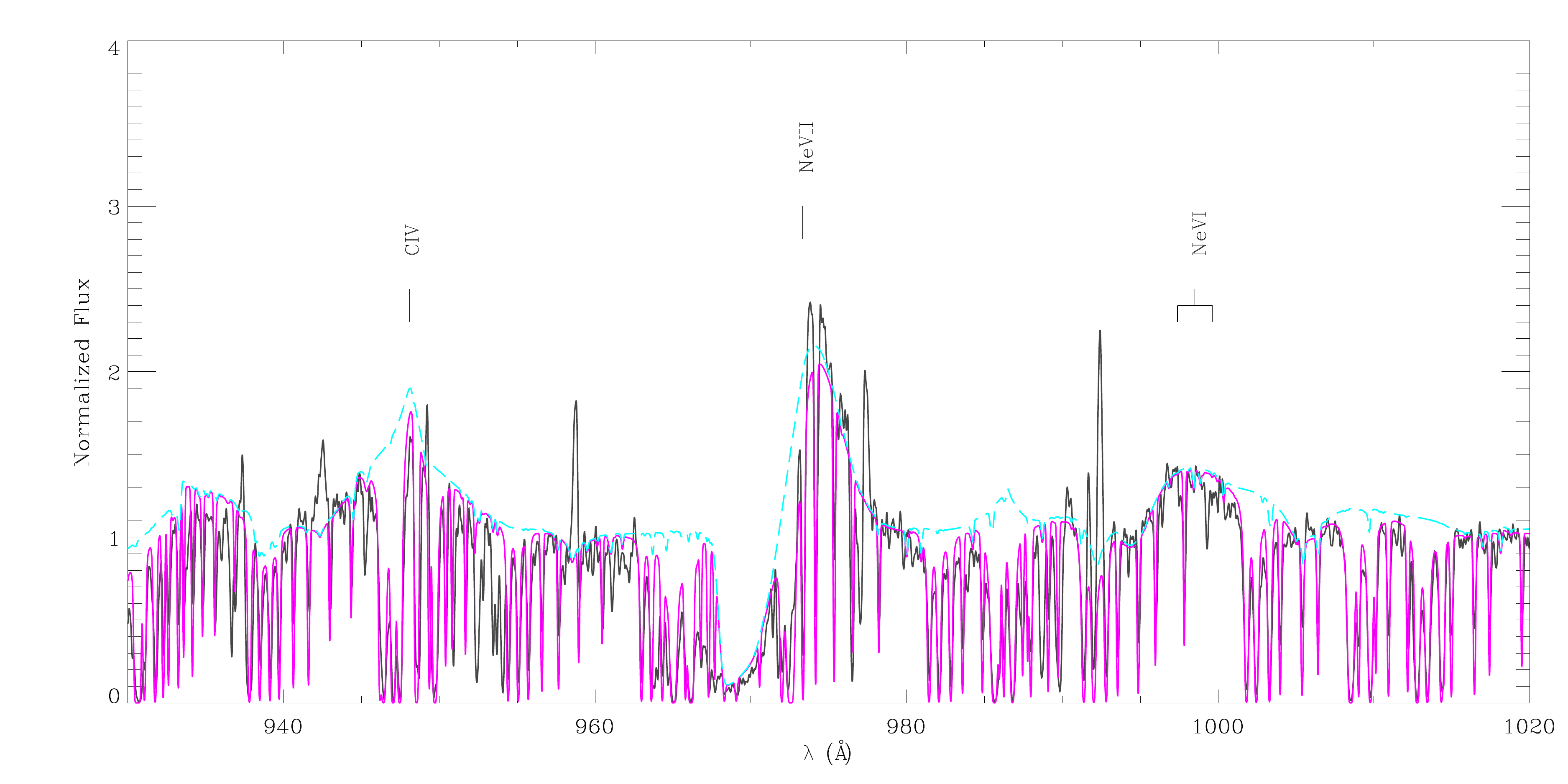}
\includegraphics[angle=0,scale=0.435]{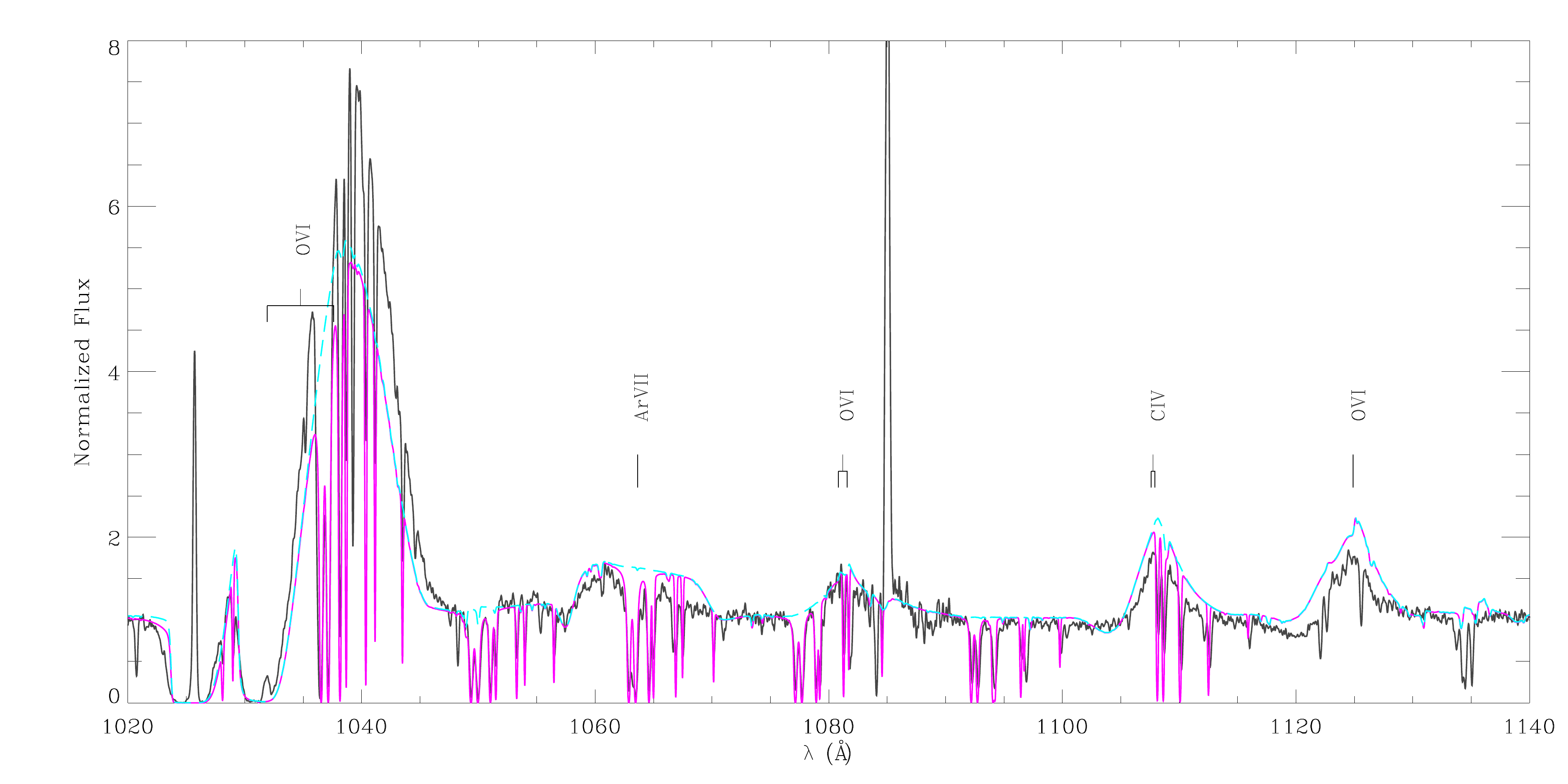}
\caption{The light-blue dashed line is our best-fitting model. In pink, we apply to this model, the effect of absorption by interstellar molecular 
hydrogen with a column density of $logN(H_{2})=19.5$ (where $N$ is given in units of cm$^{-2}$). NGC 6905's far-UV observed spectrum is shown as a continuous black line.}
\label{Habs}
\end{figure}

Figs. \ref{Halphaabs} and \ref{Habs} show our best-fitting model, which includes the additional heavy ions and an argon abundance of ten times the solar value, 
calculated adopting a turbulence velocity $v_{\mathrm{turb}}=150$ km s$^{-1}$. We applied the interstellar absorption due to atomic 
and molecular hydrogen, treated as described in \citet{Herald2002,2004ApJ...609..378H,Herald2004a}, to the synthetic spectra. We assumed that the 
temperature 
of the interstellar gas is 100 K, a typical value for the ISM. From fitting the Ly$_{\alpha}$ $\lambda$ 1215 $\mathrm{\AA}$ H I absorption profile, we derived a neutral hydrogen 
column density of $20.7 < \log N(HI) < 21.1$ (where $N$ is given in units of cm$^{-2}$). Using the numerous H$_{2}$ absorption lines in the FUSE range, we constrained the molecular hydrogen
column density to $19.3 < \log N(H_{2}) < 19.7$ (where $N$ is given in units of cm$^{-2}$). 

\begin{figure}
\centering
\includegraphics[angle=0,scale=0.435]{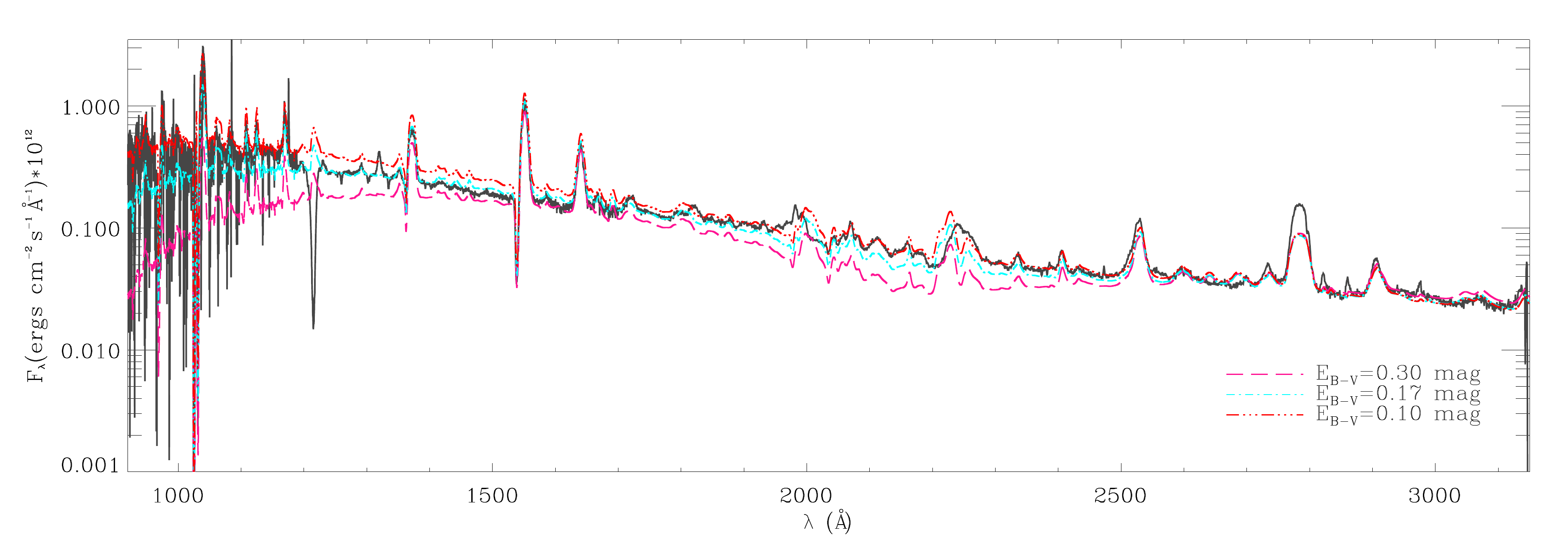}
\caption{NGC 6905's spectra (continuous black line) along with our best-fitting model reddened with 
different values of colour excess, assuming MW-type extinction with $R_{\mathrm{V}}=3.1$.}
\label{Reddening}
\end{figure}

After the stellar parameters have been constrained based on line diagnostics, the observed slope of the spectra was compared 
to the best-fitting model reddened using different values of the colour excess, as shown in Fig. \ref{Reddening}. The reddening 
of \citet{Cardelli1989} with $R_{\mathrm{V}}=3.1$ was adopted. The STIS G140L and G230L spectral regions are well matched 
using a colour excess of $E_{\mathrm{B-V}}=0.17$ mag, while a lower reddening is required in order to fit the FUSE region. Thus, we conclude 
on a colour excess of $E_{\mathrm{B-V}}=0.17 \pm 0.05$ mag. The continuum comparison suggests that the interstellar extinction may be less steep in the 
UV than the extinction curve adopted. Model continuum accuracy has 
not yet been reportedly tested versus observations and may be addressed with future work on an expanded sample.        

According to \citet{Bohlin1978}, the neutral hydrogen column density and the colour excess relate as 
$\langle N(HI)/E_{\mathrm{B-V}} \rangle = 4.8 \times 10^{21}$ atoms cm$^{-2}$ mag$^{-1}$. An $E_{\mathrm{B-V}}=0.17$ mag implies 
$\log N(HI)=20.9$ (where $N$ is given in units of cm$^{-2}$), which is within the range derived by us from the Ly$_{\alpha}$ $\lambda$ 1215 $\mathrm{\AA}$ absorption.

\subsubsection{Results from the analysis}

\begin{figure}
\centering
\includegraphics[angle=0,scale=0.435]{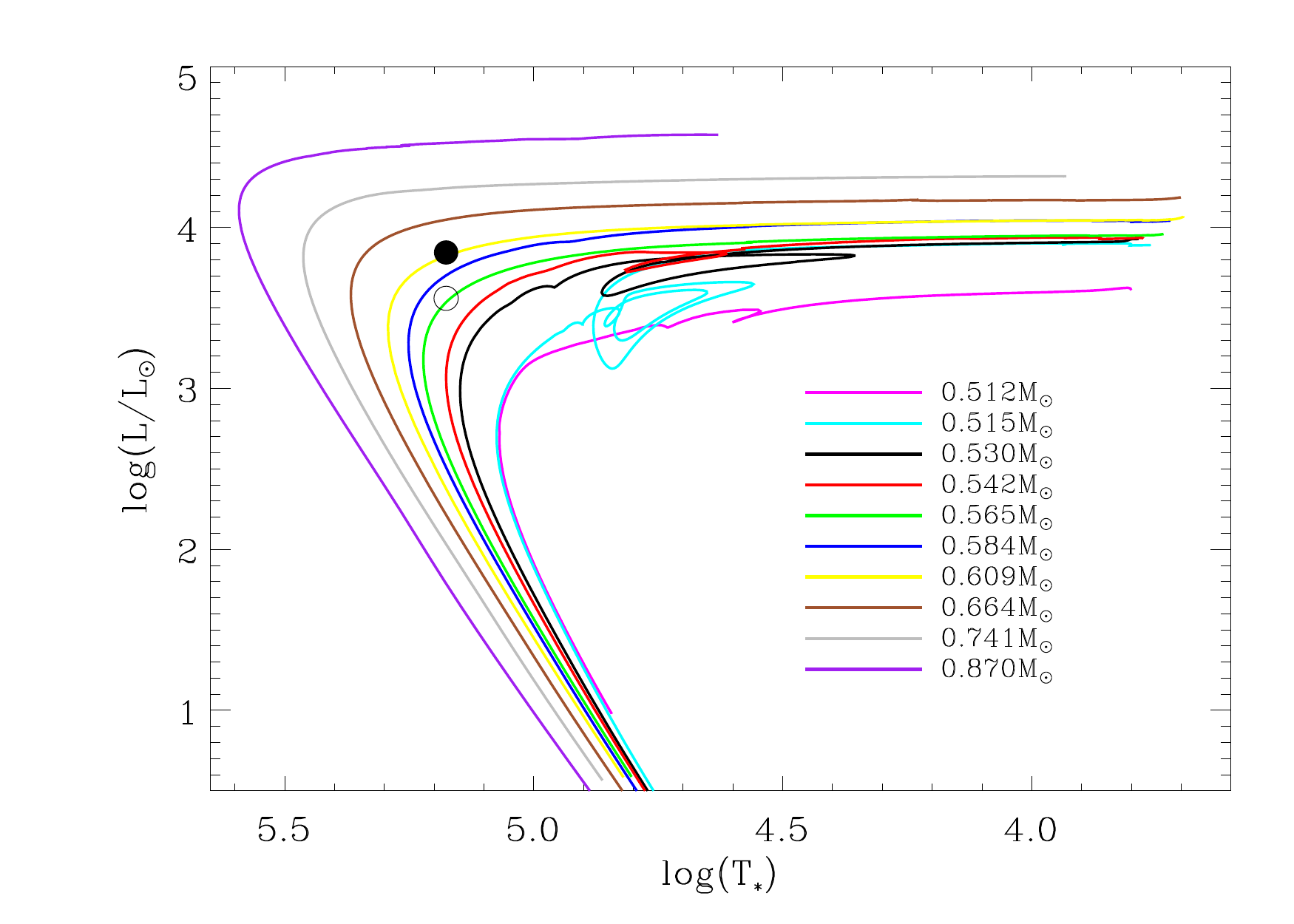}
\caption{HR diagram showing the evolutionary tracks of \citet{2006A&A...454..845M} and the positions of our 
best-fitting grid model to NGC 6905 (which would imply a distance of 2.3 kpc), as a filled circle, and 
of the model rescaled assuming a distance of 1.7 kpc, as an open circle.}
\label{reescaled}
\end{figure}

The parameters of our best-fitting grid model are $T_{\ast}=150000$ K, $\dot{M}=1 \times 10^{-7}$ M$_{\odot}$ yr$^{-1}$, 
$R_{\mathrm{t}}=10.7$ R$_{\odot}$, $R_{\ast}=0.12$ R$_{\odot}$, and $v_{\infty}=2000$ km s$^{-1}$. 
The value of the stellar radius follows the evolutionary track of \citet{2006A&A...454..845M} for $0.6$ M$_{\odot}$ CSPNe 
and a temperature of $150000$ K. Nevertheless, as discussed previously, two models sharing $R_{\mathrm{t}}$, 
$v_{\infty}$, and $T_{\ast}$, but with different values of $R_{\ast}$ and $\dot{M}$ will have similar spectral features. As a 
consequence, different combinations of parameters result in equally good fits to the observations.  
In the literature, the distances derived range between 
1.7 and 2.6 kpc \citep{Maciel1984,Cahn1992,VZ1994,Zhang1995,Stanghellini2008}. If we adopt 
the most recent and also lower value 
for the distance, a radius of $R_{\ast}=0.09$ R$_{\odot}$ is obtained. A model with this radius and $T_{\ast}=150000$ K 
would be in a lower mass evolutionary track \citep{2006A&A...454..845M} of $0.57$ M$_{\odot}$, as seen in Fig. 
\ref{reescaled}, where the filled circle is the grid model calculated by us and the open 
circle is assuming the literature distance of 1.7 kpc. By keeping the transformed 
radius constant, the radius of $R_{\ast}=0.09$ R$_{\odot}$ implies a mass-loss rate of $\dot{M}=6.2 \times 10^{-8}$ M$_{\odot}$ yr$^{-1}$.

The central star of NGC 6905 has been previously studied in the literature through spectral analyses in
optical, ultraviolet and far-ultraviolet regimes. It has been recently
analysed by \citet{2007ApJ...654.1068M}, using low resolution optical spectra, 
low resolution IUE spectra and FUSE spectra in the $1000-1175$ $\mathrm{\AA}$ interval, which excluded 
the Ne VII $\lambda$ $973.3$ $\mathrm{\AA}$ line from the analysis. They also made use of the code CMFGEN to perform the spectral analysis and 
found parameters close to the values determined by us: $T_{\ast}\sim150000$ K, 
$\dot{M}\sim7 \times 10^{-8}$ M$_{\odot}$ yr$^{-1}$, $R_{\mathrm{t}}=10.5$ R$_{\odot}$, and $v_{\infty}=1890$ km s$^{-1}$, 
considering a clumping filling factor of $\textit{f}=0.1$, the same value adopted by us. In our analysis, we 
included many ionic species not considered in theirs, which allowed us to achieve a good fit of the 
O V $\lambda$ $1371.3$ $\mathrm{\AA}$ line and improve the fit of the O V $\lambda \lambda$ $2781.0$, $2787.0$ $\mathrm{\AA}$ line. 
This example proves the usefulness of the grid for constraining the main stellar parameters, and our extended analysis resolved the so far 
inconsistent diagnostic of the O V lines.

\clearpage

\section{Summary and Conclusions}
\label{sec:conclusions}

We have used the state-of-the-art stellar atmosphere code CMFGEN to build a
comprehensive grid of synthetic spectra covering the range of stellar parameters
 appropriate for H-deficient CSPNe of temperatures above 50000 K. The models account for
 line-blanketing, non-LTE expanding atmospheres (where the velocity law is fixed and the
 mass-loss rate is a free parameter), and wind clumping. Many ionic species
neglected in previous models were included (Table \ref{GridIons}). We provide a uniform set
of models that allows systematic analysis of far-UV, UV, optical and IR observed spectra in order to constrain
photospheric and wind parameters. We
examined the far-UV, UV and optical synthetic spectra of the grid and
selected spectral features best suited for constraining values
of mass-loss rate and stellar temperature. 

We also analysed the effect of soft X-rays on the synthetic spectra. We found that, at
$T_{\ast}=165000$ K, an unlikely high X-ray luminosity was necessary in order to cause a
relevant effect in the spectra. At $T_{\ast}=100000$ K, a reasonable value of $L_{\mathrm{X}}$
strongly affects the O VI $\lambda \lambda$
$1031.9$, $1037.6$ $\mathrm{\AA}$. For the $T_{\ast}=65000$ K model, an even lower value of
X-ray luminosity is sufficient to affect the O VI doublet.

As a sample application of the grid, we used it to analyse the UV spectra of the central star of NGC 6905 showing
that the grid can be used to constrain the photospheric and wind parameters to a good extent and to greatly facilitate 
more detailed analysis. 

For NGC 6905, the temperature of our best-fitting model is also a lower limit, since any lower temperature model would produce 
a too strong O V $\lambda$ $1371.3$ $\mathrm{\AA}$ line which, as we have shown, is also affected by the inclusion 
of new ions. A slightly higher temperature for this star improves the fits of the far-UV O VI doublet and Ne VII line, while 
the other diagnostics remain acceptable. 
Thus, we constrain the temperature of the central star of NGC 6905 as being between 150000 and 165000 K, a 10 per cent 
overall uncertainty.

We extended the study beyond
the grid of models by considering
parameters that were not explored in the grid. We found higher values of the turbulence velocity 
to improve the fit of O VI $\lambda \lambda$ $1031.9$, $1037.6$ $\mathrm{\AA}$ and 
C IV $\lambda \lambda$ $1548.2$, $1550.8$ $\mathrm{\AA}$ lines. We showed the effects of different abundances of neon, 
oxygen and argon and we were able to put constraints on the nitrogen and iron abundances, both subsolar. In this object, a 
lower value of nitrogen abundance than the one chosen for the grid was necessary to fit the observed spectra. For all plausible 
temperatures, we were unable to adequately fit all neon diagnostics simultaneously with a single neon abundance.

By adding 
heavier ions to the best-fitting grid model, we were able, for the first time, to reproduce the 
observed intensity of the O V $\lambda$ $1371.3$ $\mathrm{\AA}$ line and to improve the fit of the O V 
$\lambda \lambda$ $2781.0$, $2787.0$ $\mathrm{\AA}$ line.  Finally, by fitting the several H$_{2}$ absorption lines present 
in the FUSE range and the Ly$_{\alpha}$ absorption profile, 
we constrained the molecular and neutral hydrogen column densities and, by comparing the slope of the observed spectra to 
our best-fitting model, we derived the colour excess, which is in agreement with the value expected from the H I column density 
measured.    

The grid of synthetic spectra presented here is available on-line, at
http://dolomiti.pha.jhu.edu/ planetarynebulae.html.
Additional models will be added to the grid in the future, covering
the temperature regime below $50000$ K, as well as new grids
for different surface gravity regimes, suitable for the study of PG1159 
stars that present wind signatures in their spectrum.

\textit{Acknowledgements.} G.R. Keller gratefully acknowledges financial support from the
Brazilian agencies CAPES (which supported her work at the Johns Hopkins University, Department of Physics and Astronomy) and FAPESP, 
grants 0370-09-6 and 06/58240-3. The data presented 
in this paper were obtained from the Multimission Archive at the Space Telescope Science Institute 
(MAST). STScI is operated by the Association of Universities for Research in Astronomy, 
Inc., under NASA contract NAS5-26555. Support for MAST for non-HST data is provided by the NASA 
Office of Space Science via grant NNX09AF08G and by other grants and contracts.

\bibliographystyle{pp_mnras}
\bibliography{clumping3}

\clearpage
\onecolumn

\footnotesize

\clearpage

\clearpage

\clearpage

\begin{table}
\caption{ Track A: stellar parameters of the grid models for CSPNe with 0.5$M_{\odot}$.}
\scriptsize
\begin{tabular}{|c|ccccccc|}
\hline Model & T$_{\ast}$ [kK] & $\log$(g) & R$_{\ast}$ [R$_{\odot}$] &
L$_{\ast}$ [10$^{3}$L$_{\odot}$] &
$\log(\dot{M})$ [M$_{\odot}$yr$^{-1}$]  & v$_{\infty}$ [km s$^{-1}$] & R$_{\mathrm{t}}$ [R$_{\odot}$]\\
\hline
A50.M73.V500 & 50 & 4.4 & 0.75 & 3.1 & -7.3 & 500 & 40.35 \\
A50.M73.V1000 & &&&&& 1000 & 64.04 \\
\cline{6-8}
A50.M67.V500 &&&&& -6.7 & 500 & 16.02 \\
A50.M67.V1000 &&&&&& 1000 & 25.42 \\
\cline{6-8}
A50.M63.V500 &&&&& -6.3 & 500 & 8.70 \\
A50.M63.V1000 &&&&&& 1000 & 13.80 \\
\hline
A65.M73.V500 & 65 & 4.8 & 0.47 & 3.5 & -7.3 & 500 & 25.55 \\
A65.M73.V1000 &&&&&& 1000 & 40.56 \\
\cline{6-8}
A65.M67.V500 &&&&& -6.7 & 500 & 10.14 \\
A65.M67.V1000 &&&&&& 1000 & 16.10 \\
\cline{6-8}
A65.M63.V500 &&&&& -6.3 & 500 & 5.51 \\
A65.M63.V1000 &&&&&& 1000 & 8.74 \\
\hline
A80.M73.V500 & 80 & 5.3 & 0.27 & 2.5 & -7.3 & 500 & 14.36 \\
A80.M73.V1000 &&&&&& 1000 & 22.80 \\
\cline{6-8}
A80.M67.V500 &&&&& -6.7 & 500 & 5.70 \\
A80.M67.V1000 &&&&&& 1000 & 9.05 \\
\cline{6-8}
A80.M63.V500 &&&&& -6.3 & 500 & 3.10 \\
A80.M63.V1000 &&&&&& 1000 & 4.91 \\
\hline
A100.M73.V1500 & 100 & 6.0 & 0.12 & 1.2 & -7.3 & 1500 & 13.30 \\
A100.M73.V2000 &&&&&& 2000 & 16.12 \\
A100.M73.V2500 &&&&&& 2500 & 18.70 \\
\cline{6-8}
A100.M70.V1500 &&&&& -7.0 & 1500 & 8.38 \\
A100.M70.V2000 &&&&&& 2000 & 10.15 \\
A100.M70.V2500 &&&&&& 2500 & 11.78 \\
\cline{6-8}
A100.M67.V1500 &&&&& -6.7 & 1500 & 5.28 \\
A100.M67.V2000 &&&&&& 2000 & 6.40 \\
A100.M67.V2500 &&&&&& 2500 & 7.42 \\
\hline
A125.M73.V1500 & 125 & 6.3 & 0.08 & 1.5 & -7.3 & 1500 & 9.43 \\
A125.M73.V2000 &&&&&& 2000 & 11.43 \\
A125.M73.V2500 &&&&&& 2500 & 13.26 \\
\cline{6-8}
A125.M70.V1500 &&&&& -7.0 & 1500 & 5.94 \\
A125.M70.V2000 &&&&&& 2000 & 7.20 \\
A125.M70.V2500 &&&&&& 2500 & 8.35 \\
\cline{6-8}
A125.M67.V1500 &&&&& -6.7 & 1500 & 3.74 \\
A125.M67.V2000 &&&&&& 2000 & 4.54 \\
A125.M67.V2500 &&&&&& 2500 & 5.26 \\
\hline
\end{tabular}
\label{M05}
\end{table}

\clearpage
\begin{table}
\caption{ Track B stellar parameters of the grid models for CSPNe with 0.6$M_{\odot}$.}
\scriptsize
\begin{tabular}{|c|ccccccc|}
\hline
Model & T$_{\ast}$ [kK] & $\log$(g) & R$_{\ast}$ [R$_{\odot}$] &
L$_{\ast}$ [10$^{3}$L$_{\odot}$] &
$\log(\dot{M})$ [M$_{\odot}$yr$^{-1}$]  & v$_{\infty}$ [km s$^{-1}$] & R$_{\mathrm{t}}$ [R$_{\odot}$]\\
\hline
B50.M70.V500 & 50 & 4.0 & 1.29 & 9.2 & -7.0 & 500 & 43.80 \\
B50.M70.V1000  \tablenotemark{a} &&&&&& 1000 & 69.53 \\
\cline{6-8}
B50.M63.V500 &&&&& -6.3 & 500 & 14.98 \\
B50.M63.V1000  \tablenotemark{a} &&&&&& 1000 & 23.78 \\
\cline{6-8}
B50.M60.V500 &&&&& -6.0 & 500 & 9.45 \\
B50.M60.V1000  \tablenotemark{a} &&&&&& 1000 & 14.99 \\
\hline
B65.M70.V500 & 65 & 4.4 & 0.82 & 10.5 & -7.0 & 500 & 27.86 \\
B65.M70.V1000  \tablenotemark{a} & &&&&& 1000 & 44.23 \\
\cline{6-8}
B65.M63.V500 &&&&& -6.3 & 500 & 9.53 \\
B65.M63.V1000  \tablenotemark{a} &&&&&& 1000 & 15.13 \\
\cline{6-8}
B65.M60.V500 &&&&& -6.0 & 500 & 6.01 \\
B65.M60.V1000 \tablenotemark{a} &&&&&& 1000 & 9.53 \\
\hline
B80.M70.V500 & 80 & 4.8 & 0.52 & 9.6 & -7.0 & 500 & 17.59 \\
B80.M70.V1000 \tablenotemark{a} &&&&&& 1000 & 27.92 \\
\cline{6-8}
B80.M63.V500 &&&&& -6.3 & 500 & 6.02 \\
B80.M63.V1000 \tablenotemark{a} &&&&&& 1000 & 9.55 \\
\cline{6-8}
B80.M60.V500 &&&&& -6.0 & 500 & 3.79 \\
B80.M60.V1000 \tablenotemark{a} &&&&&& 1000 & 6.02 \\
\hline
B100.M70.V1500 & 100 & 5.3 & 0.29 & 7.4 & -7.0 & 1500 & 20.54 \\
B100.M70.V2000 &&&&&& 2000 & 24.88 \\
B100.M70.V2500 \tablenotemark{a} &&&&&& 2500 & 28.87 \\
\cline{6-8}
B100.M67.V1500 &&&&& -6.7 & 1500 & 12.94 \\
B100.M67.V2000 &&&&&& 2000 & 15.67 \\
B100.M67.V2500 \tablenotemark{a} &&&&&& 2500 & 18.19 \\
\cline{6-8}
B100.M65.V1500 &&&&& -6.5 & 1500 & 9.87 \\
B100.M65.V2000 &&&&&& 2000 & 11.96 \\
B100.M65.V2500 \tablenotemark{a} &&&&&& 2500 & 13.88 \\
\hline
B125.M70.V1500 & 125 & 5.7 & 0.18 & 7.2 & -7.0 & 1500 & 12.95 \\
B125.M70.V2000 &&&&&& 2000 & 15.68 \\
B125.M70.V2500 \tablenotemark{a} &&&&&& 2500 & 18.20 \\
\cline{6-8}
B125.M67.V1500 &&&&& -6.7 & 1500 & 8.16 \\
B125.M67.V2000 &&&&&& 2000 & 9.88 \\
B125.M67.V2500 \tablenotemark{a} &&&&&& 2500 & 11.46 \\
\cline{6-8}
B125.M65.V1500 &&&&& -6.5 & 1500 & 6.22 \\
B125.M65.V2000 &&&&&& 2000 & 7.54 \\
B125.M65.V2500 \tablenotemark{a} &&&&&& 2500 & 8.75 \\
\hline
B150.M70.V1500 & 150 & 6.0 & 0.12 & 7.0 & -7.0 & 1500 & 8.85 \\
B150.M70.V2000 &&&&&& 2000 & 10.72 \\
B150.M70.V2500 \tablenotemark{a} &&&&&& 2500 & 12.44 \\
\cline{6-8}
B150.M67.V1500 &&&&& -6.7 & 1500 & 5.58 \\
B150.M67.V2000 &&&&&& 2000 & 6.76 \\
B150.M67.V2500 \tablenotemark{a} &&&&&& 2500 & 7.84 \\
\cline{6-8}
B150.M65.V1500 &&&&& -6.5 & 1500 & 4.26 \\
B150.M65.V2000 &&&&&& 2000 & 5.16 \\
B150.M65.V2500 \tablenotemark{a} &&&&&& 2500 & 5.98 \\
\hline
B165.M70.V1500 & 165 & 6.3 & 0.09 & 5.8 & -7.0 & 1500 & 6.67 \\
B165.M70.V2000 &&&&&& 2000 & 8.08 \\
B165.M70.V2500 \tablenotemark{a} &&&&&& 2500 & 9.37 \\
B165.M70.V3000 &&&&&& 3000 & 10.58 \\
\cline{6-8}
B165.M67.V1500 &&&&& -6.7 & 1500 & 4.20 \\
B165.M67.V2000 &&&&&& 2000 & 5.09 \\
B165.M67.V2500 \tablenotemark{a} &&&&&& 2500 & 5.90 \\
B165.M67.V3000 &&&&&& 3000 & 6.67 \\
\cline{6-8}
B165.M65.V1500 &&&&& -6.5 & 1500 & 3.21 \\
B165.M65.V2000 &&&&&& 2000 & 3.88 \\
B165.M65.V2500 \tablenotemark{a} &&&&&& 2500 & 4.51 \\
B165.M65.V3000 &&&&&& 3000 & 5.09 \\
\hline
B200.M73.V1500 & 200 & 7.0 & 0.04 & 2.4 & -7.3 & 1500 & 4.60 \\
B200.M73.V2000 &&&&&& 2000 & 5.57 \\
B200.M73.V2500 &&&&&& 2500 & 6.47 \\
B200.M73.V3000 &&&&&& 3000 & 7.31 \\
\cline{6-8}
B200.M70.V1500 &&&&& -7.0 & 1500 & 2.90 \\
B200.M70.V2000 &&&&&& 2000 & 3.51 \\
B200.M70.V2500 \tablenotemark{a} &&&&&& 2500 & 4.08 \\
B200.M70.V3000 &&&&&& 3000 & 4.6 \\
\cline{6-8}
B200.M67.V1500 &&&&& -6.7 & 1500 & 1.83 \\
B200.M67.V2000 &&&&&& 2000 & 2.21 \\
B200.M67.V2500 \tablenotemark{a} &&&&&& 2500 & 2.57 \\
B200.M67.V3000 &&&&&& 3000 & 2.90 \\
\cline{6-8}
B200.M65.V1500 &&&&& -6.5 & 1500 & 1.39 \\
B200.M65.V2000 &&&&&& 2000 & 1.69 \\
B200.M65.V2500 \tablenotemark{a} &&&&&& 2500 & 1.96 \\
B200.M65.V3000 &&&&&& 3000 & 2.21 \\
\hline
\end{tabular}
\label{M06}
\tablenotetext{a}{Models with different neon abundances are also available.}
\end{table}

\clearpage

\clearpage
\begin{table}
\caption{ Track C: stellar parameters of the grid models for CSPNe with 0.9$M_{\odot}$.}
\scriptsize
\begin{tabular}{|c|ccccccc|}
\hline Model & T$_{\ast}$ [kK] & $\log$(g) & R$_{\ast}$ [R$_{\odot}$] &
L$_{\ast}$ [10$^{3}$L$_{\odot}$] &
$\log(\dot{M})$ [M$_{\odot}$yr$^{-1}$]  & v$_{\infty}$ [km s$^{-1}$] & R$_{\mathrm{t}}$ [R$_{\odot}$]\\
\hline
C100.M67.V1500 & 100 & 4.8 & 0.62 & 33.5 & -6.7 & 1500 & 27.59 \\
C100.M67.V2000 &&&&&& 2000 & 33.42 \\
C100.M67.V2500 &&&&&& 2500 & 38.79 \\
\cline{6-8}
C100.M65.V1500 &&&&& -6.5 & 1500 & 21.06 \\
C100.M65.V2000 &&&&&& 2000 & 25.51 \\
C100.M65.V2500 &&&&&& 2500 & 29.60 \\
\cline{6-8}
C100.M64.V1500 &&&&& -6.4 & 1500 & 17.38 \\
C100.M64.V2000 &&&&&& 2000 & 21.06 \\
C100.M64.V2500 &&&&&& 2500 & 24.44 \\
\hline
C125.M67.V1500 & 125 & 5.3 & 0.35 & 25.8  & -6.7 & 1500 & 15.47 \\
C125.M67.V2000 &&&&&& 2000 & 18.75 \\
C125.M67.V2500 &&&&&& 2500 & 21.75 \\
\cline{6-8}
C125.M65.V1500 &&&&& -6.5 & 1500 & 11.81 \\
C125.M65.V2000 &&&&&& 2000 & 14.31 \\
C125.M65.V2500 &&&&&& 2500 & 16.60 \\
\cline{6-8}
C125.M64.V1500 &&&&& -6.4 & 1500 & 9.75 \\
C125.M64.V2000 &&&&&& 2000 & 11.81 \\
C125.M64.V2500 &&&&&& 2500 & 13.70 \\
\hline
C165.M67.V1500 & 165 & 5.7 & 0.22 & 31.2 & -6.7 & 1500 & 9.75 \\
C165.M67.V2000 &&&&&& 2000 & 11.82 \\
C165.M67.V2500 &&&&&& 2500 & 13.71 \\
C165.M67.V3000 &&&&&& 3000 & 15.48 \\
\cline{6-8}
C165.M65.V1500 &&&&& -6.5 & 1500 & 7.44 \\
C165.M65.V2000 &&&&&& 2000 & 9.03 \\
C165.M65.V2500 &&&&&& 2500 & 10.46 \\
C165.M65.V3000 &&&&&& 3000 & 11.82 \\
\cline{6-8}
C165.M64.V1500 &&&&& -6.4 & 1500 & 6.14 \\
C165.M64.V2000 &&&&&& 2000 & 7.44 \\
C165.M64.V2500 &&&&&& 2500 & 8.64 \\
C165.M64.V3000 &&&&&& 3000 & 9.75 \\
\hline
C200.M67.V1500 & 200 & 6.0 & 0.15 & 33.8 & -6.7 & 1500 & 6.90 \\
C200.M67.V2000 &&&&&& 2000 & 8.36 \\
C200.M67.V2500 &&&&&& 2500 & 9.71 \\
C200.M67.V3000 &&&&&& 3000 & 10.96 \\
\cline{6-8}
C200.M65.V1500 &&&&& -6.5 & 1500 & 5.27 \\
C200.M65.V2000 &&&&&& 2000 & 6.38 \\
C200.M65.V2500 &&&&&& 2500 & 7.41 \\
C200.M65.V3000 &&&&&& 3000 & 8.36 \\
\cline{6-8}
C200.M64.V1500 &&&&& -6.4 & 1500 & 4.35 \\
C200.M64.V2000 &&&&&& 2000 & 5.27 \\
C200.M64.V2500 &&&&&& 2500 & 6.11 \\
C200.M64.V3000 &&&&&& 3000 & 6.90 \\
\hline
\end{tabular}
\label{M08}
\end{table}

\clearpage
\begin{table}
\caption{Species considered in the calculation of the grid models.}
\begin{tabular}{cc}
\hline 
element & ions  \\
\hline
 He & I, II, III  \\
\hline
 C & II\tablenotemark{a} , III\tablenotemark{a} , IV, V\\
\hline
 N  & II\tablenotemark{a} , III\tablenotemark{a} , IV\tablenotemark{a} , V, VI\\
\hline
  O & II\tablenotemark{a} , III\tablenotemark{a} , IV\tablenotemark{a} , V, VI, VII \\
\hline
Ne & II\tablenotemark{a} , III\tablenotemark{a} , IV\tablenotemark{a} , V, VI, VII, VIII, IX\\
\hline
 Al  & III\tablenotemark{a} , IV\tablenotemark{a} , V\tablenotemark{a}\\
\hline
 Si  & III\tablenotemark{a} , IV, V\tablenotemark{a} , VI\tablenotemark{a} \\
\hline
 P  & IV\tablenotemark{a} , V, VI \\
\hline
 S  & III\tablenotemark{a} , IV\tablenotemark{a} , V\tablenotemark{a} , VI, VII \\
\hline
 Fe & IV\tablenotemark{a} , V\tablenotemark{a} , VI\tablenotemark{a} , VII, VIII, IX, X, XI \\
\hline
\end{tabular}
\label{GridIons}
\tablenotetext{a}{These ionic species are not included in the calculation of all models. They were introduced where needed based on the analysis of the ionization fractions.}
\end{table}

\clearpage

\begin{table}
\begin{center}
\caption{ Spectra of NGC 6905's central star used in the analysis.\label{spectra}}
\begin{tabular}{ccccccc}
\hline
Instrument & Data Set & Date & Resolution [$\mathrm{\AA}$] & Aperture [arcsec] & Range [$\mathrm{\AA}$] \\
\hline
FUSE & A1490202000 & 2000 Aug 11 & $\sim$0.06 & 30$\times$30 & 905-1187 \\
STIS + G140L & O52R01020 & 1999 Jun 29 & $\sim$1.20 & 52$\times$0.5 & 1150-1736 \\
STIS + G230L & O52R01010 & 1999 Jun 29 & $\sim$3.15 & 52$\times$0.5 & 1570-3180 \\
\hline
\end{tabular}
\end{center}
\end{table}

\end{document}